\documentclass[12pt,reqno]{amsart}
\usepackage{mathtools,amssymb,amsfonts,latexsym,amsthm,enumerate,mathabx}
\usepackage[T1]{fontenc}
\usepackage{lmodern}

\usepackage{xfrac}
\usepackage{mathrsfs}
\DeclareFontFamily{U}{rsfs}{\skewchar\font127 }
\DeclareFontShape{U}{rsfs}{m}{n}{%
   <-6> rsfs5
   <6-8> rsfs7
   <8-> rsfs10
}{}

\usepackage{psfrag}
\usepackage{graphicx}
\usepackage[font=footnotesize]{caption}
\usepackage[font+=scriptsize]{subcaption}

\usepackage[usenames,dvipsnames]{xcolor}
\usepackage[pdftitle={On the asymptotics of dimers on tori},pdfauthor={Richard W. Kenyon, Nike Sun, and David B. Wilson},colorlinks=true,linkcolor=NavyBlue,urlcolor=RoyalBlue,citecolor=PineGreen,bookmarks=true,bookmarksopen=true,bookmarksopenlevel=3,unicode]{hyperref}
\usepackage{bm}

\usepackage[normalem]{ulem}
\usepackage{pkg}

\newcommand{\batemanii}{MR698780}
\newcommand{\grafakos}{MR2445437}
\newcommand{\hildebrand}{MR895822}
\newcommand{\kast}{MR0253689}
\newcommand{\conwaysloane}{MR1662447}
\newcommand{\ckp}{MR1815214}
\newcommand{\kastquadratic}{Kasteleyn19611209}
\newcommand{\kenyonaihp}{MR1473567}
\newcommand{\kos}{MR2215138}
\newcommand{\ko}{MR2219249}
\newcommand{\ksdimers}{MR2099145}
\newcommand{\miller}{MR0110193}

\newcommand{\pf}{\mathrm{Pf}\,}

\DeclareMathOperator{\wind}{\text{\textup{\textsf{wind}}}}

\newcommand{\match}{\mathfrak{m}}

\newcommand{\free}{\mathbf{f}}
\newcommand{\freezero}{\free_0}
\newcommand{\freesrf}{\free_1}
\newcommand{\freecorner}{\free\hspace{-1pt}\lrcorner}

\newcommand{\lat}{\L}
\newcommand{\avg}{\mathbf{A}}
\newcommand{\avgint}{\mathbf{I}}

\newcommand{\hess}{H}
\newcommand{\taux}{\tau_{\mathrm{re}}}
\newcommand{\tauy}{\tau_{\mathrm{im}}}
\newcommand{\q}{q}
\newcommand{\qtau}{q_\tau}
\newcommand{\var}{\nu}
\newcommand{\cts}{{\text{\textup{cts}}}}

\newcommand{\old}[1]{}

\makeatletter
 \newcommand{\hypertop}[1]{\Hy@raisedlink{\hypertarget{#1}{}}}
\makeatother

\newcommand{\fscone}{\protect\hyperlink{FSC_1}{\text{\textup{\textsf{FSC}}}_1}}
\newcommand{\fsctwo}{\protect\hyperlink{FSC_2}{\text{\textup{\textsf{FSC}}}_2}}
\newcommand{\fscodd}{\protect\hyperlink{FSC_3}{\text{\textup{\textsf{FSC}}}_3}}
\newcommand{\logfscone}{\protect\hyperlink{FSC_1}{\text{\textup{\textsf{fsc}}}_1}}
\newcommand{\logfsctwo}{\protect\hyperlink{FSC_2}{\text{\textup{\textsf{fsc}}}_2}}
\newcommand{\logfscodd}{\protect\hyperlink{FSC_3}{\text{\textup{\textsf{fsc}}}_3}}
\newcommand{\cf}{\protect\hyperlink{e:cf}{\bm{\Xi}}}
\renewcommand{\vth}{\protect\hyperlink{Jacobi_theta}{\vartheta}}
\newcommand{\scrF}{\protect\hyperlink{script-R-F}{\mathscr{F}}}
\newcommand{\scrR}{\protect\hyperlink{script-R-F}{\mathscr{R}}}
\newcommand{\deta}{\protect\hyperlink{Dedekind}{\eta}}

\newcommand{\zz}{\text{\itshape\sffamily Z}}

\newcommand{\freezeroIsing}{{}^\ddagger\free_0}
\newcommand{\freesrfIsing}{{}^\ddagger\free_1}
\newcommand{\freecornerIsing}{{}^\ddagger\free\hspace{-1pt}\lrcorner}
\newcommand{\logfscIsing}{{}^\ddagger\logfsc}

\newcommand{\zzIsing}{{}^\ddagger\hspace{-2pt}{\text{\itshape\sffamily Z}}}

\newcommand{\PIsing}{{}^\ddagger{\P}}

\newcommand{\zzE}[1][]{\zz_E^{\,#1}}
\newcommand{\ZZ}{\mathcal{Z}}
\newcommand{\ZZE}[1][]{\ZZ_E^{\,#1}}

\newcommand{\End}{{\text{End}_+(\Z^2)}}
\newcommand{\SL}{{{\operatorname{SL}}_2\Z}}
\newcommand{\Graph}{\mathcal{G}}
\newcommand{\Catalan}{G}
\newcommand{\Edges}{\mathcal{E}}
\newcommand{\Vertices}{\mathcal{V}}
\newcommand{\giv}{|}
\newcommand{\Signs}{\bm{S}}

\newcommand{\TS}{\textstyle}
\newcommand{\DS}{\displaystyle}
\newcommand{\tf}[2]{\smash{\tfrac{#1}{#2}}}
\newcommand{\slf}[2]{\smash{\sfrac{#1}{#2}}}
\newcommand{\ttr}{\text{\textup{\texttt{r}}}}
\newcommand{\tts}{\text{\textup{\texttt{s}}}}
\newcommand{\zro}{\text{\textup{\texttt{0}}}}
\newcommand{\one}{\text{\textup{\texttt{1}}}}

\newcommand{\windh}{\text{\textup{\textsf{w}}}_\mathrm{h}}
\newcommand{\windv}{\text{\textup{\textsf{w}}}_\mathrm{v}}

\newcommand{\ellh}{\hyperlink{ellhv}{\ell}_{\mathrm{h}}}
\newcommand{\ellv}{\hyperlink{ellhv}{\ell}_{\mathrm{v}}}
\newcommand{\ellhv}{\hyperlink{ELLhv}{\bm{\ell}}}
\newcommand{\meanvec}{\hyperlink{MuSigma}{\bm{\mu}_E}}
\newcommand{\covmatz}{\bm{\Si}_0}
\newcommand{\covmat}{\hyperlink{MuSigma}{\bm{\Si}_E}}
\newcommand{\covmatinv}{\hyperlink{MuSigma}{\bm{\Si}_{\mathrlap{E}}}^{-1}}
\newcommand{\offset}{\operatorname{\bf{off}}}

\newcommand{\logfsc}{\text{\textsf{fsc}}}
\newcommand{\phaseh}{\ze}
\newcommand{\phasev}{\xi}

\newcommand{\matchp}{\mathfrak{n}}

\newcommand{\param}{{\bm{\al}}}

\newcommand{\paraminv}{{\bm{\theta}}}
\newcommand{\pzro}{{\bm{0}}}
\newcommand{\paramx}{\al_\mathrm{h}}
\newcommand{\paramy}{\al_\mathrm{v}}

\newcommand{\Ronkin}{\hyperlink{Ronkin}{\bm{R}}}

\newcommand{\pdah}{\pd_{\param_\mathrm{h}}}
\newcommand{\pdav}{\pd_{\param_\mathrm{v}}}

\newcommand{\As}{\bm{a}_z}
\newcommand{\Bst}{\bm{b}}
\newcommand{\At}{\bm{a}_w}
\newcommand{\Disc}{\bm{d}}
\newcommand{\gridlat}{\bm{L}}
\newcommand{\pskew}{p_\natural}
\newcommand{\prect}{p_\square}
\newcommand{\avgprect}{p^\mathrm{avg}_\square}
\newcommand{\intprect}{p^\mathrm{int}_\square}
\newcommand{\Pskew}{P_\natural}
\newcommand{\Prect}{P_\square}
\newcommand{\phimin}{\phi_\star}
\newcommand{\psimin}{\psi_\star}

\newcommand{\mx}[1]{\begin{pmatrix}#1\end{pmatrix}}

\newcommand{\usi}{\underline{\smash{\sigma}}}

\newcommand{\refmatch}{\overline{\match}}

\newcommand{\AAz}{A_z}
\newcommand{\AAw}{A_w}

\begin{document}

\title[On the asymptotics of dimers on tori]{On the asymptotics of dimers on tori}{}
\author[R.~\!W.\ Kenyon]{Richard W.\ Kenyon}
\author[N.\ Sun]{Nike Sun}
\author[D.\ B.~\!Wilson]{David B.~\!Wilson}

\thanks{Research of R.K.\ supported by NSF grant DMS-1208191 and the Simons Foundation. Research of N.S.\ supported by 
a Department of Defense NDSEG Fellowship.}

\address{Department of Mathematics, Brown University
\newline\indent 151 Thayer Street, Providence, Rhode Island 02912}
\address{Department of Statistics, Stanford University
\newline\indent Sequoia Hall, 390 Serra Mall, Stanford, California 94305}
\address{Microsoft Research
\newline\indent One Microsoft Way, Redmond, Washington 98052}
\subjclass[2010]{82B20}

\begin{abstract}
We study asymptotics of the dimer model on large toric graphs. Let $\lat$ be a weighted $\Z^2$-periodic planar graph, and let $\Z^2 E$
be a large-index sublattice of~$\Z^2$. For $\lat$ bipartite we show that the dimer partition function $\zz_E$ on the quotient $\lat/(\Z^2 E)$ has the asymptotic expansion
\[\zz=\exp\{A\,\freezero+\logfsc+o(1)\}\]
where $A$ is the area of $\lat/(\Z^2 E)$, $\freezero$ is the free energy density in the bulk, and $\logfsc$ is a finite-size correction term depending only on the conformal shape of the domain together with some parity-type information. Assuming a conjectural condition on the zero locus of the dimer characteristic polynomial, we show that an analogous expansion holds for $\lat$ non-bipartite. The functional form of the finite-size correction differs between the two classes, but is universal within each class. Our calculations yield new information concerning the distribution of the number of loops winding around the torus in the associated double-dimer models.
\end{abstract}

\maketitle

\section{Introduction}
\label{sec:intro}

Dimer systems have been studied since the 1960s when they were introduced to model close-packed diatomic molecules, and research on them has flourished with a renewed vigor since the 1990s (see e.g.~\cite{MR2198850}).

A \emph{dimer configuration\/} on a graph
$\Graph=(\Vertices,\Edges)$
is a perfect matching on~$\Graph$:
that is, a subset of edges $\match\subseteq\Edges$
such that every vertex $v\in \Vertices$ is covered by exactly one edge of $\match$;
for this reason $\match$ is also referred to as a \emph{dimer cover}.
If $\Graph$ is a finite undirected graph equipped with non-negative edge weights
$(\nu_e)_{e\in\Edges}$, a probability measure on dimer covers is given by
\[
\P_\Graph(\match)
\equiv \f{\nu_\Graph(\match)}{\zz_\Graph},\quad\text{with }
	\nu_\Graph(\match)\equiv\prod_{e\in\match}\nu_e \text{ and }
	\zz_\Graph \equiv\sum_{\match} \nu_\Graph(\match).\]
The non-normalized measure $\nu_\Graph$ is the \emph{dimer measure\/} on the $\nu$-weighted graph $\Graph$. The normalizing constant $\zz_\Graph$ is the associated \emph{dimer partition function}, with $\log\zz_\Graph$ the \emph{free energy\/} and $|\Vertices|^{-1}\log\zz_\Graph$ (free energy per vertex) the \emph{free energy density}.

An ordered pair of independent dimer configurations gives (by superposition) a \emph{double-dimer configuration}, consisting of even-length loops and doubled edges. The double-dimer partition function is $\ZZ_\Graph=(\zz_\Graph)^2$. Double-dimer configurations on planar graphs are closely related to the Gaussian free field \cite{MR1872739,arXiv:1105.4158}.

\subsection{Square lattice dimer partition function}

Kasteleyn, Temperley, and Fisher \cite{\kastquadratic,MR0136398,MR0136399} showed how to compute the dimer partition function $\zz_\Graph$
on a finite planar graph $\Graph$ as the Pfaffian of a certain signed adjacency matrix, now known as the Kasteleyn matrix.
For graphs embedded on
a torus or other low-genus surface,
$\zz_\Graph$ can be computed by combining a small number of Pfaffians \cite{\kastquadratic,galluccio-loebl,MR1750896}; we provide further background in \S\ref{ss:kast}.
Using this method,
Kasteleyn \cite{\kastquadratic} showed that on the unweighted square lattice,
both the $m\times n$ rectangle
and $m\times n$ torus
have asymptotic free energy density
\[\freezero\equiv
	\lim_{\substack{m,n\to\infty,\\mn\text{ even}}} (mn)^{-1} \log \zz_\Graph
	= \Catalan/\pi,\]
where $\smash{\Catalan\equiv\sum_{j\ge0} (-1)^j/(2j+1)^2= 0.915965594\ldots}$
is Catalan's constant.
(If $mn$ is odd, clearly $\zz_\Graph=0$.)
In the case of $m$ and $n$ both even, Fisher \cite{MR0136399} calculated the free energy of the $m\times n$ rectangle to be given more precisely by
\[\begin{array}{rl}
\log\zz \hspace{-6pt}&= mn\,\freezero - 2(m+n)\,\freesrf + O(1),\quad\text{with}\\
\freesrf \hspace{-6pt} &= \tf14\log(1+2^{1/2})
		-\tf{1}{2}\Catalan/\pi
\end{array}\]
--- the second term in the expansion of $\log\zz$
is linear in the rectangle perimeter,
so we interpret $\freesrf$ as the surface free energy density
while $\freezero$ is the bulk free energy density.

\begin{figure}[htbp]
\begin{center}
\psfrag{evenxeven}{$m$ even, $n$ even}
\includegraphics[width=.6\textwidth]{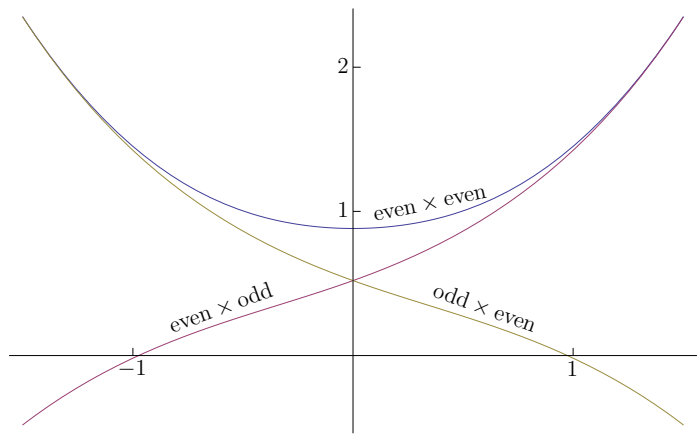}
\end{center}
\caption{\emph{Unweighted square lattice dimers.}
Finite-size corrections $\logfsc$ for
	rectilinear $m\times n$ tori,
shown as a function of logarithmic aspect ratio
	$\log(n/m)$.
Curves are labeled according to parity of $(m,n)$.
}
\label{f:torus-square-fsc}
\end{figure}

Ferdinand \cite{ferdinand} refined the calculation further
	for both rectangle and torus,
	finding a constant-order correction term
	which depends on both the ``shape'' of the region
	(the choice of rectangle or torus boundary conditions,
	as well as the aspect ratio $\smash{\tf{m}{n}}$)
	as well as the parities of $m$ and $n$.
For $mn$ even, Ferdinand found
\beq\label{e:ferdinand.unweighted}
\begin{array}{l}
\log\zz=
	mn\,\freezero
	+ \text{(perimeter)}\,\freesrf
	+ \text{(corners)}\,\freecorner
	+ \logfsc^\mathrm{topology}_{(-1)^{m+n}}(\tf{n}{m})
	+o(1) \\
=\begin{cases}
mn\,\freezero + \logfsc^\mathrm{tor}_{(-1)^{m+n}}
	(\tf{n}{m}) + o(1),
	& \text{($m\times n$ torus);}\\
mn\,\freezero + 2(m+n)\,\freesrf
	+4 \freecorner
	+\logfsc^\mathrm{rect}_{(-1)^{m+n}}(\tf{n}{m})+o(1)
	& \text{($m\times n$ rectangle)}
\end{cases}
\end{array}
\eeq
where $\freecorner$ is a constant which may be interpreted as the free energy per corner,
and the four functions
$\logfsc^\mathrm{tor}_{\pm1},\logfsc^\mathrm{rect}_{\pm1}$
are explicit analytic functions of the aspect ratio $\slf{n}{m}$.
These functions $\logfsc$ are called the \emph{finite-size corrections\/} to the free energy: they contain information about long-range properties of the dimer system (see e.g.~\cite{PhysRevLett.56.742,privman1990finite,cardy1996scaling}).
Figure~\ref{f:torus-square-fsc}
shows these finite-size corrections for the $m\times n$ torus.
We shall see (Figure~\ref{f:torus.square.fsc.seven}) that if we expand our consideration slightly to all \emph{near-rectilinear\/} tori --- tori which are rotated with respect to the coordinate axis, or which deviate slightly from being perfectly rectangular --- then in fact \emph{seven\/} $\logfsc$ curves arise in the limit.
	
Kasteleyn, Fisher, and Ferdinand
also carried out these calculations for the \emph{weighted\/}
square lattice where the horizontal edges receive weight $a$
while the vertical edges receive weight $b$.
In this setting they found (for $mn$ even)
\beq\label{e:ferdinand.weighted}
\log\zz
=mn\,\freezero|_{a,b}
	+(\text{perimeter})\,\freesrf|_{a,b}
	+(\text{corners})\,\freecorner|_{a,b}
	+\logfsc^\mathrm{topology}_{(-1)^{m+n}}(\tf{n b}{m a})
	+o(1)\eeq
where the free energy coefficients
$\freezero,\freesrf,\freecorner$
depend on the weights $a,b$
in a complicated manner,
but the finite-size correction
$\smash{\logfsc^\mathrm{topology}_{(-1)^{m+n}}(\tf{n b}{m a})}$
is the same function as appearing in the expansion \eqref{e:ferdinand.unweighted}
	for the unweighted square lattice,
	now applied to the ``effective'' aspect ratio
	$\tf{n b}{m a}$.
In this sense the finite-size corrections are seen to be robust to the particulars of the model.

Finite-size corrections for square lattice dimers have also been explicitly computed on the cylinder \cite[eq.~(46)]{MR1227270} \cite{PhysRevE.67.066114}, M\"obius band \cite[eq.~(48)]{MR1227270} \cite{PhysRevE.67.066114}, and Klein bottle \cite{PhysRevE.67.066114}. In each of these topologies, for each given choice of side length parities, the finite-size correction is an analytic function of the aspect ratio \cite{PhysRevE.67.066114}. See \cite{MR2194994,MR2280327} for a discussion of these finite-size corrections in the context of logarithmic conformal field theory.

\subsection{Characteristic polynomial and spectral curve}
\label{ss:intro.chp}

In this article we consider dimer systems defined on two broad classes of \emph{critically\/} weighted $\Z^2$-periodic planar lattices --- rather loosely, a bipartite and a non-bipartite class. We assume throughout that the lattices are connected, with each edge occurring with positive probability. Within each class, we compute an asymptotic expansion of the dimer free energy on large toric quotient graphs --- including ``skew'' or ``helical'' (non-rectilinear) tori --- and explicitly determine the finite-size correction.

On \emph{non-bipartite lattices},
the finite-size correction depends
on a single parameter~$\tau$
in the complex upper half-plane
describing the \emph{conformal shape\/} of the domain
--- $\slf{\tau}{i}$ generalizes the
``effective aspect ratio'' $\tf{nb}{ma}$ appearing in \eqref{e:ferdinand.weighted}.
On \emph{bipartite lattices},
the correction
depends further on
whether the \emph{finite torus\/} is
globally bipartite
or non-bipartite,
as well as on a \emph{phase\/} parameter
$(\ze,\xi)\in\T^2$
which generalizes the signs
$((-1)^m,(-1)^n)$ appearing in \eqref{e:ferdinand.weighted}.
The functional form of the correction is universal within each class.

More precisely, the bipartite and non-bipartite graph classes which we
consider throughout this paper are characterized by algebraic
conditions on the \emph{dimer characteristic polynomial}.  This is a
certain Laurent polynomial $P(z,w)$, whose definition depends only on
the combinatorics of the \emph{fundamental domain}, the $1\times1$
toric quotient of the $\Z^2$-periodic graph.

On the unit torus $\T^2\equiv\set{(z,w)\in\C:|z|=|w|=1}$,
the characteristic polynomial $P(z,w)$ is non-negative.
Many large-scale quantities of interest in the dimer model
can be computed from $P$:
for example the free energy per fundamental domain is
given by half the logarithmic Mahler measure
\beq\label{e:free}
\freezero
\equiv
\tf12\iint_{\T^2}
	\log P(z,w) \, \f{dz}{2\pi iz}\,\f{dw}{2\pi iw}.
\eeq
Edge-edge correlations are obtained from the Fourier transform of $P(z,w)^{-1}$ \cite{MR1473567}.

\emph{Criticality\/} in dimer models is characterized by the intersection of the \emph{spectral curve\/}
\[\{(z,w)\in\C^2:P(z,w)=0\},\]
with the unit torus $\T^2$.
Dimer models on \emph{bipartite\/} graphs have been quite deeply understood,
in part via the classification of the spectral curve as a simple Harnack curve
\cite{\kos,\ko,MR2099145}. The bipartition of the graph gives a natural factorization $P(z,w)=Q(z,w)Q(z^{-1},w^{-1})$ with $Q$ a real polynomial, so that the factors $Q(z,w)$ and $Q(z^{-1},w^{-1})$ are complex conjugates for $(z,w)\in\T^2$ (see~\S\ref{ss:prelim.bip}). It is known that if the zero set of $Q$ on $\T^2$ is non-empty, then it consists of a pair of complex conjugate zeroes --- which either are distinct, or coincide at a real root of~$Q$. In the case of distinct zeroes, or zeroes coinciding at a real root at which $Q$ has a node, the model is \emph{critical\/} or \emph{liquid}, with polynomial decay of correlations. (A \emph{node} is a point $(z_0,w_0)$ at which the polynomial is a product of two distinct lines $(b_1(z-z_0)+c_1(w-w_0))(b_2(z-z_0)+c_2(w-w_0))$ plus higher-order terms.) In all other cases the model is off-critical, and belongs (depending on the geometry of the spectral curve) either to a \emph{gaseous\/} (exponential decay of correlations) or \emph{frozen\/} (no large-scale fluctuations) phase.

Far less is known about the spectral curves of non-bipartite dimer systems. In this setting it is conjectured that the characteristic polynomial $P(z,w)$ is either non-vanishing on the unit torus, or is vanishing to second order at a single \emph{real node\/} which is one of the four points $(\pm1,\pm1)$. This conjecture has been proved for the Fisher lattice with edge weights corresponding to any bi-periodic ferromagnetic Ising model on the square lattice \cite{arXiv:1008.3936}. For lattices satisfying this condition one can show (see~\cite{\kos})
that frozen phases do not exist: when the spectral curve is disjoint from the unit torus the model is gaseous (off-critical), and when it intersects at a real node the model is liquid (critical). In this paper we assume this condition and illustrate its implications for critical dimer systems.

\subsection{Statement of results}
\label{ss:intro.results}

Let~$\lat$ be a weighted $\Z^2$-periodic quasi-transitive (that is, the quotient $\lat/\Z^2$ is finite) planar graph.
We consider dimers on large toric quotients of $\lat$, as follows:
let $\End$ be the set of integer $2\times2$ matrices
\beq\label{e:E}
E\equiv
	\bpm u & v \\ x & y \epm
\quad\quad\quad\text{with }
\det E>0.
\eeq
Any $E\in\End$ defines the toric graph $\lat_E\equiv \lat/(\Z^2 E)$, the quotient of $\lat$ modulo translation by the vectors in the lattice $\Z^2 E\equiv\set{ a(u,v) + b(x,y) : a,b\in\Z }$. We take asymptotics with $E$ tending to infinity while being ``well-shaped'' in the sense that
\enlargethispage{34pt}
\beq\label{e:e.large}
\begin{array}{l}
\text{$\det E$ tends to infinity}\\
\hspace{10pt}\text{while remaining within a constant factor of both $\|(u,v)\|^2$ and $\|(x,y)\|^2$.}
\end{array}
\eeq

\subsubsection{Finite-size correction to the characteristic polynomial}
\label{sss:intro.chp.fsc}

The $1\times1$ toric quotient $\lat_I$ (with $I$ the $2$-dimensional identity matrix) is called the \emph{fundamental domain}. We assume it has $k$ vertices with $k$ even: as a consequence (see \S\ref{ss:odd.general}), $\lat$ is equipped with a \emph{periodic Kasteleyn orientation\/} in which the contour loop surrounding each face has an odd number of clockwise-oriented edges \cite{\kast}.
(In \S\ref{s:odd} we discuss how to handle $k$ odd,
for which such orientations do not exist.)
 The dimer characteristic polynomial $P(z,w)$ is the determinant of a certain $k$-dimensional matrix $K(z,w)$ associated with the fundamental domain, which may be considered as the discrete Fourier transform of the (infinite-dimensional) weighted signed adjacency matrix of $\lat$. For a brief review and formal definitions see \S\ref{ss:kast}.

Of course for given $\lat$ there is some freedom in the choice of fundamental domain: in particular any $\lat_E$ may be regarded as the fundamental domain, with corresponding characteristic polynomial $P_E(\ze,\xi)$ which is the determinant of a $(k\det E)$-dimensional matrix $K_E(\ze,\xi)$. It can be obtained from $P(z,w)$ by the double product formula
\beq\label{e:productformula}
P_E(\ze,\xi)
= \prod_{\substack{z^u w^v=\ze \\ z^x w^y=\xi}} P(z,w),
\eeq
(see e.g.\ \cite{\ckp,\kenyonaihp,\kos}).
If the characteristic polynomial $P$
is \emph{non-vanishing\/} on the unit torus, it is easily seen from \eqref{e:productformula} (see Theorem \ref{t:z}, below) that, in the limit~\eqref{e:e.large},
$\log P_E(\ze,\xi)
	= (\det E)\,2\freezero+o(1)$
uniformly over $(\ze,\xi)\in\T^2$,
which readily implies
(using e.g.~Proposition~\ref{p:torus.z0}) the free energy expansion
$\log \zz_E=(\det E)\,\freezero+o(1)$.

In this paper we compute an asymptotic expansion of $P_E(\ze,\xi)$
	($\ze,\xi\in\T$)
in the more interesting critical case where $P(z,w)$
is vanishing to second order at \emph{nodes\/} on the unit torus.
Formally, let us say that $P$ has a \emph{positive node\/} at $(e^{i\ttr_0},e^{i\tts_0})\in\T^2$ if it is vanishing there to second order with positive-definite Hessian matrix:
\beq\label{e:hess}
\begin{array}{l}
P(e^{\pi i(\ttr_0+\ttr)},e^{\pi i(\tts_0+\tts)})
= \pi^2\ip{(\ttr,\tts)}{ H (\ttr,\tts)} + O(\|(\ttr,\tts)\|^3)
\quad\text{where }\\
\DS H = \bpm \AAz & B \\ B & \AAw \epm
\text{ with }
	\AAz,\AAw>0
	\text{ and }
	D\equiv\sqrt{\AAz \AAw-B^2}>0.
\end{array}
\eeq
In the bipartite case (see above), distinct conjugate zeroes of $Q$ correspond to positive nodes of $P$; see \eqref{e:bip.hess}. If instead $Q$ has a real node, the Harnack property implies that this node is positive (up to global sign change). We associate to $H$ the parameter
\beq\label{e:tau.H}
\tau[H]
	\equiv (-B+i D)/\AAw \in
	\H \equiv\set{z\in\C:\imag z>0}.
\eeq

\bThm\label{t:double.product}
Suppose $P(z,w)$ is an analytic non-negative function defined on
	the unit torus $\T^2$,
non-vanishing except at positive nodes $(z_j,w_j)$
($1\le j\le\ell$) with associated Hessians~$H_j$.
Then, in the limit \eqref{e:e.large}, for $\ze,\xi\in\T$ we have
\beq\label{e:double.product.fsc}
\log P_E(\ze,\xi)
= 2(\det E)\,\freezero
	+\sum_{j=1}^\ell 2\log\cf
		\Big(
		\f{\ze}{z_j^u w_j^v},
		\f{\xi}{z_j^x w_j^y}
		\,\Big|\,
		\tau_j\Big)
	+ O\Big( \f{1}{n^{2/5} \bm{r}}
	\Big)\eeq
where $\freezero$ is given by \eqref{e:free},
$\bm{r}$ is the minimum Euclidean distance between $(1,1)$
and the set of points
$(\ze/(z_j^u w_j^v),\xi/(z_j^x w_j^y))$,
$\tau_j$ is the parameter \eqref{e:tau.H} associated to the transformed Hessian $(E^t)^{-1}H_j E^{-1}$,
and $\cf$ is the explicit function \eqref{e:cf}.
\eThm

In the two settings we consider
	(see \S\ref{ss:intro.chp}),
the spectral curve of the characteristic polynomial
either intersects the unit torus at a single
positive node $(z_0,w_0)=(\pm1,\pm1)$
with Hessian $H$,
or at conjugate positive nodes
$(z_0,w_0)\ne(\overline{z}_0,\overline{w}_0)$ with the \emph{same\/} Hessian $H$ (see~\eqref{e:bip.hess}). These conjugate nodes may occur at the same point, in which case $P$ vanishes to fourth order; however in this case we can still treat each node separately in Theorem \ref{t:double.product}.
In either case we define
\beq\label{e:tau.domain.phase}
\begin{array}{rl}
\DS
\tau_E\equiv \f{x+y\,\tau[H]}{u+v\,\tau[H]}
	=\tau[(E^t)^{-1} H E^{-1}]
	\in\H
&\quad\text{the \emph{conformal shape of $\lat_E$};}\\[12pt]
(\phaseh_E,\phasev_E)
\equiv (e^{\pi i \ttr_E},e^{\pi i \tts_E})
\equiv (z_0^u w_0^v,z_0^x w_0^y)\in\T^2
&\quad\text{the \emph{domain phase of $\lat_E$}.}
\end{array}
\eeq
where $\ttr_E,\tts_E$ are chosen to lie in the interval $(-1,1]$. (In the case of two distinct nodes, for most purposes it suffices to take the phase to be defined modulo complex conjugation. For one of our results, Theorem~\ref{t:gaussian}, we specify a distinction between the nodes to have a more explicit statement.)

\subsubsection{Finite-size correction to the dimer partition function}

By the method of Pfaffians \cite{\kastquadratic,galluccio-loebl,MR1750896} (see also \cite{cimasoni-reshetikhin}),
the dimer partition function on
$\lat_E$ is a signed combination of the
four square roots $P_E(\pm1,\pm1)^{1/2}$:
\[\zz_E
= \tf12[
	\pm P_E(+1,+1)^{1/2}
	\pm P_E(+1,-1)^{1/2}
	\pm P_E(-1,+1)^{1/2}
	\pm P_E(-1,-1)^{1/2}
	]\]
(a review is given in \S\ref{ss:kast}; see in particular
Proposition~\ref{p:torus.z0}).
In \S\ref{s:fsc}
we explain how to choose the signs
to deduce from Theorem \ref{t:double.product}
the finite-size correction to the dimer partition function
for the two classes of critically weighted graphs described above:

\bThm\label{t:z}
If the spectral curve $\set{P(z,w)=0}$ is disjoint from the unit torus, then $\log\zz_E= (\det E)\,\freezero+o(1)$.
\bnm[a.]
\item \label{t:z.a}
If the spectral curve intersects the unit torus at a single real positive node with associated Hessian $H$, then
\[\log \zz_E = (\det E)\,\freezero
	+ \logfscone(\tau_E) + o(1)\]
\hypertop{FSC_1}
where $\tau_E$ is as in \eqref{e:tau.domain.phase}, and
$\logfscone\equiv\log\fscone$ with
\[\TS\fscone(\tau)
	\equiv
	\tf12
	\sum_{\ze,\xi=\pm1}\cf(\ze,\xi\giv\tau).
\]

\enlargethispage{12pt}
\item \label{t:z.b}
Suppose the fundamental domain is bipartite,
with dimer characteristic polynomial
$P(z,w)=Q(z,w)\,Q(\slf{1}{z},\slf{1}{w})$
non-vanishing on $\T^2$
except at distinct
conjugate positive nodes
$(z_0,w_0)\ne(\overline{z}_0,\overline{w}_0)$
with associated Hessian $H$.\footnote{The Hessian is necessarily the same at both nodes, see (\ref{e:bip.hess}).} Then
\[\log\zz_E
	=
	(\det E)\,\freezero
	+ \logfsctwo(\phaseh_E,\phasev_E\giv\tau_E) + o(1)\]
where $\tau_E,\phaseh_E,\phasev_E$ are as in \eqref{e:tau.domain.phase}, and $\logfsctwo\equiv\log\fsctwo$ with
\hypertop{FSC_2}
\[\TS
\fsctwo
(\ze,\xi\giv\tau)
\equiv
	\tf12
	\sum_{z,w=\pm1} \cf( z\ze,w\xi\giv\tau )^2
	\]
which has the equivalent expression
\beq\label{e:fsctwo.gaussian}
\fsctwo(e^{\pi i \ttr},e^{\pi i \tts}\giv\tau)=
\f{\sum_{\bm{e}\in\Z^2}
	\exp\{ -\tf{\pi}{2}
		g_\tau(  \bm{e}-(\tts,-\ttr) ) \}}
	{ |\deta(\tau)|^2(2\imag\tau)^{1/2} }
\eeq
where for $\tau\in\H$, $g_\tau$ is the quadratic form
\beq\label{e:qfdef}
g_\tau(\bm{e})
\equiv
(\tauy)^{-1}(e_1^2 + 2 \taux e_1 e_2 + |\tau|^2 e_2^2)
\eeq
and $\deta$ is the Dedekind eta function.

\item \label{t:z.b.1}
Suppose the fundamental domain is bipartite,
with dimer characteristic polynomial
$P(z,w)=Q(z,w)\,Q(\slf{1}{z},\slf{1}{w})$
non-vanishing on $\T^2$
except at a single (real) root at which $Q$ has a positive node
with associated Hessian $H$.  Then
\[\log\zz_E
	=
	(\det E)\,\freezero
	+ \logfsctwo(1,1\giv\tau_E) + o(1)\]
where $\tau_E$ is as in \eqref{e:tau.domain.phase}.

\item \label{t:z.c}
If the spectral curve intersects the unit torus at two real positive nodes $(z_1,w_1)$ and $(z_2,w_2)$ with the same associated Hessian $H$, then
\[\log\zz_E
=(\det E)\,\freezero
	+\logfscodd(\phaseh_E,\phasev_E\giv\tau_E)
	+o(1)\]
\hypertop{FSC_3}
where, defining $(z_0,w_0)\equiv(z_1z_2,w_1w_2)$,
the parameters $\tau_E,\phaseh_E,\phasev_E$ are as in \eqref{e:tau.domain.phase},
and $\logfscodd\equiv\log\fscodd$ with
\[\TS \fscodd(\ze,\xi\giv\tau)
\equiv
	\tf12\sum_{z,w=\pm1}
	\cf(z,w\giv\tau)\cf(z\ze,w\xi\giv\tau)
\]
We further have the simplifications
\[\begin{array}{rl}
\fscodd(+1,+1\giv\tau)
	\hspace{-6pt}&=\fsctwo(+1,+1\giv\tau)\\
\fscodd(+1,-1\giv\tau)
	\hspace{-6pt}&=\cf(-1,-1\giv\tau)\cf(-1,+1\giv\tau)
	=\cf(-1,+1\giv2\tau)\\
\fscodd(-1,+1\giv\tau)
	\hspace{-6pt}&=\cf(-1,-1\giv\tau)\cf(+1,-1\giv\tau)
	=\cf(+1,-1\giv\slf\tau2)\\
\fscodd(-1,-1\giv\tau)
	\hspace{-6pt}&=\cf(-1,+1\giv\tau)\cf(+1,-1\giv\tau)
	=\cf(+1,-1 \giv \tf{1+\tau}{2})
\end{array}
\]

\enm
\eThm

\noindent
See Figure~\ref{f:logfscs} for plots of these functions $\logfscone$, $\logfsctwo$, and $\logfscodd$.
In \cite[Thm.~5.1]{MR2215138} it is shown that for bipartite graphs on tori, case~\ref{t:z.c} does not occur.
However, for graphs on tori that are locally bipartite but not globally bipartite, such as an odd$\times$even
grid on a torus, we see in Section~\ref{ss:square} that this case does occur.

We emphasize again that the
functional form of the finite-size correction
is universal within each class:
the finite-size correction $\cf$ to the characteristic polynomial
(Theorem~\ref{t:double.product}) is an explicit function
depending only on the three parameters $\ze,\xi,\tau$.
Thus in Theorem~\ref{t:z}\ref{t:z.a}
the graph structure enters into the correction only through~$\tau$
(that is, only through the Hessian associated with the real node). In the bipartite setting
(Theorem~\ref{t:z}\ref{t:z.b}, \ref{t:z.b.1},~and~\ref{t:z.c}),
the finite-size correction depends on the graph structure only through
$\tau$ and $(\phaseh_E,\phasev_E)$.

\begin{figure}[hb]
\centering
\includegraphics[width=.5\textwidth]{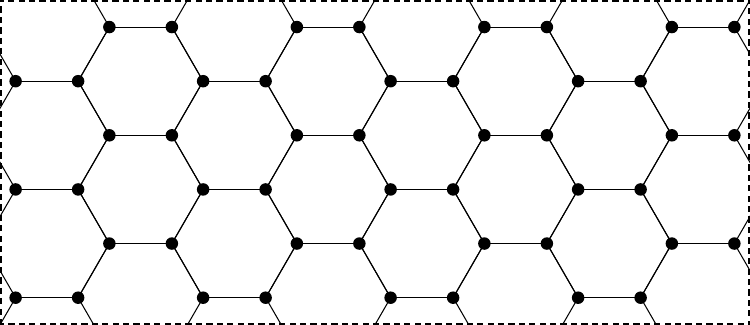}
\caption{The $4\times3$ toric quotient of the honeycomb graph, with effective aspect ratio $\rho=\slf{\sqrt{3}}{4}$ (also the actual aspect ratio of this geometric embedding) in the unweighted setting.}
\label{f:hex.four.by.three}
\end{figure}

As we explain in~\S\ref{ss:modular}, the parameter $\tau$ has a simple interpretation as the half-period ratio of the torus with respect to its ``natural'' or ``conformal'' embedding. Consequently the finite-size corrections are invariant under modular transformations.
For example, for the unweighted honeycomb graph,
the $m\times n$ torus (Figure~\ref{f:hex.four.by.three})
has $\tau = i\rho$
where $\rho=\slf{n}{(m\sqrt{3})}$
is the effective or geometric aspect ratio.

\begin{figure}[ht]
\centering
\includegraphics[width=.5\textwidth]{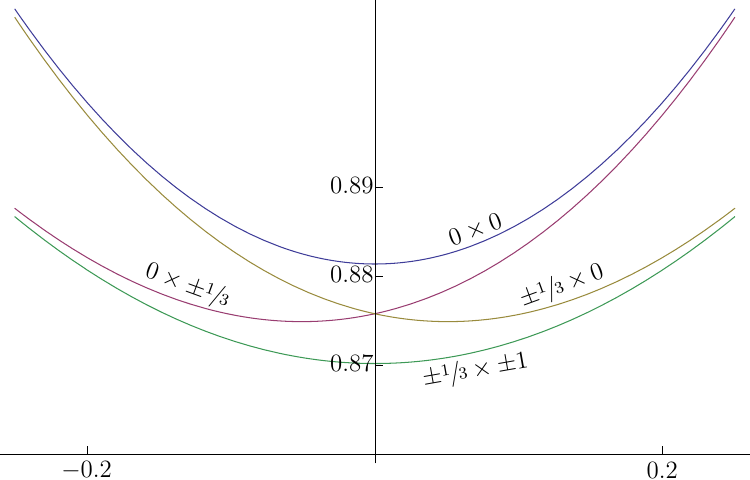}
\caption{\emph{Unweighted honeycomb graph dimers.}
Finite-size corrections
$\logfsctwo(\phaseh_E,\phasev_E\giv \tau)$
for near-rectilinear toric quotients,
shown as a function of logarithmic aspect ratio
$\log\rho$,
labelled according to value of
$(u-v,x-y)/3$ modulo $1$
(see \S\ref{ss:hex}).}
\label{f:hex.logfsc}
\end{figure}

The domain phase parameter
$(\phaseh_E,\phasev_E)$
 is of a quite different nature:
it generalizes the signs $(-1)^m,(-1)^n$ appearing in \eqref{e:ferdinand.weighted},
and depends sensitively on the entries of $E$.
For example, for dimers on the honeycomb lattice,
the finite-size correction for
$m\times n$ quotients (Figure~\ref{f:hex.four.by.three})
was computed by
Boutillier and de~Tili\`ere
in the case $n\equiv0\bmod3$ \cite{MR2561433}.
Figure~\ref{f:hex.logfsc} shows this correction for the unweighted honeycomb lattice as a function of the logarithmic effective aspect ratio
$\log\rho$,
together with three other curves
--- one showing the different correction which applies
for $n\not\equiv0\bmod3$,
and the remaining two showing corrections
which can be found on toric quotients
which are nearly but not quite rectilinear.
Some discussion of this is given in \S\ref{ss:hex}.

\begin{figure}[b]
\begin{center}
\includegraphics[width=.45\textwidth,trim=0pt 12pt 0pt 0pt]{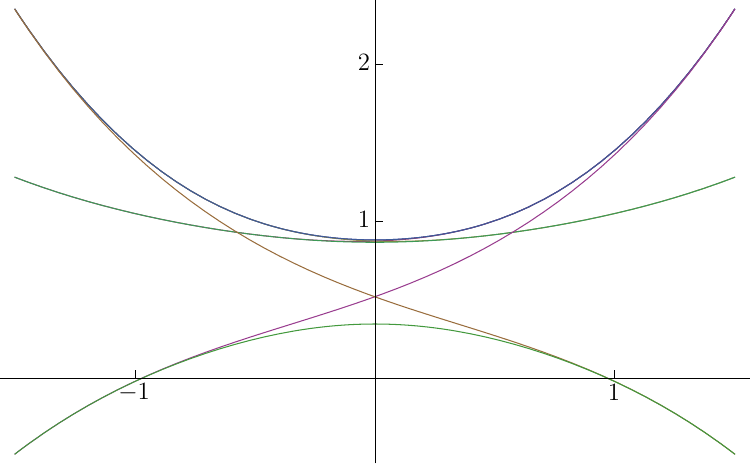}
\end{center}
\caption{\emph{Unweighted square lattice dimers.}
Finite-size corrections $\logfsc$ for near-rectilinear toric quotients,
shown as a function of the logarithmic aspect ratio $\log\rho$. There are
seven distinct curves,
depending on parities of vectors defining the torus (see~\S\ref{ss:square}).
(It is easy to distinguish only five of the curves, see Figure~\ref{f:torus.square.magnify} for a magnified view.)
}
\label{f:torus.square.fsc.seven}
\end{figure}

\begin{figure}[htp]
\centering
\begin{subfigure}[t]{.32\textwidth}
	\includegraphics[width=\textwidth,trim=15pt 20pt 45pt -11pt]{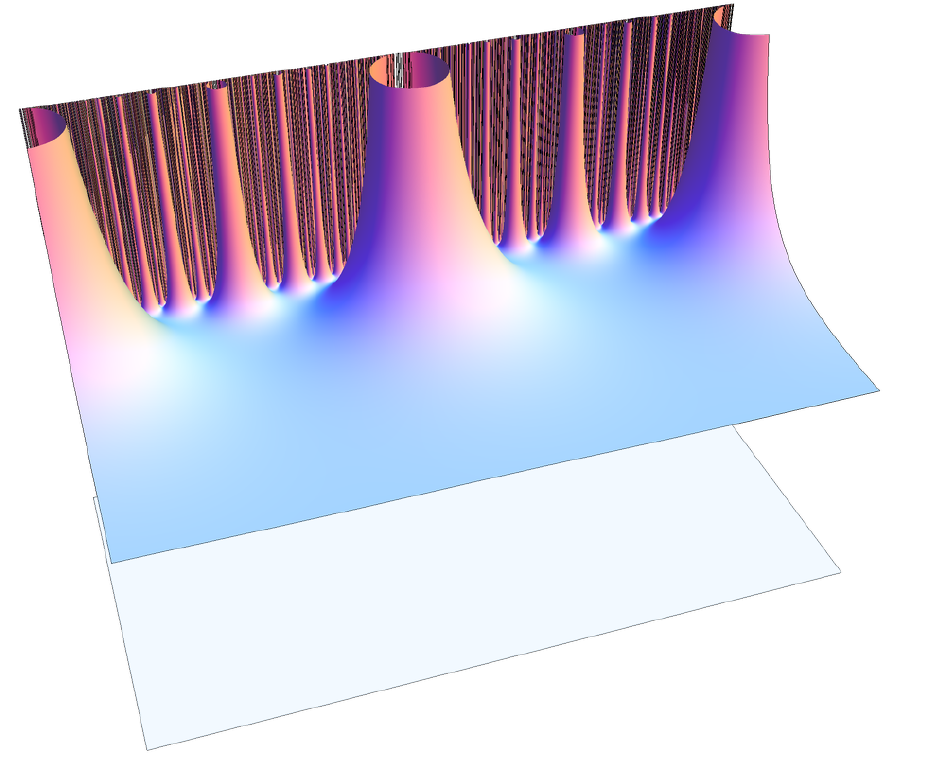}
	\caption{$\logfscone(\tau)$}
\end{subfigure}
\begin{subfigure}[t]{.32\textwidth}
	\includegraphics[width=\textwidth,trim=20pt 20pt 45pt -11pt]{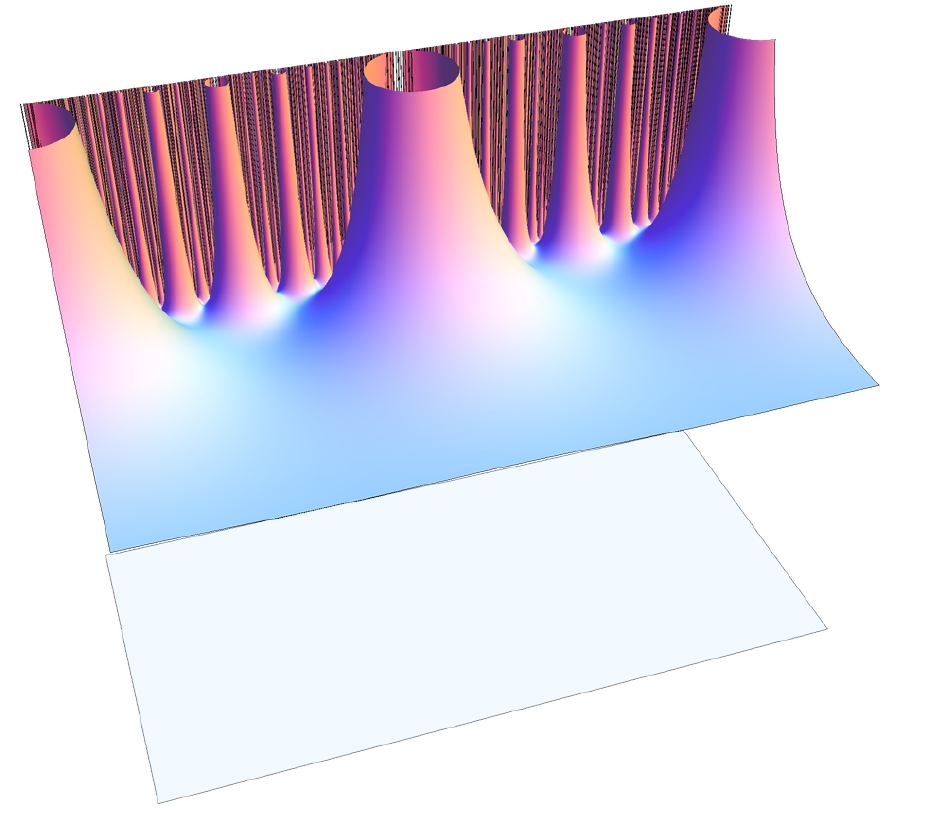}
	\caption{$\logfsctwo(+1,e^{2\pi i/6}\giv\tau)$}
\end{subfigure}
\begin{subfigure}[t]{.32\textwidth}
	\includegraphics[width=\textwidth,trim=20pt 20pt 45pt -11pt]{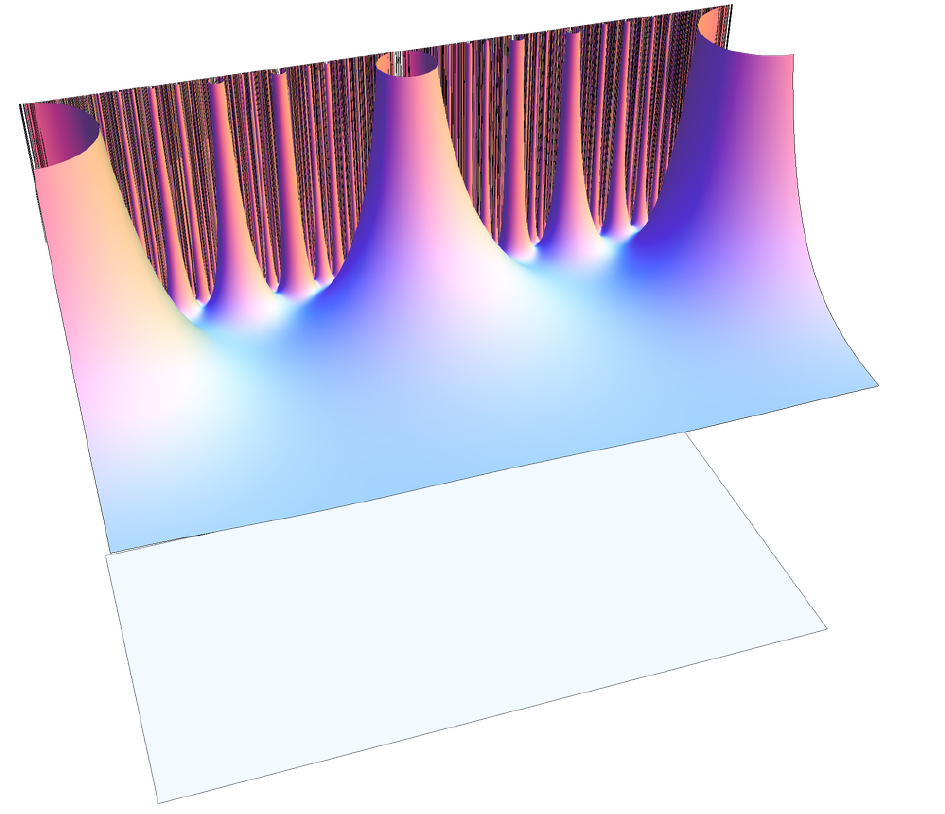}
	\caption{$\logfsctwo(i,i\giv\tau)$}
\end{subfigure}

\begin{subfigure}[t]{.32\textwidth}
	\includegraphics[width=\textwidth,trim=20pt 20pt 45pt -11pt]{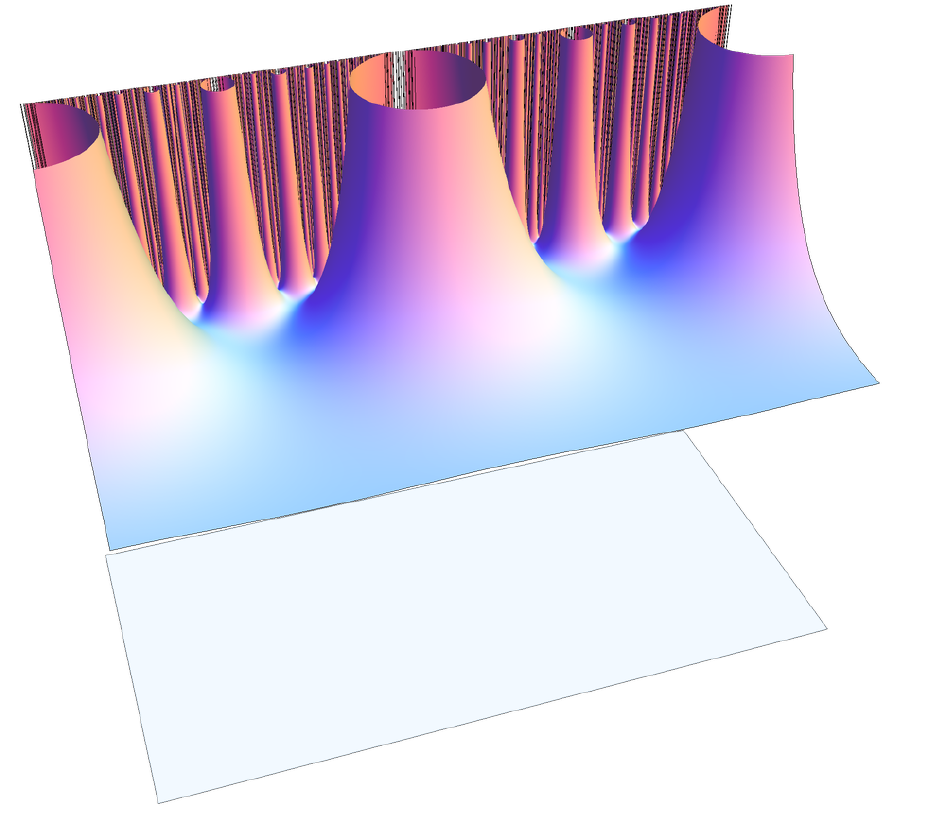}
	\caption{\rlap{\mbox{$\logfsctwo(+1,+1\giv\tau)=\logfscodd(+1,+1\giv\tau)$}}\hspace*{1.7in}}
\end{subfigure}
\begin{subfigure}[t]{.32\textwidth}
	\includegraphics[width=\textwidth,trim=20pt 20pt 45pt -11pt]{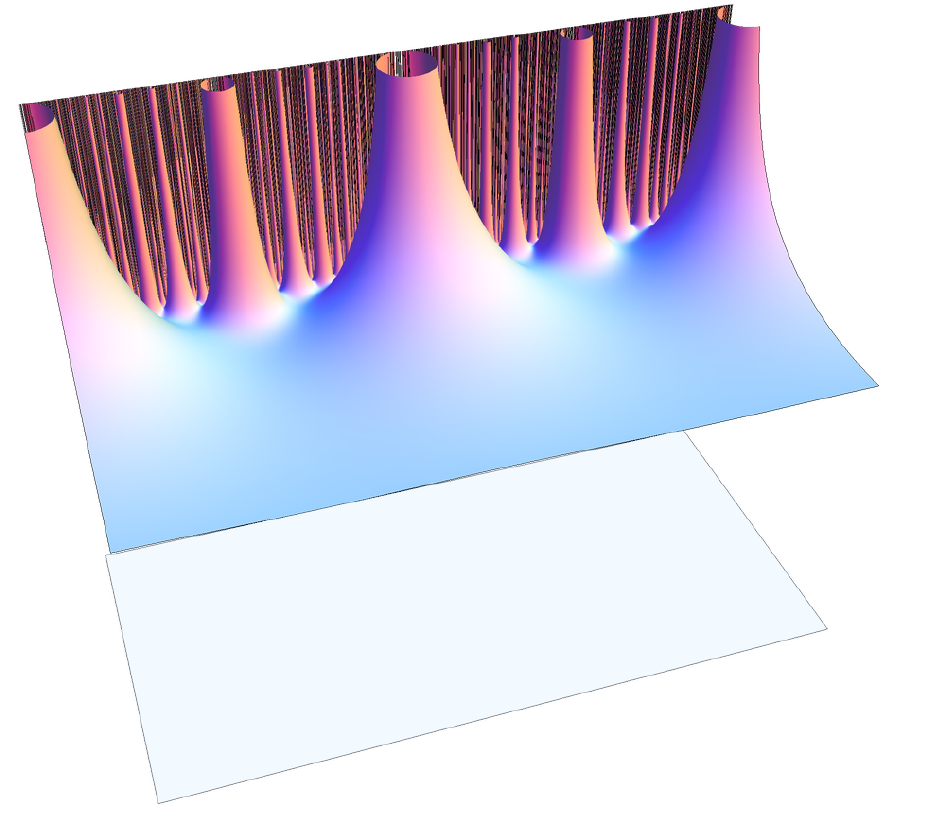}
	\caption{$\logfsctwo(i,+1\giv\tau)$}
\end{subfigure}
\begin{subfigure}[t]{.32\textwidth}
	\includegraphics[width=\textwidth,trim=20pt 20pt 45pt -11pt]{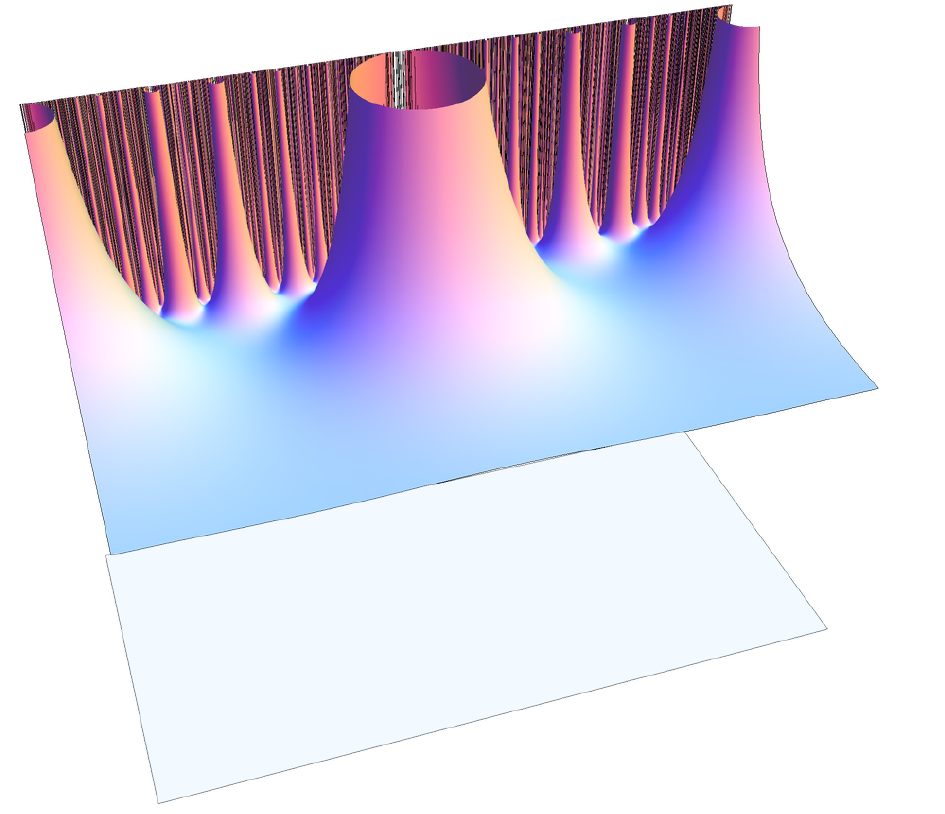}
	\caption{$\logfsctwo(+1,i\giv\tau)$}
\end{subfigure}

\begin{subfigure}[t]{.32\textwidth}
	\includegraphics[width=\textwidth,trim=13pt 130pt 30pt -21pt]{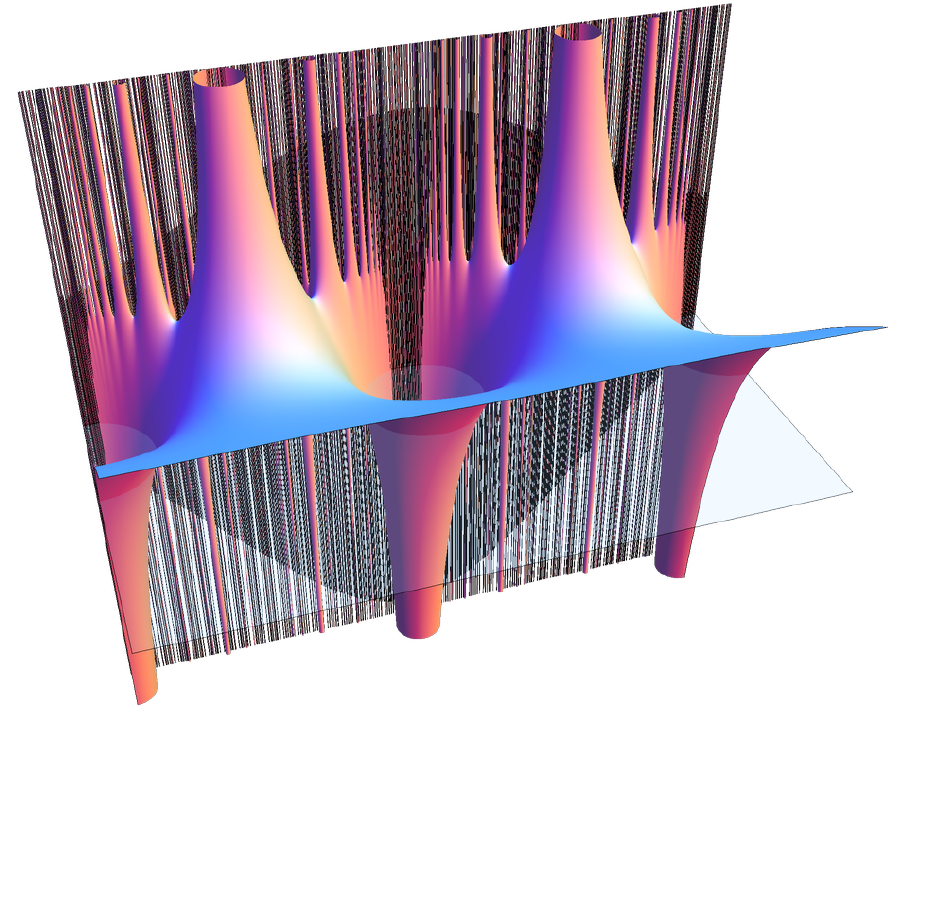}
	\caption{$\logfscodd(+1,-1\giv\tau)$}
\end{subfigure}
\begin{subfigure}[t]{.32\textwidth}
	\includegraphics[width=\textwidth,trim=13pt 130pt 30pt -21pt]{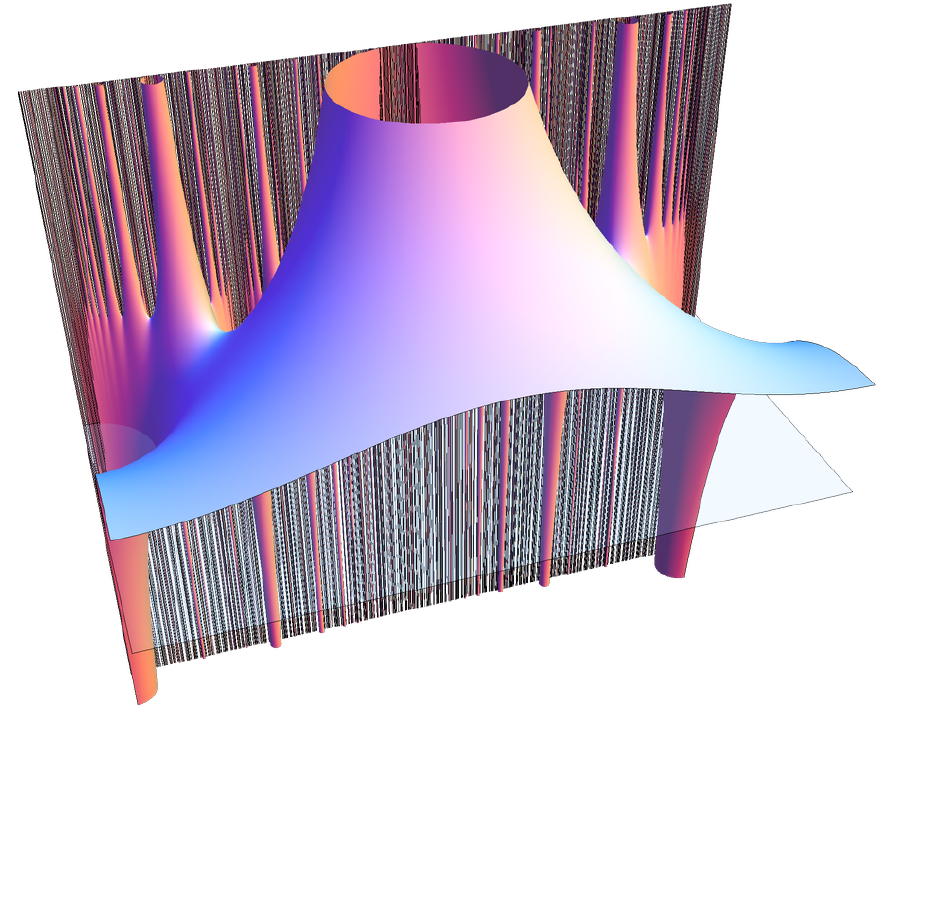}
	\caption{$\logfscodd(-1,+1\giv\tau)$}
\end{subfigure}
\begin{subfigure}[t]{.32\textwidth}
	\includegraphics[width=\textwidth,trim=13pt 130pt 30pt -21pt]{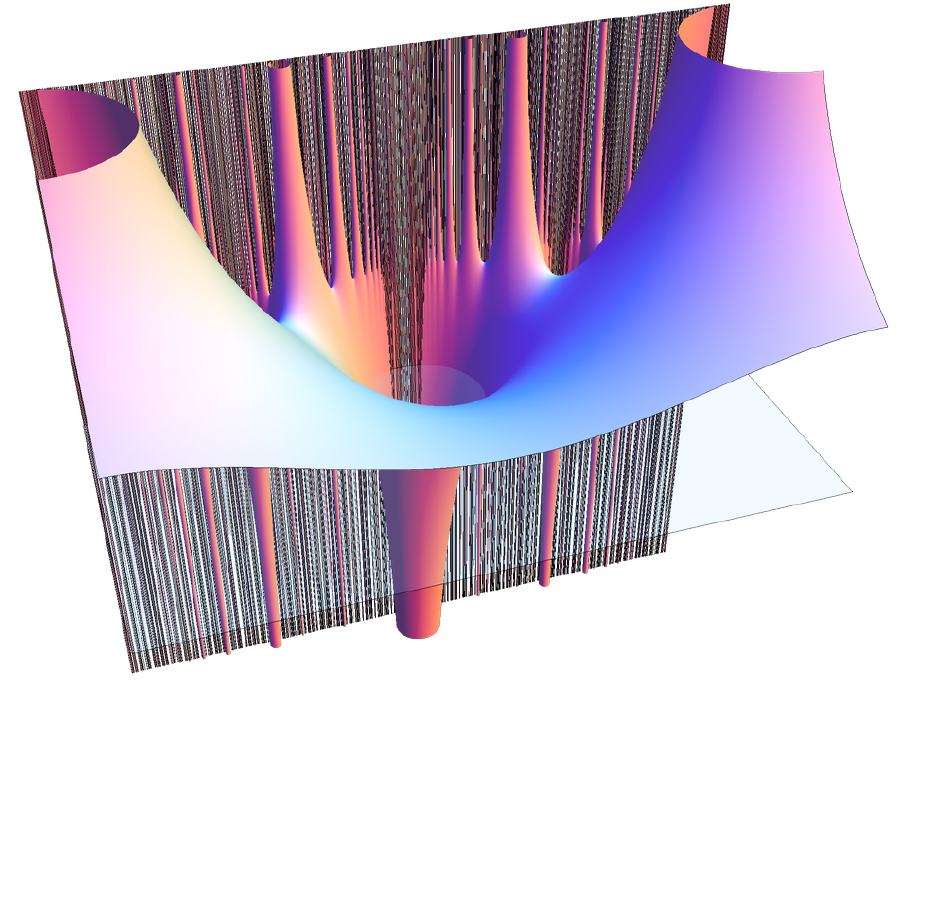}
	\caption{$\logfscodd(-1,-1\giv\tau)$}
\end{subfigure}
\caption{Free energy finite-size corrections as a function of
  $\tau=x+i y\in\H$.  In these plots, $-1\leq x\leq 1$ and $0<y\leq
  1$, and the identically zero function is indicated by the horizontal plane.
  There is one $\logfscone$ function, there are four
  $\logfscodd$ functions, and there is a
  two-parameter family of $\logfsctwo$ functions. Panels ({\sc c}) through ({\sc i}) show the seven functions for unweighted square-grid tori; their restrictions to the pure imaginary line $\tau=i\rho$ are the curves in Figure~\ref{f:torus.square.fsc.seven} (shown there as a function of $\log\rho$). Panels ({\sc b}) and ({\sc d}) show two of the functions relevant to unweighted honeycomb graph tori; see Figure~\ref{f:hex.logfsc}. The function shown in panel ({\sc a}) is relevant to the Ising model; see Figure~\ref{f:torus.ising}.}
\label{f:logfscs}
\end{figure}

\old{
\begin{figure}[htbp]
\centering
\includegraphics[width=.75\textwidth]{Figures/ford-circles}
\caption{Ford circles.  There is a circle tangent to the real axis at each rational.  If $a/b$ is a fraction in lowest terms, then
there is a circle centered at $a/b+i/(2 b^2)$ with radius $1/(2b^2)$.  There is also a ``circle'' tangent to the real axis at $\infty$, namely the horizontal line with height $1$.  The spikes in $\logfscone$, $\logfsctwo$, and $\logfscodd$ (Figure~\ref{f:logfscs}) are positioned at the Ford circles (see Sec.~\ref{sec:asympt-ford}).}
\label{f:ford}
\end{figure}
}

\enlargethispage{24pt}
In the square lattice we find a similar phase sensitivity,
but we find a dependence also on
the global bipartiteness of the torus
(for example, the $4\times3$ torus in the square lattice
is non-bipartite).
As a result,
for near-rectilinear tori
the finite-size correction lies asymptotically on
any of \emph{seven\/} curves,
Figure~\ref{f:torus.square.fsc.seven}
--- four curves for bipartite tori
and three for nonbipartite.
Further discussion of this is given in \S\ref{s:odd}.

\begin{figure}[ht]
\centering
\includegraphics[width=.4\textwidth]{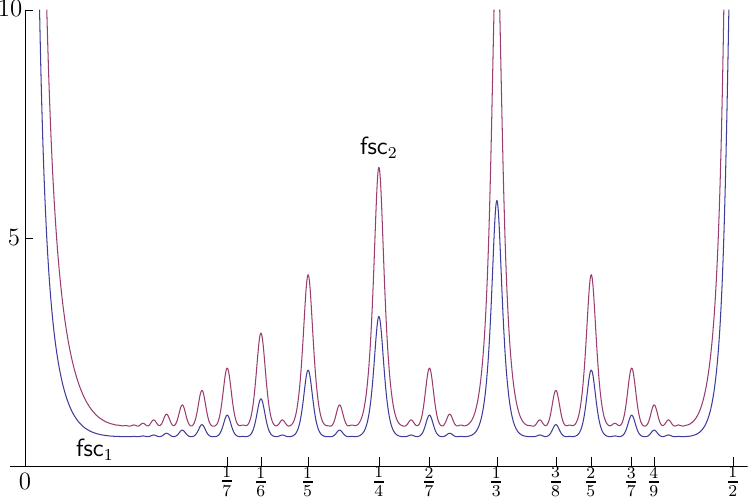}
\caption{$\logfscone(x+i/200)$ (blue) and $\logfsctwo(1,1\giv x+i/200)$ (red).
The local maxima in $\logfsctwo(1,1\giv\tau)$ are twice as high as those for $\logfscone(\tau)$.}
\label{f:fsc.one.below.two}
\end{figure}

\subsubsection{Non-contractible loops on the torus}
\label{sss:intro.loops}

Recall that the superposition of two independent dimer covers of a planar graph $\Graph$ produces a double-dimer configuration consisting
of even-length loops and doubled edges.
Alternatively, a single dimer cover of $\Graph$
may be mapped to a double-dimer configuration
by superposition with a fixed \emph{reference matching\/}
$\refmatch$.
It is of interest to study the
\emph{non-contractible\/} loops arising
from this process on toric graphs.
In addition to the finite-size corrections to the overall dimer partition functions $\zz_E$ (Theorem~\ref{t:z}), we are able to obtain some finer information on the distribution of the partition function between dimer covers of different homological types, as follows.

\medskip\noindent\emph{Non-contractible loops in the bipartite setting.}
If $\Graph$ is \emph{bipartite\/}, a double-dimer configuration resulting from the (ordered) pair $(\match,\match')$
is naturally regarded as an \emph{oriented\/}
loop configuration $\match\ominus\match'$,
with edges from $\match$ oriented black-to-white
and edges from $\match'$ oriented white-to-black.
We then let
\beq\label{e:wind}\wind\match\ominus\match'
\equiv(\windh,\windv)\in\Z^2\eeq
denote the homology class (or ``winding numbers'') of
the \emph{oriented\/} loop configuration.\footnote{If $\match\ominus\match'$ contains two loops each winding once around the torus in the $+(u,v)$ direction, then $\wind\match\ominus\match'=(2,0)$;
if the two loops wind in opposing directions
then $\wind\match\ominus\match'=(0,0)$.}

For $m\times n$ toric quotients of the unweighted honeycomb tiling
(Figure~\ref{f:hex.four.by.three}),
it was shown in \cite{MR2561433}
that for $n\in3\Z$,
the winding $\wind\match\ominus\refmatch$
of a dimer cover $\match$ with respect to a fixed reference matching $\refmatch$
is asymptotically distributed as a pair
of independent discrete Gaussians,
with variances determined by the torus aspect ratio.
The proof is based on a perturbative analysis of the finite-size correction, and we generalize their method to prove

\bThm\label{t:gaussian.intro}
In the setting of Theorem~\ref{t:z}\ref{t:z.b}, let $\match_E$ be a fixed reference matching of $\lat_E$ obtained by periodically extending a matching $\match_0$ of the fundamental domain. Then the winding $\wind\match\ominus\match_E\in\Z^2$ asymptotically fluctuates as a discrete Gaussian:
\[
\begin{array}{l}
\DS\P(\wind\match\ominus\match_E = \bm{e})
	\to
	\f{\exp\{ -\tf{\pi}{2}
		(\bm{e}-\meanvec)^t \covmatinv (\bm{e}-\meanvec) \}}
	{ \sum_{\bm{e}'\in\Z^2}
		\exp\{ -\tf{\pi}{2} (\bm{e}'-\meanvec)^t \covmatinv
			(\bm{e}'-\meanvec) \} }\quad\text{with} \vspace{4pt}\\
\DS \text{covariance }
	\covmat=\f{(E^t)^{-1}HE^{-1}}{(\det H)^{1/2}/\det E},\quad
\text{center }
	\meanvec \equiv \pm
		\tf1\pi
		(\arg\phasev_E,-\arg\phaseh_E) \mod\Z^2.
\end{array}
\]
\eThm

\noindent
A more explicit version of Theorem~\ref{t:gaussian.intro} is given as Theorem~\ref{t:gaussian}, stated and proved in \S\ref{s:torus.bip}.
Dub\'edat \cite[Thm.~7]{dubedat} proved a version of Theorem~\ref{t:gaussian.intro} for dimers on bipartite isoradial graphs.

\medskip\noindent\emph{Non-contractible loops in the non-bipartite setting.}
In the \emph{non-bipartite\/} setting,
the loop configuration $\match\oplus\match'$ is not oriented,
and we take the winding $\wind\match\oplus\match'$ to be defined only as an element of $(\Z/2\Z)^2$.
In the setting of Theorem~\ref{t:z}\ref{t:z.a},
we also compute
(Proposition~\ref{p:torus.nonbip})
the finite-size corrections to the
partition functions $\zzE[\ttr\tts]$
of the four homology classes
indexed by $(\ttr,\tts)\in\set{\zro,\one}^2$.

To note one particular motivation, we remark that this winding
is of particular interest in the context of Ising models.
On a graph $\Graph=(\Vertices,\Edges)$
with real-valued parameters
$(\be_e)_{e\in\Edges}$
(\emph{coupling constants}),
we define the associated \emph{Ising model\/}
to be the probability measure on spin configurations
$\usi\in\set{\pm1}^{\Vertices}$ given by
\[\PIsing_\Graph(\usi)
\equiv\frac1{\zzIsing_\Graph}\prod_{e=(uv)\in\Edges}
	\exp\{ \be_e\si_u\si_v \}.\]
On the square lattice with
vertical and horizontal coupling constants $\be_a$ and $\be_b$
(``Onsager's lattice''),
the bulk free energy density $\freezeroIsing$
was first calculated by Onsager
\cite{MR0010315}.
Kasteleyn \cite{MR0153427} and Fisher \cite{fisher1966dimer}
rederived this result
by exhibiting a correspondence between the \emph{Ising\/} model on a
planar (weighted) graph $\Graph$
and the \emph{dimer\/} model on various ``decorated'' versions $\Graph'$ of $\Graph$.

\begin{figure}
\includegraphics[width=.4\textwidth]{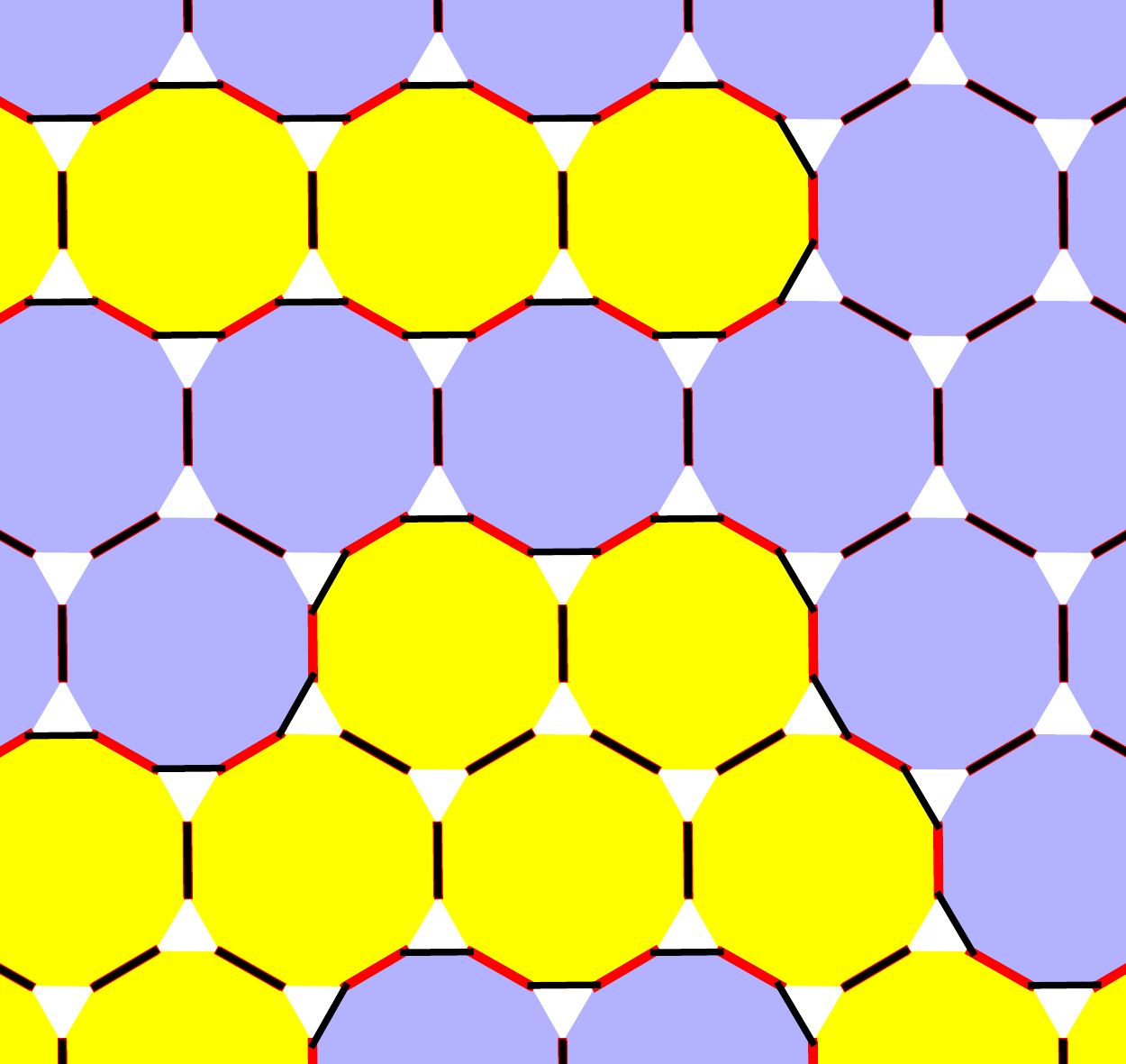} \caption{\emph{Fisher correspondence\/}
between the Ising model on the triangular lattice and the dimer model on the Fisher lattice.
The Ising spins $\pm1$ (yellow and blue) live on the dodecagonal faces of the Fisher lattice.
The Fisher lattice is given the reference matching $\refmatch$ consisting of all between-triangle edges (black).
Take the unique dimer configuration $\match$ which contains a between-triangle edge if and only if it separates like spins (red):
then the loops of $\match\oplus\match$ trace the spin domain boundaries in the low-temperature expansion of the Ising model.
The Ising and dimer partition functions are related in \eqref{e:ising.dimers}.
}
\label{f:ising.low.temp}
\end{figure}

For instance, the Ising model on the triangular lattice
with coupling constants $\be_a,\be_b,\be_c$
corresponds
--- via its low-temperature expansion
--- to the dimer model on the \emph{Fisher lattice\/}
with unit weights on the within-triangle edges, and weights
$(a,b,c)=(e^{2\be_a},e^{2\be_b},e^{2\be_c})$
on the edges between triangles
(Figure~\ref{f:ising.low.temp}).
To calculate the Ising partition function $\zzIsing_{m,n}$
on the $m\times n$ torus in the triangular lattice,
take the $m\times n$ torus in the Fisher lattice,
and fix the reference matching $\refmatch$ consisting of all $(a,b,c)$-edges. Let $\zz^{\,\zro\zro}[a,b,c]$
 denote the partition function of
dimer configurations $\match$
with $\wind\match\oplus\refmatch=(\zro,\zro)$:
then
\beq\label{e:ising.dimers}
\zzIsing_{m,n} =
	\f{2\cdot \zz^{\zro\zro}_{m,n}
		[e^{2\be_a},e^{2\be_b},e^{2\be_c}]}
		{\prod_e e^{\be_e}}.\eeq
At $\be_c=0$ ($c=1$),
the Ising model on the triangular lattice
reduces to the Ising model on Onsager's lattice.
Criticality for $\Z^2$-periodic Ising models
has been characterized in terms of the
intersection of the
Fisher lattice spectral curve with the unit torus
(\cite{LiCMP,arXiv:1008.3936}, see also \cite{CDCIsing}).

\begin{figure}[ht]
\includegraphics[width=.6\textwidth]{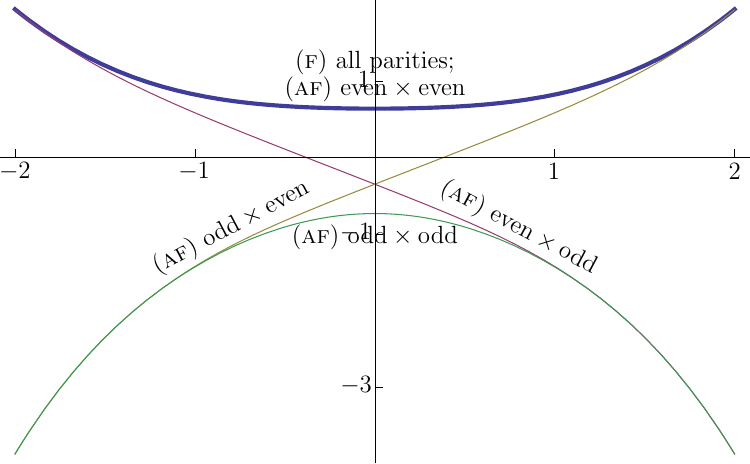}
\caption{\emph{Square lattice critical Ising
	($\be=\tf12\log(\sqrt{2}+1)$).}
Finite-size corrections for
$m\times n$ tori
as a function of logarithmic aspect ratio $\log\rho$.}
\label{f:torus.ising}
\end{figure}

Using \eqref{e:ising.dimers} and similar correspondences,
the asymptotic expansion of the Ising partition function
has been computed in numerous contexts
\cite{PhysRev.185.832,Bugrij1990171,O'Brien199663,PhysRevE.65.036103,MR1690485,PhysRevE.63.026107,PhysRevE.67.065103,MR2000227}.
In particular,
for Onsager's lattice
on the \emph{ferromagnetic\/} critical line
\beq\label{e:onsager.critical}
a+b+1=ab
\quad\text{with }a=e^{2\be_a}
\text{ and }b=e^{2\be_b},
\eeq
the Ising free energy on $m\times n$ graphs has the expansion
	(compare~\eqref{e:ferdinand.weighted})
\[
\log \zzIsing_{m,n}
	= mn\,\freezeroIsing + (\text{perimeter})\,\freesrfIsing
		+ (\text{corners})\,\freecornerIsing
		+ \logfscIsing^{\mathrm{topology}}
		 (\tf{n}{m} \tf{a^2-1}{2a})
		+o(1),
\]
where $\logfscIsing$ is an explicit analytic function depending on the topology (rectangle, torus, cylinder, etc.)
---
but not on the parity of $(m,n)$.
On the \emph{anti-ferromagnetic\/} critical line
\[a^{-1}+b^{-1}+1=(ab)^{-1},\]
the finite-size correction depends also on the parity of $(m,n)$.
Figure~\ref{f:torus.ising}
shows the finite-size corrections for $m\times n$
toric quotients of the homogeneous
Onsager's lattice
($\be_a=\be_b=\be$) at the critical points
\[\be=\pm\tf12\log(\sqrt{2}+1),\]
where $\be$ positive is ferromagnetic
and $\be$ negative is anti-ferromagnetic.

The following proposition
characterizes criticality for
the Fisher lattice,
as well as for a superficially similar lattice,
the so-called rhombitrihexagonal tiling
(Figure~\ref{f:hst}).
The latter graph has no known correspondence with the Ising model,
yet its dimer systems exhibit some similar features.
Though the proposition is easy to prove and various special cases appear in the literature, we include a detailed proof in the appendix (\S\ref{s:fisher.hst}) for completeness.
Combined with
Theorem~\ref{t:z}\ref{t:z.a},
it gives the finite-size correction for
general (critical) Ising models on
large toric quotients
(including skew tori) of the
triangular lattice and Onsager's lattice.

\bppn\label{p:fisher.crit}
For the Fisher graph (Figure~\ref{f:fisher})
	or the 3.4.6.4 graph (Figure~\ref{f:hst}),
the spectral curve can only intersect the unit torus at a real node,
characterized by the vanishing of one of the four quantities
\beq\label{e:kappa}
\bm{c}
\bpm
-\pf K(+1,+1)\\
+\pf K(+1,-1)\\
+\pf K(-1,+1)\\
+\pf K(-1,-1)\\
\epm
=\bpm
+a+b+c-a b c\\
-a+b+c+a b c\\
+a-b+c+a b c\\
+a+b-c+a b c\\
\epm
\equiv
\bpm \ka_\circ \\ \ka_a \\ \ka_b \\ \ka_c \epm
\eeq
where $\bm{c}$ is $1$ for the Fisher graph, and
$\slf12$ for the 3.4.6.4 graph.
\eppn

For dimers coming from the Ising model, such as on the Fisher graph,
the node coincides with the Ising model's critical temperature \cite{LiCMP,CDCIsing}.

We summarize the relevant background in Section~\ref{s:prelim}.
Theorem~\ref{t:z} is proved in Section~\ref{s:fsc}.
In Section~\ref{s:torus.bip} we prove Theorem~\ref{t:gaussian},
which is a stronger version of Theorem~\ref{t:gaussian.intro}.
In Section~\ref{s:odd} we consider lattices with odd-sized fundamental domain,
which provide examples for some of the cases in Theorem~\ref{t:z}.
We postpone the proof of Theorem~\ref{t:double.product} until Section~\ref{s:torus.pf},
even though the proofs of Theorem~\ref{t:z} and~\ref{t:gaussian} depend on it, since its proof is somewhat technical.
Proposition~\ref{p:fisher.crit} is proved in Appendix~\ref{s:fisher.hst}.

\subsection*{Acknowledgements}
We thank C\'edric Boutillier and B\'eatrice de Tili\`ere for several interesting conversations. This research
was conducted and completed during visits of R. K. and N. S. to Microsoft Research.

\section{Preliminaries}
\label{s:prelim}

Throughout this paper, $\lat$ denotes a $\Z^2$-periodic quasi-transitive planar graph equipped with positive edge weights.

\subsection{Kasteleyn orientation and characteristic polynomial}
\label{ss:kast}

The Kasteleyn orientation is a way of computing the dimer and double-dimer partition functions via matrix Pfaffians and determinants.
The \emph{Pfaffian\/} of a $2n\times 2n$ skew-symmetric matrix is given by
\beq\label{e:pfaffian}
\pf K\equiv
	\f{1}{n!\,2^n}
	\sum_{\si\in S_{2n}} (\sgn\si)
	\prod_{j=1}^n K_{\si(2j-1),\si(2j)},
\eeq
and satisfies
$(\pf K)^2=\det K$. If $K$ is the (skew-symmetric) weighted adjacency matrix of a finite directed graph~$\Graph$, then
each non-zero term in \eqref{e:pfaffian} corresponds to a dimer cover of~$\Graph$.
All the $n!\,2^n$ permutations $\si\in S_{2n}$ corresponding to the same dimer cover $\match$ appear with the same sign
$\sgn\match$ in \eqref{e:pfaffian}, so that we may write
$\pf K= \sum_\match\,(\pf K)_\match$
where each matching contributes
$(\pf K)_\match\equiv (\sgn\match)\prod_{(ij)\in\match}
	|K_{ij}|$.

Every finite planar graph $\Graph$ can be equipped with a \emph{Kasteleyn\/} or \emph{Pfaffian orientation}, in which all dimer covers $\match$
appear with the \emph{same\/} sign $\sgn\match$
in \eqref{e:pfaffian} --- that is, for which $|\pf K|$ is the dimer partition function $\zz_\Graph$ of~$\Graph$, and $\det K$ is the double-dimer partition function $(\zz_\Graph)^2$.
A Kasteleyn orientation is given by arranging each
(non-external) face to be clockwise odd, i.e.\ with an odd number of edges oriented in the clockwise direction;
see \cite[\S V-D]{\kast} for details.\footnote{It is sometimes useful to allow some edges of $\Graph$ to have imaginary weights,
in which case $K$ is no longer real-valued (but still skew-symmetric). In this setting a Kasteleyn orientation of a planar graph is given
by taking the product of signed edge weights going clockwise (that is, edge $e=(u\to v)$ contributes $+\nu_e=K_{u,v}$ or $-\nu_e=K_{v,u}$
to the product according to whether it is traversed in the positive ($u\to v$) or negative ($v\to u$) direction while going clockwise around the face) around each (non-external) face to be negative real. We say that an oriented loop has \emph{sign\/} $\ze\in\T$ to mean that the product of signed edge weights along the loop equals a positive real number times $\ze$.}

Returning to the setting of \S\ref{ss:intro.results}, let $\lat$ be a planar $\Z^2$-periodic lattice,
with an even number $k$ of vertices per fundamental domain. $\lat$ can be equipped with a \emph{periodic Kasteleyn orientation\/} in which every face is clockwise odd (see \S\ref{ss:odd.general}); this defines an infinite-dimensional weighted signed adjacency matrix
(\emph{Kasteleyn matrix}) $K$, with entries $K_{ij}\equiv (\I_{i\to j}-\I_{j\to i})\nu_{ij}$ for $i,j\in\lat$.
For $z,w\in\C$ and \[E=\begin{pmatrix}u&v\\x&y\end{pmatrix}\in\End\,,\]
define a $(z,w)$-periodic function to be a function 
$f:\lat\to\C$ satisfying
$f( p+a(u,v)+b(x,y) ) = f(p) z^a w^b$ for $p\in\lat$ and $a,b\in\Z$.
We let
\beq\label{e:KEzw}
K_E(z,w)
\quad\text{the ``Fourier transform of $K$ with respect to
	$E$''}
\eeq
denote the action of $K$ on the (finite-dimensional) space of
$(z,w)$-periodic functions.
We write $K(z,w)\equiv K_I(z,w)\in\C^{k\times k}$ (where $I$ is the identity matrix) and call
\[P(z,w)\equiv \det K(z,w)
\quad\text{the \emph{characteristic polynomial\/} of $\lat$.}\]
Note that $K(z,w)^{\operatorname{t}}=-K(\slf{1}{z},\slf{1}{w})$, so $P(z,w)=P(\slf{1}{z},\slf{1}{w})$.

\subsection{Bipartite characteristic polynomial}
\label{ss:prelim.bip}

Note that in \eqref{e:KEzw}
the linear map $K(z,w)$
was defined without reference to a basis,
which is unnecessary for defining the determinant.
To consider Pfaffians of $K(z,w)$, however, we must fix a basis: from the relation
$\pf (MKM^t) = (\det M)(\pf K)$
it is clear that even an orthogonal change of basis can change the sign of the Pfaffian.
We therefore assume a fixed ordering $1,\ldots,k$
of the vertices of the fundamental domain, and take the basis
$(f_1,\ldots,f_k)$
where
$f_p(q) = z^a w^b$
if $q$ is the vertex corresponding to $p$ in the $(a,b)$-translate
of the fundamental domain,
and $f_p(q)=0$ for all other $q$.
For the action of $K_E(z,w)$
fix any ordering of the fundamental domains and
take the basis
\beq\label{e:pfaffian.basis}
(f^1_1,\ldots,f^1_k,\ldots,f^{\det E}_1,\ldots,f^{\det E}_k)
\eeq
where $\smash{f^{\bm{e}}_p(q)}$
is the $(z,w)$-periodic function (with period $E$)
corresponding to the $p$-vertex in the $\bm{e}$-th fundamental domain.\footnote{Since the number of vertices per fundamental domain is even, the arbitrary ordering of fundamental domains within $\lat_E$ will not affect the Pfaffian.}

If a planar graph $\Graph$ (with positive edge weights) is bipartite with parts $B$ (black) and $W$ (white), an equivalent characterization of a Kasteleyn orientation is that the boundary of each non-external face has an odd or even number of edges $B\to W$ according to whether its length is $0$ or $2$ modulo $4$.\footnote{More generally, if imaginary weights are allowed, the condition is that the product of signed $B\to W$ edge weights is negative or positive real according to whether the length is $0$ or $2$ modulo $4$.}

Suppose $\lat$ has bipartite fundamental domain, with
$\slf{k}{2}$ vertices of each color; and
for $E\in\End$ let $b_E\equiv (\det E)\slf{k}{2}$.
The action of $K_E(z,w)$ interchanges the $(z,w)$-periodic functions supported on $B$ with those supported on $W$:
from the basis \eqref{e:pfaffian.basis},
there is an orthogonal change-of-basis matrix $O$
with $\det O = (-1)^{b_E(b_E-1)/2}$ such that
\[O\,K_E(z,w)\,O^t
=
\begin{pmatrix}
	0 & \textsc{k}_E(z,w)\\
	-\textsc{k}_E(\slf{1}{z},\slf{1}{w})^t&0
\end{pmatrix}
\equiv \wt{K}_E(z,w),\]
with $\textsc{k}_E(z,w)$ the action of $K_E(z,w)$
from $W$-supported to $B$-supported functions.
For $z,w\in\set{\pm1}$ the matrix
$K_E(z,w)$ is skew-symmetric, with Pfaffian
\beq\label{e:bip.pfaffian}
\pf K_E(z,w)
= (\det O) (\pf \wt{K}_E(z,w))
= \det \textsc{k}_E(z,w)
\equiv Q_E(z,w).\eeq
The \emph{bipartite characteristic polynomial\/} is
$Q(z,w)\equiv Q_I(z,w)$.
In this setting it is known
that $Q(z,w)$ either has no roots on the unit torus or two roots, which are necessarily complex conjugates;
it is possible for the roots to coincide \cite{\ksdimers}.  Simple zeroes of $Q(z,w)$ are nodes of $P(z,w)\equiv Q(z,w)Q(\slf{1}{z},\slf{1}{w})=\abs{Q(z,w)}^2$ with associated positive-definite Hessian
\beq\label{e:bip.hess}
H
=\bpm
\abs{z \pd_z Q}^2 & \real[ z \pd_z Q \, \ol{w \pd_w Q} ] \\
\real[ z \pd_z Q \, \ol{w \pd_w Q} ] & \abs{w \pd_w Q}^2
\epm\bigg|_{(z,w)=(z_0,w_0)}
=\bpm \AAz & B \\ B & \AAw \epm.
\eeq
In particular, distinct conjugate nodes
of $P$ must have the same Hessian matrix. If instead $Q$ has a real node then $P$ vanishes there to fourth order, but the finite-size corrections to $\zz_E$ can be determined using the second-order expansion of $Q$.

\subsection{Pfaffian method for toric graphs}

For \emph{non-planar\/} graphs Kasteleyn orientations do not in general exist. Instead the dimer partition function of the toric graph $\lat_E$ can be computed as a linear combination of four Pfaffians,
as follows (cf.\ \cite{\kastquadratic}).

Fix arbitrarily a \emph{reference matching\/} $\match_0$ of the fundamental domain, and ``unroll'' the matching to obtain a periodic reference matching $\match_\infty$ of $\lat$.
Assume that no edges of the reference matching cross
	between different fundamental domains
(which can be achieved by deforming the domain boundaries
	in a periodic manner), so that
$\match_0$
occurs with the same sign in $\pf K(z,w)$ for all $z,w\in\set{\pm1}$.
This sign can be switched by reversing the orientation of all edges incident to any single vertex,
and we hereafter take it to be $+1$.
If $\match_E$ is the projection of $\match_\infty$ to $\lat_E$,
then for the basis
\eqref{e:pfaffian.basis} we have
$(\pf K_E(+1,+1))_{\match_E}
=(\pf K(+1,+1))_{\match_0}^{\det E}$
--- thus $\match_E$ appears with sign $+1$
in $\pf K_E(z,w)$ for
all $E\in\End$ and all $z,w\in\set{\pm1}$.

Next, say that an even-length cycle on $\lat_E$ is \emph{$\match_E$-alternating\/} if every other edge comes from $\match_E$. All $\match_0$-alternating cycles on the fundamental domain with the same homology must occur with the same sign: to see this, let $C_1,C_2$ be two $\match_0$-alternating cycles of the same homology type. Then we can transform $C_1$ to $C_2$ by deforming the cycle across planar faces one at a time (the intermediate cycles need not have even length). Switching $C_1$ with $C_2$ as needed, we may assume that each face traversed by this process has boundary partitioned into a segment $\gm_-$ (containing $\ell_-$ edges) which is traveled in the negative direction by the cycle just before the face is traversed, and another segment $\gm_+$ (containing $\ell_+$ edges) which is traveled in the positive direction by the cycle just after the face is traversed. Since the face is clockwise odd (i.e., has negative sign in the counterclockwise direction), $\sgn(\gm_-)\sgn(\gm_+) = -(-1)^{\ell_-}$. The deformation from $\gm_-$ to $\gm_+$ ``crosses'' $\ell_--1$ vertices in the sense that it brings $\ell_--1$ more vertices (strictly) to the left of the cycle. Thus the total sign change between $C_1$ and $C_2$ is $(-1)^\ell$ with $\ell$ the total number of vertices crossed.  Since $C_1$ and $C_2$ are both $\match_0$-alternating, $\match_0$ must restrict to a perfect matching of the $\ell$ vertices crossed: therefore $\ell$ must be even, and so $\sgn(C_1)=\sgn(C_2)$ as claimed.

Appropriately reversing edges along horizontal or vertical ``seams'' (boundaries separating adjacent copies of the fundamental domain) produces a periodic Kasteleyn orientation of $\lat$ such that in any $\lat_E$ with the inherited orientation, \emph{every\/} $\match_E$-alternating cycle has sign $+1$. We hereafter assume that the lattice $\lat$ has been ``pre-processed'' such that all these sign conditions hold, that is:

\bdfn\label{d:oriented}
Fix $\match_0$ a reference matching of the fundamental domain $\lat_I$, let $\match_\infty$ denote its periodic extension to $\lat$. We say that $\lat$ is \emph{$\match_0$-oriented\/} if
(i) no edges of $\match_\infty$ cross boundaries separating different copies of the fundamental domain,
(ii) $\match_0$ occurs with positive sign in $\pf K(+1,+1)$
(hence in all four Pfaffians
	$\pf K(\pm1,\pm1)$),
and
(iii) every $\match_0$-alternating cycle in the fundamental domain has sign $+1$.
\edfn

For $\ttr,\tts\in\set{0,1}$ let $\zzE[\ttr\tts]$ denote the partition function of matchings $\match$ such that the superposition of $\match$ with $\match_0$ is of homology $(\ttr,\tts)$ modulo~2. For \emph{any\/} periodic Kasteleyn orientation of $\lat$, it is easily seen that
\beq\label{e:torus.sign.relations}
\f{(\pf K_E(z,w))_\match\,(\pf K_E(z,w))_{\match_0}}
{(\pf K_E(1,1))_\match\,(\pf K_E(1,1))_{\match_0}}
= z^\ttr w^\tts
\quad
\begin{array}{l}
\text{for $z,w\in\set{\pm1}$;}\\
\text{$\match$ contributing to $\zzE[\ttr\tts]$}.
\end{array}
\eeq
Specializing to the case that $\lat$ is $\match_0$-oriented, the argument of \cite{\kastquadratic} (also explained in \cite[Ch.~4]{McCoy-Wu})
gives the following

\bppn\label{p:torus.z0}
If lattice $\lat$ is $\match_0$-oriented, then
\beq\label{e:torus.table}
\bpm
-\pf K_E( +1,+1 )\\
+\pf K_E( +1,-1 )\\
+\pf K_E( -1,+1 )\\
+\pf K_E( -1,-1 )\\
\epm
=\underbrace{\bpm -1&1&1&1\\
	1&-1&1&1\\
	1&1&-1&1\\
	1&1&1&-1 \epm}_{\Signs}
\bpm \zzE[\zro\zro]\\
	\zzE[\one\zro]\\
	\zzE[\zro\one]\\
	\zzE[\one\one]
\epm.
\eeq
In particular, the dimer partition function of $\lat_E$ is
\[\zzE
=\tf12[
	-\pf K_E(+1,+1)
	+\pf K_E(+1,-1)
	+\pf K_E(-1,+1)
	+\pf K_E(-1,-1)
	].\]
\eppn

We shall also define $\ZZE[\ttr\tts]$ to be the partition function of double-dimer configurations $\match_1\oplus\match_2$ with homology $(\ttr,\tts)$ modulo 2. It can be seen from \eqref{e:torus.sign.relations} that
\beq\label{e:z.dd}
\TS\ZZE[\ttr\tts]
\equiv
\sum_{\ttr',\tts'} \zzE[\ttr'\tts']
	\zzE[(\ttr'+\ttr)(\tts'+\tts)].
\eeq
The double-dimer partition function on $\lat_E$ is given by the sum $\ZZE=\sum_{\ttr,\tts}\ZZE[\ttr\tts]$.

\subsection{Special functions and Poisson summation}
\label{ss:jacobi}

For dimer systems on tori we find finite-size corrections which can be expressed in terms of the \emph{Jacobi theta functions\/}
 $\vth_{\ttr\tts}$ ($\ttr,\tts\in\set{0,1}$), whose definition we now briefly recall (for further information see e.g.\ \cite{\batemanii}). These are functions $\vth_{\ttr\tts}(\var\giv\tau)$ of complex variables $\var$ and $\tau\equiv\taux+i\tauy$, with $\tauy>0$, expressed equivalently as functions $\vth_{\ttr\tts}(\var,\qtau)$ of $\var$ and the
 \emph{nome}
\beq\label{e:nome}
\qtau\equiv e^{\pi i\tau}
= e^{\pi i\taux}|\qtau|.
\eeq
Note that $|\qtau|<1$.
Each theta function is given by an infinite sum:
\\[-6pt]\hypertarget{Jacobi_theta}\beq\label{e:jacobi.theta}
\begin{array}{rl}
\vth_{\zro\zro}(\var\giv\tau)
	\hspace{-6pt}&\equiv
	\sum_{j\in\Z}e^{2\pi i j\var}\qtau^{j^2},\\
\vth_{\zro\one}(\var\giv\tau)
	\hspace{-6pt}&\equiv
	\sum_{j\in\Z} (-1)^j e^{2\pi i j\var}\qtau^{j^2},\\
\vth_{\one\zro}(\var\giv\tau)
	\hspace{-6pt}&\equiv
	\sum_{j\in\Z+1/2}e^{2\pi i j\var}\qtau^{j^2},\\
\vth_{\one\one}(\var\giv\tau)
	\hspace{-6pt}&\equiv
	i \sum_{j\in\Z+1/2} (-1)^{j-1/2} e^{2\pi i j\var}\qtau^{j^2}.
\end{array}
\eeq
The theta functions also have infinite product expressions, as follows:
\beq\label{e:jacobi.product}
\begin{array}{l}
\vth_{\zro\zro}(\var\giv\tau)
	=G(\qtau) \prod_{\ell\in\N-1/2}
	(1+2\qtau^{2\ell} (\cos2\pi\var) + \qtau^{4\ell})\\
\vth_{\zro\one}(\var\giv\tau)
	= G(\qtau) \prod_{\ell\in\N-1/2}
	(1-2\qtau^{2\ell} (\cos2\pi\var) + \qtau^{4\ell})\\
\vth_{\one\zro}(\var\giv\tau)
=\smash{2\qtau^{1/4}
	(\cos \pi\var) G(\qtau)
	\prod_{\ell\ge1} (1
		+2\qtau^{2\ell}(\cos 2\pi\var) + \qtau^{4\ell})}\\
\vth_{\one\one}(\var\giv\tau)
=\smash{
	-2\qtau^{1/4} (\sin \pi\var) G(\qtau)
	\prod_{\ell\ge1} (1-2\smash{\qtau^{2\ell}} (\cos 2\pi\var)
	+ \qtau^{4\ell})},
\end{array}
\eeq
\hypertop{Dedekind}
where $G(q)\equiv\prod_{j\ge1} (1-q^{2j})$;
this is also the $q$-Pochhammer symbol
$(q^2;q^2)_\infty$. The 
\emph{Dedekind $\eta$ function\/} is
\beq\label{e:eta}\deta(\tau)\equiv \qtau^{1/12} G(\qtau).\eeq
We also write $\vth_{\ttr\tts}(\tau)\equiv\vth_{\ttr\tts}(0\giv\tau)$; the function $\vth_{\one\one}(\tau)$ is identically zero.\footnote{Another standard notation is given by $\vth_1\equiv-\vth_{\one\one}$, $\vth_2\equiv\vth_{\one\zro}$, $\vth_3\equiv\vth_{\zro\zro}$, $\vth_4\equiv\vth_{\zro\one}$.}

Many useful theta function identities may be found in \cite{\conwaysloane,\batemanii} (see in particular \cite[p.~356]{\batemanii}). The theta functions satisfy the relations
\beq\label{e:theta.quasiperiodic}
\vth_{\zro\zro}(\var+1\giv\tau)
	=\vth_{\zro\zro}(\var\giv\tau),\quad
\vth_{\zro\zro}(\var+\tau\giv\tau)
	=e^{-\pi i(2\var+\tau)}\vth_{\zro\zro}(\var\giv\tau).
\eeq
The four theta functions are related by the transformations
\beq\label{e:between.thetas}
\begin{array}{rl}
\vth_{\zro\zro}(\var+\tf12\giv\tau)
	\hspace{-6pt}&=\vth_{\zro\one}(\var\giv\tau),\\
\vth_{\zro\zro}(\var+\tf12\tau\giv\tau)
	\hspace{-6pt}&=e^{-\pi i(\var+\tau/4)}\vth_{\one\zro}(\var\giv\tau)\\
\vth_{\zro\zro}(\var+\tf12(1+\tau)
	\giv\tau)
	\hspace{-6pt}&=-ie^{-\pi i(\var+\tau/4)}\vth_{\one\one}(\var\giv\tau).
\end{array}
\eeq

\subsection{Finite-size correction and Gaussian sum formulas}
\label{ss:cf.relns}

The correction appearing in \eqref{e:double.product.fsc}
is expressed in terms of these special functions as follows:
for $\phi,\psi\in\R$ and $\ttr,\tts\in\set{0,1}$ we define
\\[-\baselineskip]\hypertarget{e:cf} \beq\label{e:cf}
\cf^{\ttr\tts}(-e^{2\pi i\phi},-e^{2\pi i \psi}\giv\tau)
\equiv
\bigg|
	\f{\vth_{\ttr\tts}(\phi\tau
	-\psi\giv\tau)
	e^{\pi i \tau \phi^2}}{\deta(\tau)}
\bigg|,
\quad\text{and }\cf\equiv\cf^{\zro\zro}.
\eeq
From the relations \eqref{e:theta.quasiperiodic}, the evaluation of $\cf$ does not depend on the integer parts of $\phi$ or $\psi$. It then follows from \eqref{e:between.thetas} that the same holds for any $\cf^{\ttr\tts}$, and further
\beq\label{e:xi.a}
\cf( (-1)^\ttr \ze,
	(-1)^\tts \xi \giv\tau)
=\cf^{\ttr\tts}(\ze,\xi\giv\tau)
\quad\text{for all }\ze,\xi\in\T.
\eeq
Recalling $\vth_{\ttr\tts}(\tau)\equiv\vth_{\ttr\tts}(0\giv\tau)$,
we define $\cf^{\ttr\tts}(\tau)\equiv\cf^{\ttr\tts}(-1,-1\giv\tau)$,
with $\cf^{\one\one}\equiv0$.

Expressions involving theta functions can often be transformed in a useful way using the Poisson summation formula:
for $f\in L^1(\R^d)$, denote its Fourier transform
$\smash{\hat f(k) = \int_{\R^d} e^{-2\pi i \ip{k}{x}} f(x) \d{x}}$.
With this normalization, the Gaussian function $e^{-\pi x^2}$ is preserved by the Fourier transform.
If both $f,\hat{f}$ satisfy
$|f(x)|+|\hat f(x)| \lesssim (1+|x|)^{-d-\de}$ for some $\de>0$,
then they are both continuous functions with
\[
\sum_{k\in\Z^d} \hat f(k) e^{2\pi i \ip{k}{x}}
= \sum_{n\in\Z^d} f(x + n)
\quad\text{for all $x\in\R^d$.}\]
(see e.g.\ \cite{\grafakos}).
Our typical application of this formula is to transform expressions involving theta functions into partition functions for discrete Gaussian distributions:

\bdfn\label{d:disc.gauss}
For $\mu\in\R^d$, $\covmatz\in\R^{d\times d}$ positive-definite, and $L$ any discrete subset of $\R^d$,
the \emph{discrete Gaussian on $L$\/}
with center parameter $\bm{\mu}$ and scale parameter $\covmatz$
is the $L$-valued random variable $X$ with
\[\P(X=\bm{e}) =
\f{\exp\{ -\tf{\pi}{2}
	[(\bm{e}-\bm{\mu})\covmatz^{-1}(\bm{e}-\bm{\mu})] \}}
{ \sum_{ \bm{e}'\in L }
	\exp\{ -\tf{\pi}{2}
		[(\bm{e}'-\bm{\mu})\covmatz^{-1}(\bm{e}'-\bm{\mu})] \} }\quad
\text{for }\bm{e}\in L.\]
\edfn
Only two-dimensional discrete Gaussians arise in this paper.

Theta functions and discrete Gaussian distributions
are related as follows.
Recalling the quadratic form $g_\tau(\bm{e})$ from \eqref{e:qfdef},
if $\tau=\tau[H]$ as in \eqref{e:tau.H}, then
\beq \label{e:qfdef2}
g_\tau(\bm{e}) = \bm{e}^t\covmatz^{-1}\bm{e}
	\text{ with } \covmatz = (\det H)^{-1/2} H.
\eeq

\blem\label{l:theta.abs2}
For $\ttr,\tts\in\set{\zro,\one}$ and $\phi,\psi\in\R$,
\beq\label{e:theta.abs2}
\cf^{\ttr\tts}(-e^{2\pi i\phi},-e^{2\pi i\psi}\giv\tau)^2
	=\f{
	\sum_{j,k\in\Z}
		(-1)^{(\ttr+k)(\tts+j)}
		\exp\{-\tf{\pi}2
			g_\tau(j-2\psi,k+2\phi)  \}}
	{|\deta(\tau)|^2 (2\tauy)^{1/2}},
\eeq

\bpf
Write $\var\equiv\phi\tau-\psi$.
Recalling \eqref{e:nome},
write $q\equiv\abs{\qtau}\equiv e^{-\pi\tauy}$.
Use \eqref{e:jacobi.theta} to expand
\[\begin{array}{rl}
&\abs{\vth_{\zro\zro}(\var\giv\tau)}^2
= \sum_{x,y\in\Z}
	\q^{x^2 + y^2 + 2\phi(x+y)}
	\exp\{2\pi i [ (\phi\taux-\psi) + \tf12\taux (x+y) ] (x-y)\}
	\vspace{4pt}\\
&= \tf12\sum_{j\in\Z}
	\q^{j^2/2+2\phi j}
	\sum_{k\in\Z}
	\q^{k^2/2} \exp\{2\pi i [(\phi\taux-\psi) + \tf12\taux j]k\}
	\vspace{4pt}\\
&\qquad+
	\tf12\sum_{j\in\Z}(-1)^j
	\q^{j^2/2+2\phi j}
	\sum_{k\in\Z}(-1)^k
	\q^{k^2/2}
	\exp\{2\pi i [(\phi\taux-\psi) + \tf12\taux j]k\}
\end{array}\]
where the second equality follows by the change of variables
$j\equiv x+y$, $k\equiv x-y$.
For both of the double sums appearing in the last expression,
apply Poisson summation over $k$ for each fixed $j$
to obtain (with some rearranging, and recalling \eqref{e:xi.a})
\[
\cf^{\zro\zro}(-e^{2\pi i\phi},-e^{2\pi i\psi}\giv\tau)^2
=\left| \f{q^{\phi^2}\vth_{\zro\zro}(\var\giv\tau)}{\deta(\tau)} \right|^2
= \f{
	\sum_{j,k\in\Z}
	(-1)^{jk}
	\exp\{-\tf{\pi}{2}
		g_\tau( j-2\psi,k+2\phi )\}
	}
	{|\deta(\tau)|^2 (2\tauy)^{1/2}},
\]
proving \eqref{e:theta.abs2} for $(\ttr,\tts)=(0,0)$. The formulas for the remaining values of $(\ttr,\tts)$ follow from \eqref{e:xi.a}.
\epf
\elem

\blem\label{l:theta.cross}
The zero-argument correction factors
$\cf^{\ttr\tts}(\tau)\equiv\cf^{\ttr\tts}(-1,-1\giv\tau)$
satisfy the cross product identities
\beq\label{e:cf.cross.product}
\begin{array}{l}
\cf^{\zro\zro}(\tau)\cf^{\zro\one}(\tau)
	= \cf^{\zro\one}(2\tau),\\
\cf^{\zro\zro}(\tau)\cf^{\one\zro}(\tau)
	= \cf^{\one\zro}(\slf\tau2),\\
\cf^{\zro\one}(\tau)\cf^{\one\zro}(\tau)
	= \cf^{\one\zro}(\tf{1+\tau}{2} ).
\end{array}
\eeq
For distinct pairs $\ttr_1\tts_1$ and $\ttr_2\tts_2$ belonging to $\set{\zro\zro,\zro\one,\one\zro}$,
\beq\label{e:theta.cross.prod}
\cf^{\ttr_1\tts_1}(\tau)
\cf^{\ttr_2\tts_2}(\tau)
=\f
{ \sum_{\bm{e}\in \Z^2} \exp\{
	-\tf{\pi}{4}
	g_\tau( 2\bm{e} + (\tts_1,\ttr_1)+(\tts_2,\ttr_2) )
	\} }
{|\deta(\tau)|^2(\tauy)^{1/2}}
\eeq

\bpf
We have the identities
(see e.g.\ \cite[\S4.1]{\conwaysloane})
\[\begin{array}{rlrl}
\vth_{\zro\zro}(\tau)\vth_{\zro\one}(\tau)
	\hspace{-6pt}&=\vth_{\zro\one}(2\tau)^2,
	\quad&
	2 g_{2\tau}(\bm{e})
	\hspace{-6pt}&=
	g_\tau(e_1,2e_2)\\
2\vth_{\zro\zro}(\tau)\vth_{\one\zro}(\tau)
	\hspace{-6pt}&=\vth_{\one\zro}(\slf\tau2)^2,
	\quad&
	2 g_{\tau/2}(\bm{e})
	\hspace{-6pt}&=
	g_\tau(2e_1,e_2)\\
2 e^{i\pi/4}
\vth_{\zro\one}(\tau)\vth_{\one\zro}(\tau)
	\hspace{-6pt}&= \vth_{\one\zro}( \tf{1+\tau}{2} )^2,
	\quad&
	2g_{(1+\tau)/2}(\bm{e})
	\hspace{-6pt}&=
	g_\tau(2e_1+e_2,e_2)
\end{array}
\]
Straightforward manipulations
using the product formulas \eqref{e:jacobi.product}
give \eqref{e:cf.cross.product}.
Combining with \eqref{e:theta.abs2} gives \eqref{e:theta.cross.prod}:
for example, recalling $\cf^{\one\one}(\tau)=0$,
\[
\cf^{\zro\zro}(\tau)\cf^{\zro\one}(\tau)
=\f{\cf^{\zro\one}(2\tau)^2+\cf^{\one\one}(2\tau)^2}
	{ | \deta(\tau)/\deta(2\tau) |^2 }
= \f{
	\sum_{\bm{e}\in(2\Z)^2 +(\one,\zro)}
	\exp\{ -\tf{\pi}{2} g_{2\tau}(e_1,\slf{e_2}{2}) \}}
	{ |\deta(\tau)|^2 (\tauy)^{1/2} },
\]
and combining with the above identity for $g_{2\tau}$ gives the formula.
\epf
\elem

\section{Finite-size correction to the torus partition function}
\label{s:fsc}

In this section we prove Theorem~\ref{t:z}, determining the finite-size corrections to the dimer partition function $\zzE$ as well as to its four components $\zzE[\ttr\tts]$. The critical non-bipartite setting (single real node) is treated in \S\ref{ss:torus.real.node}, while the critical bipartite setting (distinct conjugate nodes) is treated in \S\ref{ss:torus.two.nodes}. In both cases the asymptotic expansion of the \emph{absolute value\/} $|\pf K_E(\ze,\xi)|=P_E(\ze,\xi)^{1/2}$ is given by Theorem~\ref{t:double.product}, and we explain how to determine the sign of $\pf K_E(\ze,\xi)$. Applying $\Signs^{-1}$ to both sides of \eqref{e:torus.table} then gives expressions for the finite-size corrections to the quantities $\zzE[\ttr\tts]$ as signed combinations of absolute values or squared absolute values of the functions $\cf^{\ttr\tts}$. In some cases, we can apply
Lemma~\ref{l:theta.abs2} to obtain Gaussian sum formulas
for the finite-size corrections.
Lastly, in \S\ref{ss:modular}
we explain the interpretation of $\tau$ as the shape of the torus in its ``conformal'' embedding.

We take the standard branch of the logarithm, which is holomorphic on $\C\setminus\R_{\le0}$,
and continuous in $z$ as it approaches the negative real half-line
from the upper half-plane.
If for $z\in\T$ we write
$e^{i\te}\equiv z$,
unless otherwise specified we mean that
$-\pi<\te\le\pi$ equals
$\arg z$, the imaginary part of (the chosen branch of) $\log z$.

\subsection{Finite-size correction in presence of real nodes}
\label{ss:torus.real.node}

\subsubsection*{Calculations of the finite-size corrections}

\bppn[implies Theorem~\ref{t:z}\ref{t:z.a}]
\label{p:torus.nonbip}
Suppose $\lat$ is $\match_0$-oriented
with characteristic polynomial $P(z,w)$
which is non-vanishing on $\T^2$
except at a positive node
$(z_0,w_0)\in\set{\pm1}^2$
where it has expansion
\eqref{e:hess} with Hessian matrix $H$.
Then, in the limit \eqref{e:e.large},
\[\log\zzE[\ttr\tts]
	=(\det E)\,\freezero
	+ \logfscone^{\ttr+\tts_E,\tts+\ttr_E}(\tau_E) + o(1)\]
where $\tau_E\in\H$
and the indices $\ttr_E,\tts_E\in\set{0,1}$
are as in \eqref{e:tau.domain.phase}, and
$\logfscone^{\ttr\tts}\equiv\log\fscone^{\ttr\tts}$ with
\[\TS\fscone^{\ttr\tts}(\tau)
	\equiv
	\tf14 \sum_{\ttr',\tts'}
		(-1)^{(\tts+\ttr')(\ttr+\tts')}
		\cf^{\ttr'\tts'}(\tau).\]
Summing over $\ttr,\tts$ gives the statement of Theorem~\ref{t:z}\ref{t:z.a}.

\bpf
The asymptotic expansions of absolute values of the Pfaffians $\pf K_E(\ze,\xi)$ are given by Theorem~\ref{t:double.product}, so the issue is to determine their signs for $\ze,\xi\in\set{\pm1}^2$. The location of the node at $(z_0,w_0)$ implies
$\pf K_E(\phaseh_E,\phasev_E)=0$. If $\phaseh_E=\phasev_E=1$ then \eqref{e:torus.table} gives
\beq\label{e:z.twice.z00}
0
=-\zzE[\zro\zro]
	+\zzE[\one\zro]
	+\zzE[\zro\one]
	+\zzE[\one\one],
\eeq
implying that the other three entries on the left-hand side of \eqref{e:torus.table} must be non-negative.
A similar argument applies for the other three possibilities for the location of $(\phaseh_E,\phasev_E)$. It follows from Theorem~\ref{t:double.product} and \eqref{e:xi.a} that for $\ttr,\tts\in\set{0,1}$,
\[
\f{P_E((-1)^\ttr,
	(-1)^{\tts})^{1/2}}
{e^{(\det E)\,\freezero + o(1)}}
=
\cf^{\ttr\tts}(\phaseh_E,\phasev_E\giv\tau_E)
=\cf^{(\ttr+1,\tts+1)+(\ttr_E,\tts_E)}(\tau_E).\]
The matrix $\Signs$ in \eqref{e:torus.table} satisfies $\Signs=4\Signs^{-1}$, so
\[\bpm
\zzE[\zro\zro]\\
\zzE[\one\zro]\\
\zzE[\zro\one]\\
\zzE[\one\one]
\epm
=\Signs^{-1}
	\bpm
P_E(+1,+1 )^{1/2}\\
P_E(+1,-1)^{1/2}\\
P_E(-1,+1)^{1/2}\\
P_E(-1,-1)^{1/2}\\
\epm
=
e^{(\det E)\,\freezero + o(1)}\,
\tf14\Signs
\bpm
\cf^{(\one,\one)+(\ttr_E,\tts_E)}(\tau_E)\\
\cf^{(\one,\zro)+(\ttr_E,\tts_E)}(\tau_E)\\
\cf^{(\zro,\one)+(\ttr_E,\tts_E)}(\tau_E)\\
\cf^{(\zro,\zro)+(\ttr_E,\tts_E)}(\tau_E)
\epm,\]
implying the result.
\epf
\eppn

\bpf[Proof of Theorem~\ref{t:z}\ref{t:z.b.1}]
Suppose $Q$ has a single real node at $(z_0,w_0)=(\pm1,\pm1)$.  Let $(\phaseh_E,\phasev_E)$ be as in \eqref{e:tau.domain.phase}; $(\phaseh_E,\phasev_E)=(\pm1,\pm1)$.   A very slight modification to the proof of Theorem~\ref{t:double.product} gives
\[\log P_E(\ze,\xi)^{1/2}
=(\det E)\,\freezero
	+ 2 \log
	\cf(\ze/\phaseh_E,\xi/\phasev_E\giv\tau_E)
	+ o(1).\]
(The proof of Theorem~\ref{t:double.product}
approximates $P$ near a positive node by a certain polynomial $\Prect$ determined from the Hessian; see \eqref{e:Prect}. In this case we instead determine $\Prect$ from the Hessian associated to the node of $Q$, and approximate $P$ by the square of $\Prect$; the rest of the proof remains essentially unchanged.)
The finite-size correction to $\zz_E$ is then computed by the same argument as for Proposition~\ref{p:torus.nonbip}, except with $\cf$ replaced with~$\cf^2$.
Finally we observe that $\fsctwo(\pm1,\pm1\giv\tau)=\fsctwo(1,1\giv\tau)$.
\epf

\begin{figure}[t]
\begin{subfigure}[h]{.4\textwidth}
\includegraphics[width=\textwidth]{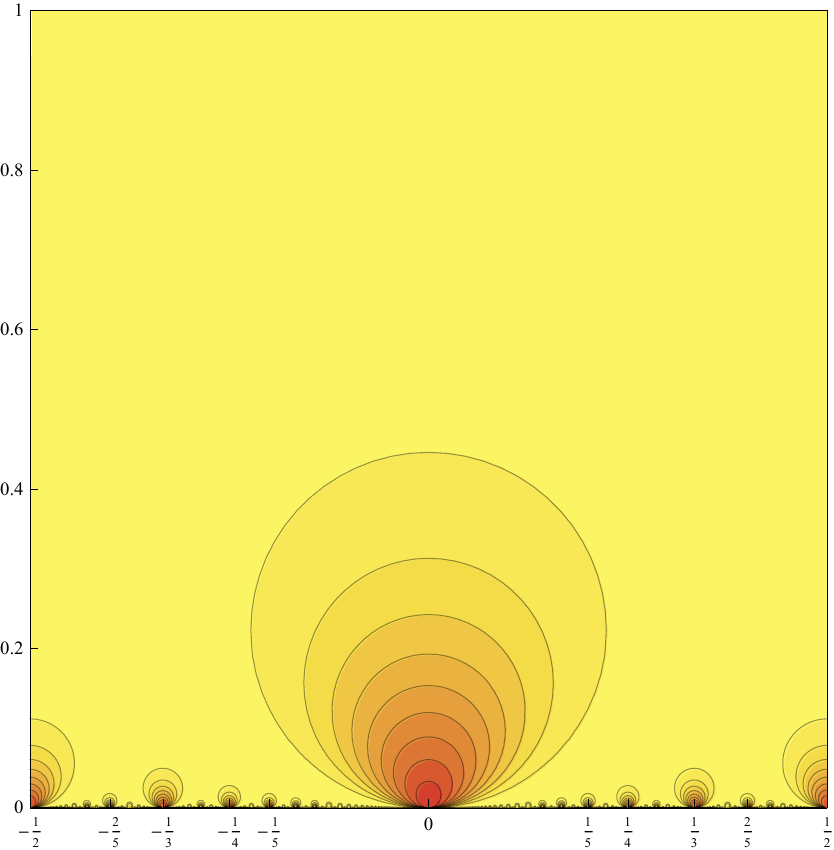}
\caption{$\logfscone(\tau)\equiv\log\fscone(\tau)$}
\end{subfigure}
\begin{subfigure}[h]{.4\textwidth}
\includegraphics[width=\textwidth]{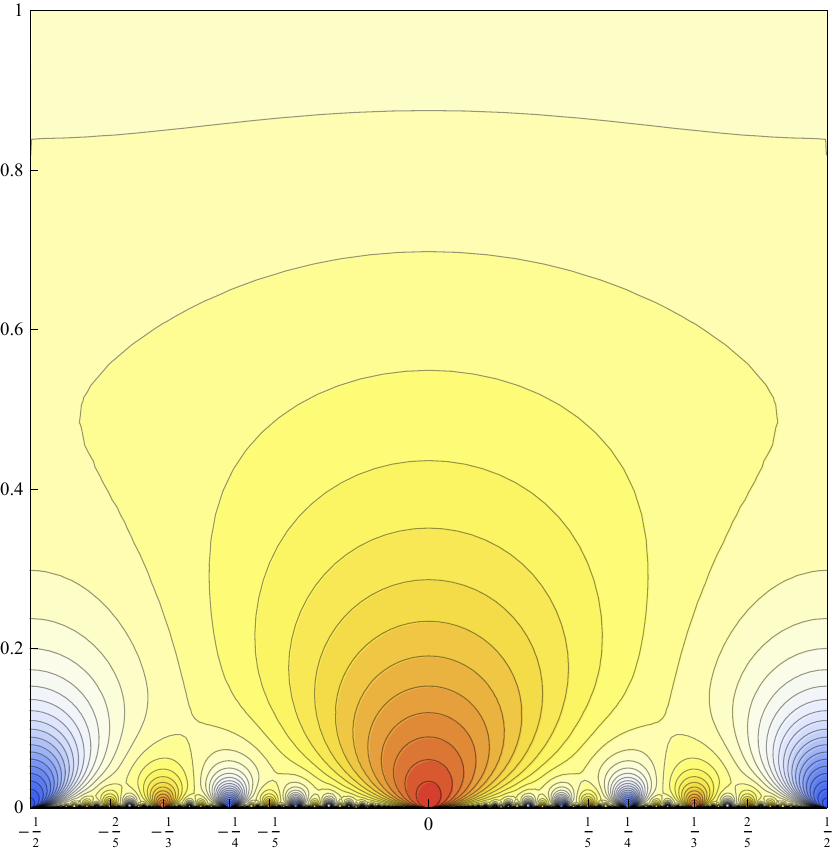}
\caption{$\log\cf(+1,-1\giv\tau)$}
\end{subfigure}
\begin{subfigure}[h]{.4\textwidth}
\includegraphics[width=\textwidth]{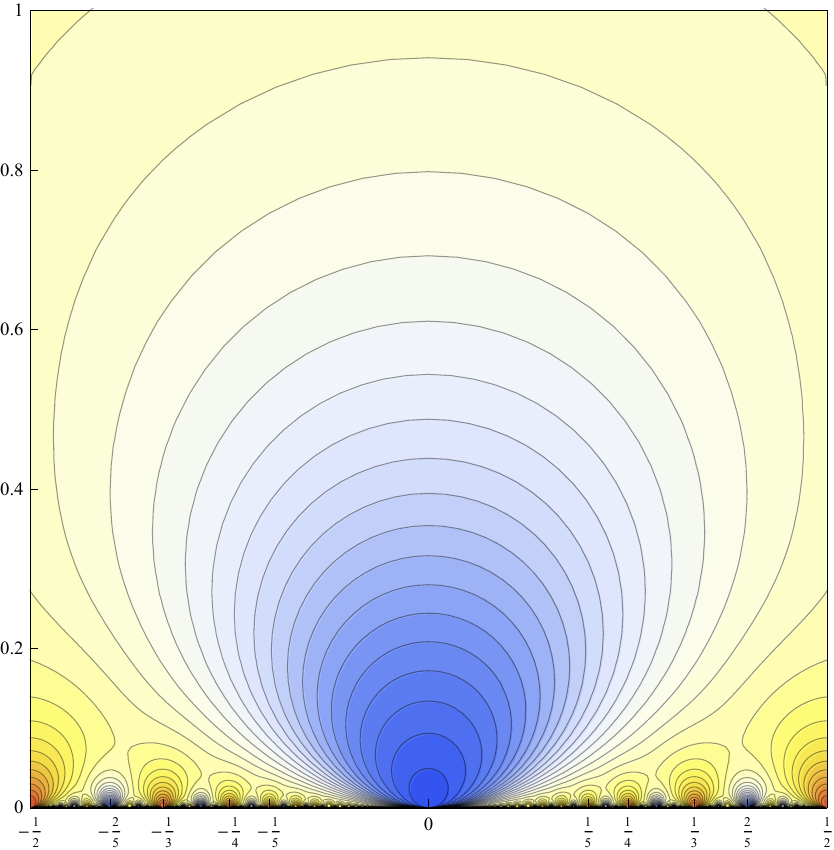}
\caption{$\log\cf(-1,+1\giv\tau)$}
\end{subfigure}
\begin{subfigure}[h]{.4\textwidth}
\includegraphics[width=\textwidth]{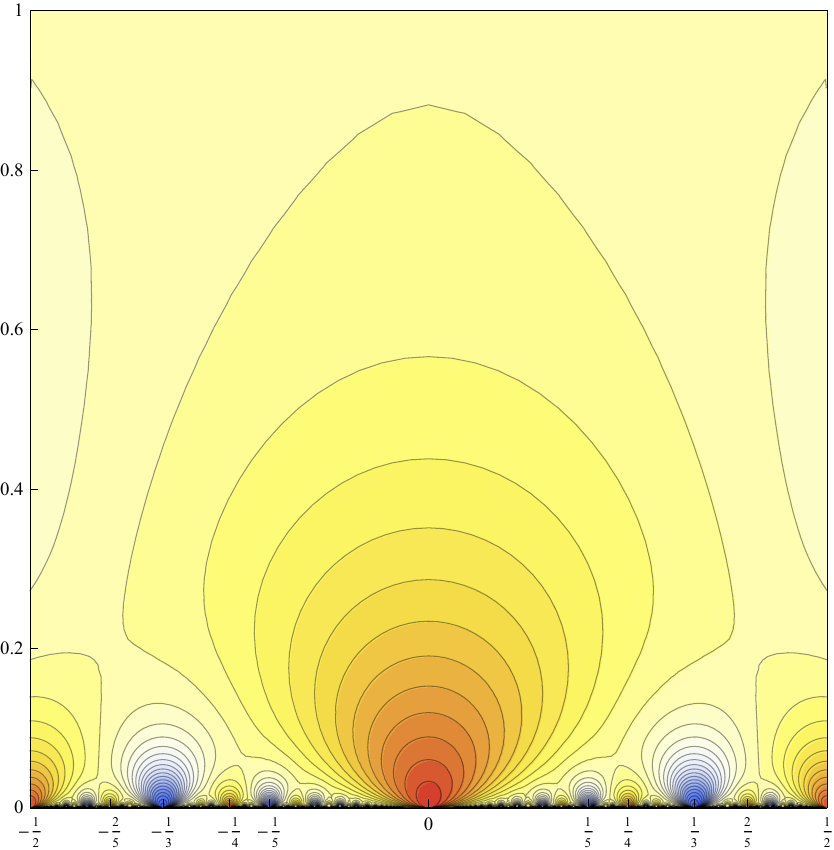}
\caption{$\log\cf(-1,-1\giv\tau)$}
\end{subfigure}
\caption{(Theorem~\ref{t:z}\ref{t:z.a})
Finite-size corrections
$\logfscone\equiv\log\fscone$
and $\log\cf$,\\
shown as a function of $\tau=x+iy$
for
$-\slf12\le x\le \slf12$, $0<y\le 1$.\\
(Recall $\fscone(\tau)=\tf12\sum_{\ze,\xi=\pm1}
	\cf(\ze,\xi\giv\tau)$,
	with $\cf(+1,+1\giv\tau)$
	identically zero.)
}
\label{f:logfscone}
\end{figure}

\begin{figure}[htbp]
\begin{subfigure}[h]{.4\textwidth}
\centering
\includegraphics[width=\textwidth]{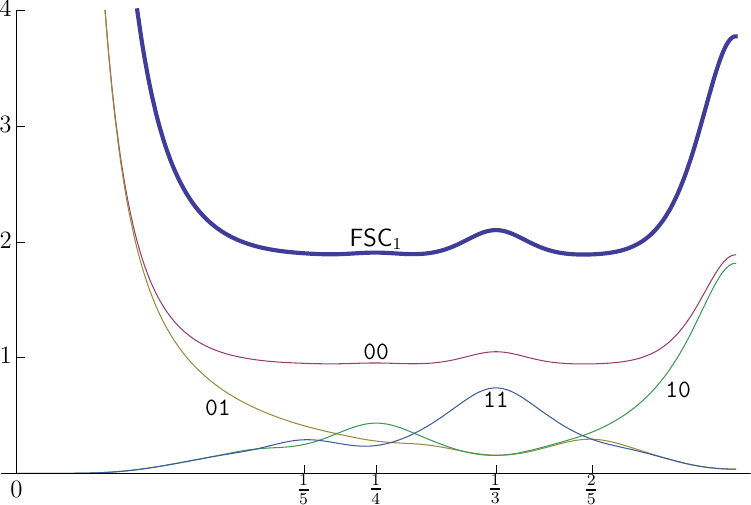}
\caption{$y=\slf{1}{20}$}
\label{f:fsc.single.node.y.one.div.twenty}
\end{subfigure}
\quad
\begin{subfigure}[h]{.4\textwidth}
\centering
\includegraphics[width=\textwidth]{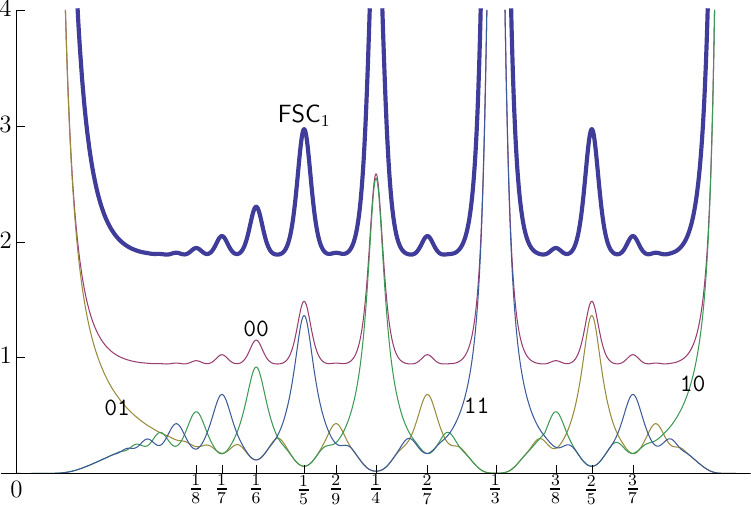}
\caption{$y=\slf{1}{100}$}
\label{f:fsc.single.node.y.one.div.hundred}
\end{subfigure}
\caption{(Proposition~\ref{p:torus.nonbip})
$\fscone(x+iy)$ (bold) and $\fscone^{\ttr\tts}(x+iy)$ with $y$ fixed}
\end{figure}

Clearly, the most straightforward application of Proposition~\ref{p:torus.nonbip} is when $z_0=w_0=+1$:
in this case $\pf K_E(+1,+1)=0$,
implying that $\zzE=2\zzE[\zro\zro]$ and
\[
\log\zzE[\ttr\tts] = (\det E)\,\freezero+\logfscone^{\ttr\tts}(\tau_E)+o(1).\]
Figure~\ref{f:logfscone}
shows $\logfscone(x+iy)$ as a function of $(x,y)$. Figures~\ref{f:fsc.single.node.y.one.div.twenty}~and~\ref{f:fsc.single.node.y.one.div.hundred} show $\fscone(x+iy)$ together with the individual contributions $\fscone^{\ttr\tts}(x+iy)$ as a function of $x$ with $y$ fixed at $\slf{1}{20}$
and $\slf{1}{100}$ respectively.

If $(z_0,w_0)\ne(1,1)$, then the correspondence between $\zzE[\ttr\tts]$ and $\fscone^{\ttr'\tts'}$ depends on the parities of the entries of $E$.
On the other hand, the following corollary shows that finite-size corrections to the \emph{double}-dimer partition function do not depend on the location of the node $(z_0,w_0)$:

\bcor
In the setting of Proposition~\ref{p:torus.nonbip},
\begin{align*}
\f{\ZZE[\ttr\tts]}
	{ e^{(\det E)\,2\freezero+o(1)} }
&=
\f{\sum_{\bm{e}\in\Z^2}
	\exp\{ -\tf{\pi}{4} g_\tau(2\bm{e}+(\ttr,\tts))\}}
	{2\abs{\deta(\tau)}^2(\tauy)^{1/2}}
\quad\text{if }(\ttr,\tts)\ne(0,0),\\
\f{\ZZE[\zro\zro]}
	{ e^{(\det E)\,2\freezero+o(1)} }
&=
\f{ \sum_{\bm{e}\in\Z^2}
	\exp\{ -\tf{\pi}{2} g_\tau(\bm{e}) \}}
	{ 2 |\deta(\tau)|^2 (2\tauy)^{1/2} }
	=
	\f{\sum_{\bm{e}\in\Z^2}
	\exp\{ -\tf{\pi}{2} g_\tau(2\bm{e}) \}}
	{ |\deta(\tau)|^2 (2\tauy)^{1/2} }
\end{align*}

\bpf
From the definition \eqref{e:z.dd} of $\ZZE[\ttr\tts]$ and from Proposition~\ref{p:torus.nonbip} we find
\[
\TS\ZZE[\ttr\tts]
=e^{(\det E)\,2\freezero+o(1)}\,
	\tf14 \sum_{\ttr'\tts'}
	\cf^{\ttr'\tts'}(\tau)
	\cf^{(\ttr',\tts')+(\tts,\ttr)}(\tau),
\]
\emph{regardless\/} of where the real node is located.
Recalling $\cf^{\one\one}(\tau)\equiv0$ gives
\[\bpm
\ZZE[\zro\one]\\
\ZZE[\one\zro]\\
\ZZE[\one\one]
\epm
= e^{(\det E)\,2\freezero+o(1)}\,
	\tf12
\bpm
\cf^{\zro\zro}(\tau)\cf^{\one\zro}(\tau)\\
\cf^{\zro\zro}(\tau)\cf^{\zro\one}(\tau)\\
\cf^{\zro\one}(\tau)\cf^{\one\zro}(\tau)
\epm
\]
and applying \eqref{e:theta.cross.prod} gives the result for $(\ttr,\tts)\ne(\zro,\zro)$.
Using
$\cf^{\one\one}(\tau)\equiv0$ again gives
\[
\f{\ZZE[\zro\zro]}{ e^{(\det E)\,2\freezero+o(1)} }
=\tf14[\cf^{\zro\zro}(\tau)^2
		+\cf^{\zro\one}(\tau)^2
		+\cf^{\one\zro}(\tau)^2
		\pm \cf^{\one\one}(\tau)^2]
\]
and combining with
\eqref{e:theta.abs2}
gives the two expressions for the case $(\ttr,\tts)=(\zro,\zro)$.
\epf
\ecor

A similar calculation give the finite-size correction
in the presence of two real nodes:

\bpf[Proof of Theorem~\ref{t:z}\ref{t:z.c}]
There are two domain phases,
$(\phaseh^j_E,\phasev^j_E) \equiv (z_j^u w_j^v,z_j^x w_j^y)$ ($j=1,2$),
where we must have
$\smash{\pf K_E(\phaseh^j_E,\phasev^j_E)=0}$.
It follows that all four entries on the left-hand side of
\eqref{e:torus.table}
must be non-negative, therefore
\[
\bpm \zzE[\zro\zro]\\
	\zzE[\one\zro]\\
	\zzE[\zro\one]\\
	\zzE[\one\one]
\epm
= e^{(\det E)\,\freezero+o(1)}
\Signs^{-1}
\bpm
\cf^{\one\one}(\phaseh^1_E,\phasev^1_E\giv\tau_E)
	\cf^{\one\one}(\phaseh^2_E,\phasev^2_E\giv\tau_E)\\
\cf^{\one\zro}(\phaseh^1_E,\phasev^1_E\giv\tau_E)
	\cf^{\one\zro}(\phaseh^2_E,\phasev^2_E\giv\tau_E)\\
\cf^{\zro\one}(\phaseh^1_E,\phasev^1_E\giv\tau_E)
	\cf^{\zro\one}(\phaseh^2_E,\phasev^2_E\giv\tau_E)\\
\cf^{\zro\zro}(\phaseh^1_E,\phasev^1_E\giv\tau_E)
	\cf^{\zro\zro}(\phaseh^2_E,\phasev^2_E\giv\tau_E)
\epm.
\]
Thus,
with $(\phaseh_E,\phasev_E)
\equiv (\phaseh^1_E\phaseh^2_E,\phasev^1_E\phasev^2_E)
\equiv ( (-1)^{\ttr_E},(-1)^{\tts_E} )$,
we conclude
\[\TS\zzE
= e^{(\det E)\,\freezero+o(1)}
\tf12 \sum_{\ttr,\tts}
	\cf^{\ttr\tts}(\tau_E)
	\cf^{(\ttr,\tts)+(\ttr_E,\tts_E)}(\tau_E)
= e^{(\det E)\,\freezero+o(1)}
	\fscodd(\phaseh_E,\phasev_E
	\giv\tau_E)
\]
as claimed.
The alternative expressions given
in the theorem statement for the four
$\fscodd(\pm1,\pm1\giv\tau)$
follow from \eqref{e:cf.cross.product}.
\epf

\subsubsection*{Ising model on
triangular lattice and Onsager's lattice}

By way of example we discuss Proposition~\ref{p:torus.nonbip}
in the context of the Ising model.
Recall \eqref{e:ising.dimers}
the correspondence between the low-temperature expansion of the Ising model on the triangular lattice
and the dimer model on the Fisher lattice.
Criticality for the latter model
(vanishing of the Fisher characteristic polynomial
on the unit torus)
is characterized in Proposition~\ref{p:fisher.crit}: in particular,
$P(+1,+1) = \ka_\circ^2$,
so $P(+1,+1)=0$ includes the cases
\[
\begin{array}{rl}
a=b=c=\sqrt{3} &
	\text{(critical ferromagnetic Ising on triangular lattice),}\\
a+b+1=ab &
	\text{(critical ferromagnetic Ising on Onsager's lattice),}
\end{array}
\]
(in particular the homogeneous Onsager's lattice with
$\be_a=\be_b=\tf12\log(\sqrt{2}+1)$).
Meanwhile,
$P(-1,-1)=\ka_c^2$,
so $P(-1,-1)=0$
includes the case
\[\begin{array}{rl}
a^{-1}+b^{-1}+1=(ab)^{-1}
&
\text{(critical anti-ferromagnetic Ising on Onsager's lattice).}
\end{array}\]
Thus, as expected, we see parity dependence in the finite-size correction to \eqref{e:ising.dimers} only in the anti-ferromagnetic case.

We also comment briefly on the observation
(cf.~\eqref{e:z.twice.z00})
that
\beq\label{e:z.twice.zrs}
(\phaseh_E,\phasev_E) = ((-1)^{\ttr_E},(-1)^{\tts_E})
\quad\text{implies}\quad
\zzE = 2\zzE[\tts_E\ttr_E].
\eeq
In the setting of Onsager's lattice
with ferromagnetic coupling constants $\be_a,\be_b>0$,
this can be understood in terms of a duality transformation
(see \cite{LiCMP}):
in addition to the low-temperature correspondence
\eqref{e:ising.dimers} we also have
\[\TS
\begin{array}{rl}
\text{\small (Onsager's lattice Ising)}
	&\zzIsing_{m,n} \\
\text{\small (Fisher lattice
	dimers, homology $(n,m)$)}
	&= (\prod_e e^{\be_e})
		\cdot 2
		\cdot \zz^{n,m}_{m,n}[e^{-2\be_a},e^{-2\be_b},1]\\
\text{\small (Fisher lattice dimers,
	all homologies)}
	&= 2^{mn} (\prod_e \cosh\be_e)
		\zz_{m,n}[\tanh\be_a,\tanh\be_b,1].
\end{array}
\]
The first equivalence is
again obtained through the low-temperature expansion, except that instead of considering the
polygonal configuration on the dual lattice
formed by the spin domain boundaries,
we take the complementary polygonal configuration
which includes each dual edge separating like spins.
On the $m\times n$ torus, the partition function
 is restricted to dimer covers
$\match$ with $\wind\match\oplus\refmatch=(n,m)\bmod2$
because the original spin domain boundaries
must have homology $(\zro,\zro)$
(cf.~\eqref{e:ising.dimers}).
The second equivalence is obtained through the usual high-temperature expansion, with no restriction on the homology of the dimer cover.

From \eqref{e:z.twice.zrs},
the finite-size correction to
$\zz^{n,m}_{m,n}[e^{-2\be_a},e^{-2\be_b},1]$ at criticality
will be sensitive to the parity of $(m,n)$
\emph{unless\/} $P(-1,-1)=\ka_c=0$,
which corresponds precisely to the ferromagnetic
critical line \eqref{e:onsager.critical}.
On this line,
\[\zz_{m,n}[e^{-2\be_a},e^{-2\be_b},1]
=2\cdot \zz^{m,n}_{m,n}[e^{-2\be_a},e^{-2\be_b},1]
=\zz_{m,n}[\tanh\be_a,\tanh\be_b,1]\]
where the first identity is from
\eqref{e:z.twice.zrs},
and the second follows by comparing the low- and high-temperature expansions
and observing that
$2(\cosh\be_a)(\cosh\be_b)
=e^{\be_a}e^{\be_b}$.
In particular, in the homogeneous case $\be_a=\be_b=\be=\tf12\log(\sqrt{2}+1)$, we have $e^{-2\be}=\tanh\be$, so that the Fisher lattice is self-dual under the above transformations. Thus the Ising--Fisher correspondences give an alternate proof of \eqref{e:z.twice.zrs} in this special instance; we emphasize however that \eqref{e:z.twice.zrs} holds in a much more general setting.

\subsection{Finite-size correction in presence of distinct conjugate nodes}\label{ss:torus.two.nodes}

\bppn[implies Theorem~\ref{t:z}\ref{t:z.b}]\label{p:torus.cf.bip}
Suppose the $\match_0$-oriented lattice $\lat$ has bipartite fundamental domain,
with bipartite characteristic polynomial $Q(z,w)$ which is non-vanishing on $\T^2$ except at distinct conjugate zeroes $(z_0,w_0)\ne\smash{(\overline{z}_0,\overline{w}_0)}$. The two zeroes are positive nodes of the characteristic polynomial $P(z,w)=Q(z,w)\,Q(\slf{1}{z},\slf{1}{w})$ with the same Hessian matrix $H$.
Then, in the limit \eqref{e:e.large},
\[\log\zzE[\ttr\tts]
= (\det E)\,\freezero
	+ \logfsctwo^{\ttr\tts}(\phaseh_E,\phasev_E\giv\tau_E)
	+ o(1)\]
where $\tau_E,\phaseh_E,\phasev_E$ are as in \eqref{e:tau.domain.phase}, and
$\logfsctwo\equiv\log\fsctwo$ with
\[\TS
\fsctwo^{\ttr\tts}(\ze,\xi\giv\tau)
\equiv \tf14
\sum_{\ttr',\tts'}
	(-1)^{(\ttr+\tts')(\tts+\ttr')}
	\cf^{\ttr'\tts'}(-\ze,-\xi\giv\tau)^2.\]
Summing over $\ttr,\tts\in\set{0,1}$ gives
	the statement of Theorem~\ref{t:z}\ref{t:z.b}.
\eppn

\blem\label{l:bip.arg.principle}
In the setting of Proposition~\ref{p:torus.cf.bip},
the four quantities
\[Q_+^{\ttr\tts}
	\equiv
	(-1)^{\ttr\tts}Q( -(-1)^\ttr,-(-1)^\tts ),\quad
	\ttr,\tts\in\set{0,1}\]
must all be positive.

\bpf
Since all the $\zzE[\ttr\tts]$ are non-negative,
we see on the right-hand side of
\eqref{e:torus.table}
that either
all four entries are non-negative,
or exactly one is negative while the rest are positive.
In the case that $E$ is the two-dimensional identity matrix,
recalling \eqref{e:bip.pfaffian}
shows that (since $Q$ has no real zeroes on the unit torus)
\begin{equation}\label{eq:sign-Q}
\text{\textit{either three or four $Q_+^{\ttr\tts}$ are positive}}.
\end{equation}
\hypertop{ellhv}
For $z,w\in\C$ define the counts
\beq\label{e:ellh.ellv}
\begin{array}{rl}
\ellh(w)\hspace{-6pt}&\equiv
	\text{number of zeroes in $z$
	of $Q(z,w)$ inside unit circle;}\\
\ellv(z)\hspace{-6pt}&\equiv
	\text{number of zeroes in $w$
	of $Q(z,w)$ inside unit circle}
\end{array}
\eeq
(taken with multiplicity), where for the moment we leave the count undefined in the case that there is a zero exactly on the unit circle.

Now suppose that $z_0,w_0\notin\R$.
By the assumption of distinct conjugate nodes,
$\ellv(z)$ stays constant as $z$ travels around the unit circle
except that it jumps by one when $z$ crosses $z_0$ or $\overline{z}_0$:
thus
\beq\label{e:ell.diff.by.one}
|\ellh(+1)-\ellh(-1)|=|\ellv(+1)-\ellv(-1)|=1.
\eeq
By the argument principle, $\ellv(z)$ is also the total winding of the closed curve $Q(z,\T)$ around the origin,
or equivalently the total change in $(2\pi)^{-1} \arg Q(z,w)$
as $w$ travels around $\T$.
For $z=\pm1$,
the curve is symmetric about the real line,
so $\ellv(z)$ is given by twice the change in
$(2\pi)^{-1} \arg Q(z,w)$ as $w$ travels halfway around $\T$,
from $+1$ counter-clockwise to $-1$. Thus
\[\begin{array}{l}
\sgn [ Q(-1,+1) Q(-1,-1)] = (-1)^{\ellv(-1)},\quad
\sgn [ Q(+1,+1) Q(+1,-1)] = (-1)^{\ellv(+1)},\\
\sgn [ Q(+1,-1) Q(-1,-1)] = (-1)^{\ellh(-1)},\quad
\sgn [ Q(+1,+1) Q(-1,+1)] = (-1)^{\ellh(+1)}.
\end{array}\]
From \eqref{e:ell.diff.by.one}, two of the signs above are $+1$
while the other two are $-1$.

If instead $z_0\in\R$ or $w_0\in\R$,
the argument principle can be used only to determine three of the four signs, but
the fourth is then also determined since the product of the signs must be $+1$.
Thus it must always be the case that
$Q$ has the same sign at three points in $\set{\pm1}^2$,
and takes the opposite sign at the last point.
Therefore $Q_+^{\ttr\tts}= (-1)^{\ttr\tts}Q( -(-1)^\ttr,-(-1)^\tts )$ is positive
at an even number of points.  Together with \eqref{eq:sign-Q} this implies
that the $Q^{\ttr\tts}_+$ are all positive.
\epf
\elem

\bpf[Proof of Proposition~\ref{p:torus.cf.bip}]
Again the issue is to determine the signs of the four Pfaffians
$\pf K_E(\pm1,\pm1)$ which here can be all non-zero
in contrast to the setting of Proposition~\ref{p:torus.nonbip}.
From \eqref{e:bip.pfaffian},
$\pf K_E(\pm1,\pm1)
= Q_E(\pm1,\pm1)$,
where $Q_E$ can also be computed
recursively from $Q(z,w)$ as in \eqref{e:productformula}:
it then follows from Lemma~\ref{l:bip.arg.principle} that
for $\ttr,\tts\in\set{0,1}$,
\[\sgn (-1)^{\ttr\tts} \pf K_E(-(-1)^\ttr,-(-1)^\tts)
= \sgn Q^{\ttr\tts}_+ = +1,\]
and so
it follows from
 Theorem~\ref{t:double.product}
together with \eqref{e:torus.symmetries} that
\[
(-1)^{\ttr\tts} \pf K_E(-(-1)^\ttr,-(-1)^\tts)
= e^{(\det E)\,\freezero+o(1)}\,
\cf(-(-1)^\ttr\phaseh_E,-(-1)^\tts\phasev_E)^2\]
It then follows from \eqref{e:torus.table} that
\[\bpm \zzE[\zro\zro]\\
	\zzE[\one\zro]\\
	\zzE[\zro\one]\\
	\zzE[\one\one]
\epm
= e^{(\det E)\,\freezero+o(1)}
\tf14
\Signs
\bpm
\cf^{\one\one}(-\ze_E,-\xi_E\giv\tau_E)^2\\
\cf^{\one\zro}(-\ze_E,-\xi_E\giv\tau_E)^2\\
\cf^{\zro\one}(-\ze_E,-\xi_E\giv\tau_E)^2\\
\cf^{\zro\zro}(-\ze_E,-\xi_E\giv\tau_E)^2
\epm,
\]
implying the expansion of $\zzE[\ttr\tts]$
in the statement of the proposition.
Summing over $\ttr,\tts\in\set{0,1}$
and recalling Lemma~\ref{l:theta.abs2}
concludes the proof of Theorem~\ref{t:z}\ref{t:z.b}.
\epf

Figure~\ref{f:fsc.one.below.two} shows $\logfscone(x+iy)$ versus $\logfsctwo(x+iy)$ for $y=\smash{\tf1{200}}$;
note that $\logfscone$ resembles $\logfsctwo$ but lies below it.

\subsection{Modular transformation of finite-size correction}
\label{ss:modular}

From the results presented so far it is clear that the asymptotic behavior of dimer systems on large toric graphs $\lat_E$ is governed by the
conformal shape parameter $\tau_E$
defined in \eqref{e:tau.domain.phase}.
We now explain the interpretation of this parameter
as the half-period ratio of the torus with respect to its ``natural'' or ``conformal'' embedding.

Recall that any $E\in\End$ defines the fractional linear transformation
\[\gm^E:\C\to\C,\quad \gm^E(z)\equiv\f{x+zy}{u+zv};\]
note $\gm^{EE'}=\gm^E\circ\gm^{E'}$. The associated lattice \emph{half-period ratio\/} $\tau^E$
is the evaluation of this fractional linear transformation at $i$,
$\tau^E\equiv\gm^E(i)$,
a point in the upper half-plane.
The parameter $\tau$ associated to the transformed Hessian
$H_E\equiv (E^t)^{-1}HE^{-1}$ is simply the half-period ratio of a certain matrix
square root of the inverse Hessian:
\[
\tau[H_E]
= \tau^{E\bm{h}^{-1}},\quad\text{where }
	\bm{h}\equiv \AAw^{-1/2}\bpm D&0\\B&\AAw \epm
\text{ satisfies }(E\bm{h}^{-1})(E\bm{h}^{-1})^t
	= (H_E)^{-1}.
	\]
In particular we have the symmetries
\beq\label{e:torus.symmetries}
\left.\begin{array}{l}
\cf^{\ttr\tts}(\ze,\xi\giv\tau)
=\cf^{\ttr\tts}(\overline\ze,\xi\giv-\overline\tau)
=\cf^{\ttr\tts}(\ze,\overline\xi\giv-\overline\tau),\\
\text{therefore }\cf^{\ttr\tts}(\ze,\xi\giv\tau)
=\cf^{\ttr\tts}(\overline\ze,\overline\xi\giv\tau)
\end{array}\right\}
\quad\text{for $\ze,\xi\in\T$}\eeq
(these relations are also straightforward to prove directly
from the definitions of the special functions).

Two matrices $E,E'\in\End$ specify the same lattice of vectors $\Z^2 E=\Z^2 E'$ if and only if $E=TE'$ for $T\in\SL$. The half-period ratio transforms under left multiplication by $\SL$ via the group $\Gam$ of \emph{modular transformations\/} $(\gm^T)_{T\in\SL}$. The Jacobi theta and Dedekind eta functions transform naturally under the modular group.
To understand the transformations of
$\cf,\logfscone,\logfsctwo,\logfscodd$
under the modular group,
it suffices to describe
their transformations under the generating transformations
\[\begin{array}{rl}
\gm^A:\tau\mapsto\tau+1
	&\text{corresponding to }A\equiv
		(\begin{smallmatrix} 1&0\\1&1\end{smallmatrix});\\
\gm^B:\tau\mapsto -1/\tau
	&\text{corresponding to }B\equiv
		(\begin{smallmatrix} 0&1\\-1&0 \end{smallmatrix}).
\end{array}\]
From the relations
\[\begin{array}{rlrl}
\vth_{\ttr\tts}(\var\giv\tau+1)
\hspace{-6pt}&=(e^{\pi i/4})^\ttr
	\vth_{\ttr,\tts+1-\ttr}(\var\giv\tau),\,&
\deta(\tau+1)
\hspace{-6pt}&=e^{\pi i/12}\deta(\tau),\\
\vth_{\ttr\tts}(\var/\tau \mid -1/\tau)
\hspace{-6pt}&=(-i)^{\ttr\tts}(-i\tau)^{1/2}
	e^{\pi i\var^2/\tau}
	\vth_{\tts\ttr}(\var\giv\tau),\,&
\deta(-1/\tau)
\hspace{-6pt}&=(-i\tau)^{1/2}\deta(\tau),
\end{array}\]
it is straightforward to prove the following

\bppn\label{p:modular}
The functions $\cf^{\ttr\tts}$ satisfy the modular relations
\[\begin{array}{l}
\cf^{\ttr\tts}(\ze,\xi\giv\tau)
=\cf^{\ttr,\ttr+\tts}(\ze,\ze\xi\giv\tau+1),\\
\cf^{\ttr\tts}(\ze,\xi\giv\tau)
=\cf^{\tts\ttr}(\bar\xi,\ze\giv -1/\tau)
=\cf^{\tts\ttr}(\xi,\ze\giv 1/\bar\tau),\end{array}\]
implying for all $T\in\SL$ that, with $(\phaseh_E,\phasev_E)\equiv(z_0^u w_0^v,z_0^x w_0^y)$, we have
\[\begin{array}{l}
\logfscone^{\ttr\tts}(\tau)
	= \logfscone^{\smash{BTB(\ttr,\tts)}}(\gm^T(\tau))
	\text{ and }
	\logfsctwo^{\ttr\tts}(\phaseh_E,\phasev_E\giv\tau)
	= \logfsctwo^{\smash{BTB(\ttr,\tts)}}(
		\phaseh_{TE},\phasev_{TE}\giv\gm^T(\tau)),\\
\text{hence }
	\logfscone(\tau)=\logfscone(\gm^T(\tau))
	\text{ and }
	\logfsctwo(\phaseh_E,\phasev_E\giv\tau)
	=\logfsctwo(\phaseh_{TE},\phasev_{TE}\giv\gm^T(\tau)).
\end{array}\]
\eppn

\noindent
These results indicate that at criticality, the
second-order
behavior of $P(z,w)$ at its nodes
determines a ``natural'' or ``conformal'' geometric embedding of $\lat$ into the complex plane
--- that is, the embedding in which the lattice is invariant under
translations by $\bm{h}^{-1}\Z^2$, so that the matrix $E\bm{h}^{-1}$ describes $\lat_E$ in Cartesian coordinates.

\subsection{Asymptotic behavior of the scaling functions}
\label{sec:asympt-ford}
\renewcommand{\Im}{\operatorname{Im}}

In this subsection we describe the asymptotics of the finite-size correction functions
as the imaginary part $\tauy$ of $\tau$ tends to $+\infty$
(corresponding to
the situation that one of the lattice vectors defining the torus
becomes much longer than the other,
with $\det E$ still within a constant factor of $\|(u,v)\|\|(x,y)\|$).
As we shall see, when $\tauy\to\infty$, the corrections to the free energy become linear in $\tauy$, while the dependence on the twist in the torus given by $\taux$ becomes negligible.

Consider first the function
$\cf$ of~\eqref{e:cf}.
Recalling \eqref{e:jacobi.product} and \eqref{e:eta}, we express
\[\cf(-e^{2\pi i \phi},-e^{2\pi i \psi}\giv\tau)
= \bigg|
	\f{
	\prod_{\ell\in\N-1/2}
	\big[(1+ e^{2\pi i((\ell+\phi)\tau-\psi)} )
		(1+ e^{2\pi i((\ell-\phi)\tau+\psi)} ) \big]}
	{ e^{\pi i \tau (1/12-\phi^2)}}
	\bigg|.\]
We may assume $-\slf12<\phi\le\slf12$;
then, in the limit $\tauy\to\infty$,
it is clear that all terms $\ell\ne\slf12$ in the infinite product
give a negligible contribution:
\[
\f{\cf(-e^{2\pi i \phi},-e^{2\pi i \psi}\giv\tau)}{1+O(e^{-2\pi\tauy})}
= \bigg|
	\f{
	(1+ e^{2\pi i((1/2+\phi)\tau-\psi)} )
	(1+ e^{2\pi i((1/2-\phi)\tau+\psi)} )
	}
	{ e^{\pi i \tau (1/12-\phi^2)}}
	\bigg|.
\]
If $\phi$ is bounded away from $\pm\slf12$ in the limit $\tauy\to\infty$,
then
\[\cf(-e^{2\pi i \phi},-e^{2\pi i \psi}\giv\tau)
= \exp\{-\pi\tauy(\phi^2-\tf{1}{12})+o(1)\}.\]
At $\phi=\slf12$ we instead find
\[\cf(+1,-e^{2\pi i \psi}\giv\tau)
= 2\cos(\pi\psi) \exp\{  - \pi\tauy \cdot \slf16
	+ O(e^{-2\pi\tauy}) \}.\]
We therefore find in the limit $\tauy\to\infty$ that
\begin{align*}
\logfscone(\tau)
	&=
	\pi\tauy / 12 + O(e^{-\pi\tauy}),\\
\logfsctwo(\pm i,\xi\giv\tau)
	&=
	\pi\tauy /24 + \log2 + O(e^{-\pi\tauy/2})\\
\logfsctwo(e^{\pm 2\pi i\phi},\xi\giv\tau)
	&=
	\pi\tauy \cdot ( \slf1{6}-2(|\phi|\wedge(\slf12-|\phi|))^2 )
	+ o(1)\\
	&\,\quad\quad
	\quad\text{(for $|\phi|$ bounded away from $\slf14$)}\\
\logfscodd(+1,\pm1\giv\tau)
	&= \pi \tauy / 6 + O(e^{-\pi\tauy}),\\
\logfscodd(-1,\pm1\giv\tau)
	&= -\pi\tauy / 12
		+\log2 + O(e^{-\pi\tauy}).
\end{align*}
These estimates hold uniformly over $\taux\in\R$.

\old{
We return now to the Ford circles (Figure~\ref{f:ford}). For each
real rational number $\slf{a}{b}\in\Q$
(expressed in lowest terms), there is a circle centered at $\slf{a}{b}+\slf{i}{(2b^2)}$ of radius $\slf{1}{(2b^2)}$; additionally a Ford circle is given by the line $\R+i$.  The modular group maps Ford circles to Ford circles (though the centers are not preserved).  Proposition~\ref{p:modular} characterizes how the modular group acts on the scaling functions, so we can use their behavior in the limit $\tauy\to\infty$ to determine their behavior near the rationals $\Q$.  In particular, since $\logfscone(\tau)$ is invariant under the modular group and
is approximately linear in $\tauy$ (uniformly over $\taux$),
inside the $\Q$-tangent Ford circles
it has contour lines which are approximately circular and tangent to the real line. The other scaling functions are not invariant \textit{per se\/} under the modular group due to the phase shifts, but it is similarly clear that they have spikes at each rational, with near-circular contours tangent to the real line.
}

\section{Loop statistics on bipartite graphs}
\label{s:torus.bip}

In this section we show that in the bipartite setting with distinct
conjugate zeroes, the dimer winding numbers \eqref{e:wind} have
asymptotically discrete Gaussian fluctuations.  Earlier results
showing that the dimer height distribution is a discrete Gaussian were
obtained by Kenyon and Wilson \cite{KW0} for the square lattice on a
cylinder and Boutillier and de Tili\`ere \cite{MR2561433} for the
honeycomb lattice on a rectilinear torus.

\bThm[implies Thm.~\ref{t:gaussian.intro}]\label{t:gaussian}
In the setting of Theorem~\ref{t:z}\ref{t:z.b},
suppose $\lat$ is $\match_0$-oriented,
and let $\match_E$
be the reference matching of $\lat_E$
obtained by periodically extending~$\match_0$.
Let \[\hypertop{ELLhv}\ellhv\equiv(\ellh(-1),\ellv(-1))\]
as defined by \eqref{e:ellh.ellv}.
Up to switching black with white, $\wind\match\ominus\match_E$
is asymptotically \hypertop{MuSigma}distributed
in the limit \eqref{e:e.large}
as a discrete Gaussian on~$\Z^2$ with
center and scale parameters $\meanvec$ and $\covmat$ given by
\beq\label{e:gaussian.mean}
\meanvec\equiv
\tf1\pi
\bpm
x \arg z_0 + y \arg w_0\\
-u\arg z_0 - v \arg w_0
\epm-(\det E) (E^t)^{-1} \ellhv,
\eeq
\beq\label{e:gaussian.covariance}
\covmat\equiv\f{(E^t)^{-1} H E^{-1}}{(\det H)^{1/2}/\det E},
\eeq
with $(z_0,w_0)\in\T^2$ the distinguished root of $Q(z,w)$
specified by \eqref{e:decreasing.root}~below.
\eThm

\brmk
The choice of reference matching $\match_E$ is not particularly important to the result.
For an arbitrary reference matching $\matchp_E$ of $\lat_E$,
since $\match\ominus\matchp_E=(\match\ominus\match_E)\ominus(\matchp_E\ominus\match_E)$,
the limiting distribution of $\wind\match\ominus\matchp_E$ is a discrete Gaussian
with the same covariance as for $\wind\match\ominus\match_E$, but with a mean that is shifted.
\ermk

The proof of Theorem~\ref{t:gaussian} is via perturbative analysis of the expansion
of Theorem~\ref{t:z}\ref{t:z.b}, which we repeat here for convenience:
	\beq\label{e:bip.z.expand}
	\f{\zzE}{ e^{(\det E)\,\freezero+o(1)} }
	=\fsctwo(\phaseh_E,\phasev_E\giv\tau_E)
	\eeq
with $\tau_E,\ttr_E,\tts_E$ as in \eqref{e:tau.domain.phase}, and
\[
	\fsctwo(\phaseh_E,\phasev_E\giv\tau)
	=\f{ \sum_{\bm{e}\in\Z^2}
			\exp\{ -\tf{\pi}{2}g_\tau(
				\bm{e}-(\tts_E,-\ttr_E)
				) \} }
		{ |\deta(\tau)|^2(2\tauy)^{1/2} }.
\]
For $z\in\T$, if $w\mapsto Q(z,w)$ has a root at $w(z)\in\C$, then $w\mapsto Q(z^{-1},w^{-1})$ has a root at $w=1/\overline{w}(z)$. Since switching black and white simply reverses the roles of $Q(z,w)$ and $Q(z^{-1},w^{-1})$, 
recalling \eqref{e:ell.diff.by.one}
we may hereafter assume that
\beq\label{e:zeroes.inside}
\ellv(+1) = \ellv(-1)-1.
\eeq
If $z_0$ is real, then
 one of the
$\ellv(\pm1)$ is 
defined by \eqref{e:ellh.ellv}
while the other is not,
in which case we define it by~\eqref{e:zeroes.inside}.

We distinguish between the conjugate roots of $Q(z,w)$
by taking $(z_0,w_0)$ to be the root such that
on a small neighborhood of $z_0$ in $\T$,
there is a smooth function $w(z)$ such that
\beq\label{e:decreasing.root}
\text{$w(z_0)=w_0$,
$Q(z,w(z))=0$, and
$|w(z_0 e^{2\pi i r})|$
is decreasing in $r$ for $|r|$ small.}
\eeq

\begin{figure}[htbp]
\begin{subfigure}[h]{0.5\textwidth}
\centering
\includegraphics[height=1.8in]{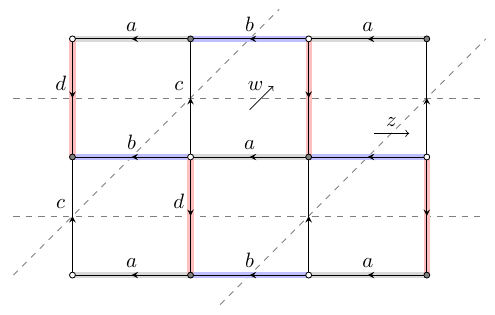}
\caption{Fundamental domain in dashed lines;
	reference matching $\match_0$ in gray;
	$\match_0$-alternating double-dimer loops
	(blue-gray horizontally,
	red-gray vertically)
	}
\end{subfigure}
\quad
\begin{subfigure}[h]{0.45\textwidth}
\centering
\includegraphics[height=1.8in]{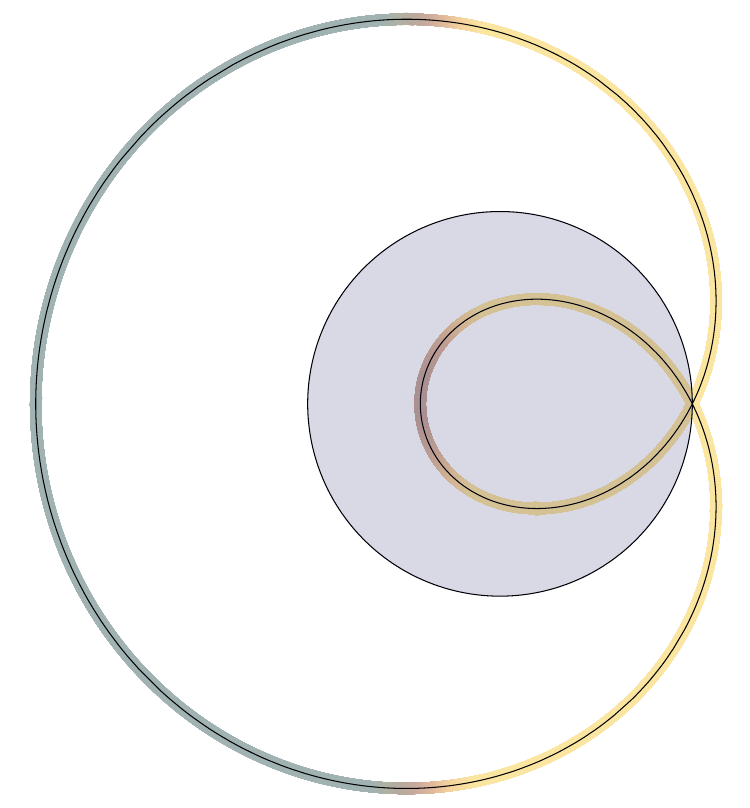}
\caption{For unweighted square lattice:
radial plot $z\mapsto z|w(z)|$
	with $z\in\T$,
colored according to $\imag w(z)$
(unit disk shaded)}
\end{subfigure}
\caption{Square lattice, with
	$Q(z,w)=a-bz-dw-\slf{cz}{w}$}
\label{f:square}
\end{figure}

\subsection{Moment-generating function of winding numbers}

Let us briefly review the notion of a dimer height function in this
bipartite setting (see \cite{\kos} and references therein). A dimer
configuration $\match$ on $\lat_E$ may be regarded as a black-to-white
unit flow.  If $\match_E$ is the reference matching on $\lat_E$, then
$\match\ominus\match_E$ is a divergence-free flow (in addition to
being an oriented loop configuration), and gives rise to a
\emph{height function\/} $h$ which is defined on the faces of the
graph as follows:

For any two faces $f_0,f_1$ in the graph, $h(f_1)-h(f_0)$ gives the flux of
$\match\ominus\match_E$ across a path in the dual graph joining $f_0$ to $f_1$.
This definition depends on the choice of the path from $f_0$ to $f_1$, but homologous paths give the same value.
In particular, for a closed path which can be deformed to a point, the height change is zero.
The horizontal winding $\windh$ of the oriented loop configuration $\match\ominus\match_E$ is
the height change along a vertical closed loop oriented downwards,
and the vertical winding $\windv$ is the height change along a horizontal closed loop oriented rightwards.

For $\param\equiv(\paramx,\paramy)\in\R^2$,
consider the lattice $\lat$ with weights modified periodically as follows:
for each edge
joining a black vertex
	in the $\bm{x}\in\Z^2$ copy of the fundamental domain
to a white vertex
in the $\bm{x}+\bm{e}\in\Z^2$ copy of the fundamental domain,
multiply the edge weight by
	$\exp\{ \ip{\param}{ \bm{e} } \}$
(regardless of the Kasteleyn orientation of the edge).
Let $\nu^\param\equiv \nu^\param_E$ and $\zz^\param\equiv \zzE[\param]$ denote the associated
(non-normalized) dimer measure and dimer partition function on $\lat_E$.
The associated bipartite characteristic polynomial is
\[Q^\param(z,w)\equiv Q( e^{\paramx} z, e^{\paramy }w ).\]
\hypertop{Ronkin}
The free energy $\free_0$ associated with the $\param$-perturbed lattice weighting is the evaluation at $\param$ of the
 \emph{Ronkin function\/} $\Ronkin$ associated to $Q(z,w)$:
\beq\label{e:ronkin}\Ronkin(\param)
	\equiv\iint_{\T^2} \log|Q(e^{\paramx}z,e^{\paramy}w)|
		\,\f{dz}{2\pi iz}\f{dw}{2\pi iw}
\eeq
(see \cite{\kos} for more information).

Recall that $\match_E$ is the periodic extension of $\match_0$. 
For a dimer configuration~$\match$ for which
$\wind\match\ominus\match_E=(\windh,\windv)$,
\[
\f{\nu^\param(\match)/\nu^\param(\match_E)}{\nu(\match)/\nu(\match_E)}
=\exp\{ \windh (u\paramx +v\paramy)+ \windv ( x\paramx+y\paramy ) \}
=\exp\{ (\windh,\windv) E\param \}.
\]
Since no edges of the reference matching are reweighted,
$\nu^\param(\match_E) = \nu(\match_E)$,
and so
\beq\label{e:bip.mgf.wind}
\zz_E^\param
= \f{\nu^\param(\match_E)}{\nu(\match_E)}
 \sum_\match
	\f{\nu^\param(\match)/\nu^\param(\match_E)}{\nu(\match)/\nu(\match_E)}
	\nu(\match)
= 
\zz_E\times \E\big[ \exp\{ (\wind \match\ominus\match_E) E\param \}\big]\,.
\eeq
where the expectation is with respect to the original normalized dimer measure on~$E$.

Given $\zz_E^\param/\zz_E$, assuming it is a sufficiently well-behaved function of $\param$,
we can determine the distribution of the winding of the double-dimer configurations on the torus~$E$.
It is enough to evaluate $\zz_E^\param/\zz_E$ (sufficiently precisely)
for $\param$ in a neighborhood of $\pzro$.
From \eqref{e:bip.mgf.wind}, \eqref{e:bip.z.expand} and \eqref{e:ronkin},
\\[-6pt]\hypertarget{script-R-F}
\beq\label{e:bip.z.perturb}
\f{\zz_E^\param}{\zz_E}
=
\underbrace{
	\f{\exp\{ \Ronkin(\param) \det E\}}{\exp\{ \Ronkin(\pzro) \det E\}}
	}_{ \scrR(\param) }
\cdot
	\underbrace{\f
	{\fsctwo(\phaseh_E^\param,\phasev_E^\param\giv\tau_E^\param)}
	{\fsctwo(\phaseh_E^\pzro,\phasev_E^\pzro\giv\tau_E^\pzro)}}_{\scrF(\param)}
\cdot (1+o(1))
\eeq
where $\tau_E^\param,\phaseh_E^\param,\phasev_E^\param$ are all evaluated with respect to
the $\param$-perturbed weights.
We compute the first factor on the right-hand side
in \S\ref{ss:bip.ratio.free}, the second factor in \S\ref{ss:bip.ratio.fs},
and then complete the proof in \S\ref{ss:bip.ratio.fs}.

\subsection{Perturbation of free energy}
\label{ss:bip.ratio.free}

In this section we estimate $\scrR(\param)$ from (\ref{e:bip.z.perturb}) by computing the gradient and Hessian of the Ronkin function $\Ronkin$.

A version of the gradient calculation also appears in \cite[Theorem~5.6]{\kos}. For $|\param|$~small, $Q^\param$ also has distinct conjugate zeroes on the unit torus, and to compute the gradient we must understand how the zeroes change with $\param$. To this end let
$\ttr_0(\param), \tts_0(\param)$
be the unique smooth real-valued functions
such that
\beq\label{e:arguments.smooth.in.perturbation}
Q^\param(e^{\pi i \ttr_0(\param)},e^{\pi i \tts_0(\param)})=0,\quad
(e^{\pi i \ttr_0(\param)},e^{\pi i \tts_0(\param)})|_{\param=\pzro}
=(z_0,w_0),
\eeq
and $1<\ttr_0(\pzro),\tts_0(\pzro)\leq 1$,
where $(z_0,w_0)$ denotes the root of $Q$ at $\param=\pzro$
which was distinguished in \eqref{e:decreasing.root}.

\blem\label{l:bip.free}
In the setting of Proposition~\ref{p:torus.cf.bip}
with \eqref{e:zeroes.inside},
\[
\nabla \Ronkin(\param)
 = \bpm\pdah\Ronkin(\param) \\ \pdav\Ronkin(\param)\epm
=\bpm
	\ellh(-1)+\tts_0(\param)\\
	\ellv(-1)-\ttr_0(\param)
	\epm
\quad\quad
	\text{ for small $\param$;}
\]
and
\[
\operatorname{Hess}\Ronkin(\param)|_{\param=\pzro} =
\left.\bpm
	\pdah^2\Ronkin(\param) & \pdav\pdah\Ronkin(\param) \\
	\pdah\pdav\Ronkin(\param) & \pdav^2\Ronkin(\param)
\epm\right|_{\param=\pzro}
= \f{1}{\pi D} \bpm \AAz & B \\ B & \AAw \epm,
\]
where $A_z$, $A_w$, $B$, and $D$ are defined in  \eqref{e:hess} and \eqref{e:bip.hess}.
Consequently,
\[
\Ronkin(\param)-\Ronkin(\pzro)
= \ip{\ellhv}{\param}
	-(-\tts_0\paramx+\ttr_0\paramy)
	+ (2\pi D)^{-1}
	(\param^t H \param)
	+O(\param^3),\]
where the constants in big-$O$ term depend on the fundamental domain (but not $E$).
\bpf
By the argument principle,
\[\pdav\Ronkin(\param)
=\real \oint_{|z|=1}\oint_{|w|=e^{\paramy}}
		\f{
		(w\pd_w Q) (e^{\paramx}z,w)}
		{Q(e^{\paramx}z,w)}
		\f{dw}{2\pi i w}\f{dz}{2\pi i z}
=\oint_{|z|=1}
	\ellv^\param(z) \f{dz}{2\pi iz},\]
where $\ellv^\param(z)$ counts
the number of zeroes in $w$ of $Q^\param(z,w)$
inside the unit circle.
It follows from the condition
\eqref{e:zeroes.inside}
and from our definition \eqref{e:arguments.smooth.in.perturbation}
of $\ttr_0,\tts_0$ that
\[\pdav\Ronkin(\param) = \ellv^\param(-1)-\ttr_0(\param),\]
where $\ellv^\param(-1)=\ellv(-1)$ for $|\param|$ small.
Therefore $\pdav^2\Ronkin(\param) = -\pdav\ttr_0(\param)$,
which is positive
due to \eqref{e:decreasing.root}.

Differentiating the relation $Q^\param(z,w)=0$ in $\param$
and evaluating at $\param=\pzro$ gives
\[\bpm \pdah \ttr_0(\param) & \pdah \tts_0(\param) \\
	\pdav \ttr_0(\param) & \pdav \tts_0(\param) \epm
	\bpm z \pd_zQ \\ w \pd_wQ \epm
= \f{i}{\pi}\bpm z \pd_zQ \\ w \pd_wQ \epm.\]
Since $\ttr_0$ and $\tts_0$ are real-valued, separating the other terms into real and imaginary parts ($z Q_z\equiv x_z + i y_z$ and $w Q_w\equiv x_w + i y_w$) gives a system of four equations for the four variables
$\pdah\ttr_0(\param)$, $\pdah\tts_0(\param)$, $\pdav\ttr_0(\param)$, $\pdav\tts_0(\param)$.
We use \eqref{e:bip.hess} to solve these and find
\[
\bpm \pdah\ttr_0(\param) & \pdah\tts_0(\param) \\
	\pdav\ttr_0(\param) & \pdav\tts_0(\param) \epm
=\f{ \sgn( x_z y_w-y_z x_w ) }
	{\pi D} \bpm -B & \AAz \\ -\AAw & B \epm
= \f{1}{\pi D}\bpm -B & \AAz \\ -\AAw & B \epm,
\]
where the last equality follows from
	the preceding observation that
	$\pdav\ttr_0(\param)<0$.

This then implies $\pdah \tts_0(\param)>0$,
so a similar line of reasoning as above gives
\[\pdah\Ronkin(\param)
=\oint_{|w|=1} \ellh^\param(w) \f{dw}{2\pi i w}
= \ellh^\param(-1)+\tts_0(\param),\]
with $\ellh^\param(-1)$ defined to be $\ellh^\param(+1)-1$
in the case $w_0=-1$. Therefore
\beq\label{e:root.derivatives}
\hspace{-12pt}
\bpm
	\pdah^2\Ronkin(\param) & \pdav\pdah\Ronkin(\param)\\
	\pdah\pdav\Ronkin(\param) & \pdav^2\Ronkin(\param)
\epm
= \bpm
	+\pdah \tts_0(\param) & +\pdav \tts_0(\param)\\
	-\pdah \ttr_0(\param) & -\pdav \ttr_0(\param)
\epm
= \f{1}{\pi D} \bpm \AAz & B \\ B & \AAw \epm.
\qedhere
\eeq
\epf
\elem

\subsection{Perturbation of finite-size correction}
\label{ss:bip.ratio.fs}

We now compute the effect of the $\param$-perturbation on the finite-size correction.
Recall the quadratic form $g_\tau$ 
which was defined in \eqref{e:qfdef}.
\blem\label{l:bip.cf.ratio}
In the setting of Proposition~\ref{p:torus.cf.bip}
with \eqref{e:zeroes.inside},
assuming $|\param|(\det E)^{1/2}\lesssim 1$,
the second factor in \eqref{e:bip.z.perturb} is
\[
\scrF(\param)
=
\f{
\sum_{\bm{e}\in\Z^2}
	\exp\{-\tf{\pi}{2}g_{\tau_E^\pzro}(\bm{e}-\meanvec) +
	 (E\param)^t (\bm{e} -\meanvec) \}
	}
{
\exp\left\{ \frac{\det E}{2\pi D}\param^t H \param \right\}
\sum_{\bm{e}\in\Z^2}
	\exp\{-\tf{\pi}{2}g_{\tau_E^\pzro}(\bm{e}-\meanvec)\}}
\exp\{O(|\param|(\log(1/|\param|))^{1/2})\}
\]
with $\meanvec$ as in \eqref{e:gaussian.mean}.

\bpf
The parameter $\tau^\param_E$ varies smoothly with $\param$, so we find
\[
\scrF(\param)
= e^{O(|\param|)}
\f
{\fsctwo(\phaseh_E^\param,\phasev_E^\param\giv\tau^\pzro_E)}
{\fsctwo(\phaseh_E^\pzro,\phasev_E^\pzro\giv\tau^\pzro_E)}\,.\]
In the expression \eqref{e:fsctwo.gaussian} for $\fsctwo(\phaseh_E^\param,\phasev_E^\param\giv\tau^\pzro_E)$, let $\offset_E^\param$
be the ``offset vector'':
\[\offset_E^\param\equiv\bpm\tts_E^\param\\-\ttr_E^\param\epm\]
where $\ttr_E^\param,\tts_E^\param$ are as in \eqref{e:tau.domain.phase}. Then
\begin{equation}
\label{e:F.as.ratio.of.sums}
\scrF(\param)
=e^{O(|\param|)}
\f{
\sum_{\bm{e}\in\Z^2}
	\exp\{ -\tf{\pi}{2}g_{\tau^\pzro_E}( \bm{e}-\offset_E^\param ) \}
}
{ \sum_{\bm{e}\in\Z^2}
	\exp\{ -\tf{\pi}{2}g_{\tau^\pzro_E}( \bm{e}-\offset_E^\pzro ) \} } .
	\end{equation}
The above is invariant
under the addition of integer vectors to $\offset_E^\param$,
so for convenience we take the unperturbed offset vector
$\offset_E^\pzro$ to have norm $\lesssim1$,
and let the $\param$-perturbed offset vector
be defined by
\[
\offset_E^\param-\offset_E^\pzro
= (\det E) (E^t)^{-1}
\bpm \tts_0^\param-\tts_0^\pzro\\
	-\ttr_0^\param+\ttr_0^\pzro
	\epm.\]
The Taylor series expansion with \eqref{e:root.derivatives} gives
\begin{align}\nonumber
\offset_E^\param-\offset_E^\pzro
&= \f{\det E}{\pi D}
	(E^t)^{-1}
	[ H\param  + O(|\param|^2) ]\\
	\label{e:offa}
&= \f{\det E}{\pi D}
	(E^t)^{-1}
	H\param  + O(|\param|^2 (\det E)^{1/2}),
\end{align}
where the second equation holds for well-shaped tori \eqref{e:e.large}.
Recall that the quadratic form $g_\tau$ can also be expressed using \eqref{e:qfdef2}
and \eqref{e:gaussian.covariance}:
 \beq\label{e:gHE}
g_{\tau^\pzro_E}(\bm{e})
=\bm{e}^t \covmatinv \bm{e} = \bm{e}^t \f{E H^{-1} E^t}{(\det H)^{-1/2} \det E}\bm{e}\,.\eeq
Since $\covmat$ is symmetric, we can rewrite
\begin{align*}
&g_{\tau^\pzro_E}( \bm{e}-\offset_E^\param )
	-g_{\tau^\pzro_E}( \bm{e}-\offset_E^\pzro )
=  g_{\tau^\pzro_E}(\offset_E^\param-\offset_E^\pzro)
  - 2(\offset_E^\param-\offset_E^\pzro)^t\covmatinv (\bm{e}-\offset_E^\pzro)\,.
\end{align*}
Combining \eqref{e:gHE} and \eqref{e:offa} and recalling that $D=\sqrt{\det H}$,
and using the well-shaped torus assumption \eqref{e:e.large}, the terms on the right-hand side are
\begin{align*}
g_{\tau^\pzro_E}(\offset_E^\param-\offset_E^\pzro) 
&=  \frac{\det E}{\pi^2 D}\param^t H \param
	+ O(|\param|^3 \det E)\,,\\
- 2(\offset_E^\param-\offset_E^\pzro)^t\covmatinv (\bm{e}-\offset_E^\pzro)
  &=  - \frac{2}{\pi} \param^t E^t (\bm{e} - \offset_E^\pzro)
       + O(|\param|^2(\det E)^{1/2} (|\bm{e}| + 1))\,.
\end{align*}
Combining these gives
\begin{multline}
\exp\{ -\tf{\pi}{2}g_{\tau^\pzro_E}( \bm{e}-\offset_E^\param ) \} =
\f{\exp\{-\tf{\pi}{2}g_{\tau^\pzro_E}(\bm{e}-\offset_E^\pzro) +
	 (E\param)^t (\bm{e} -\offset_E^\pzro) \}
	}
{\exp\left\{ \frac{\det E}{2\pi D}\param^t H \param \right\}} \\
\times 
\exp\{
	|\param|^2
	(\det E)^{1/2}
	[ |\bm{e}|+1
	+ |\param|(\det E)^{1/2} ]\}\,.
\label{e:perturb.summary.estimate}
\end{multline}
To apply this estimate
in \eqref{e:F.as.ratio.of.sums},
we truncate the sum in the numerator so that
 $|\bm{e}|$ will not be too large. For convenience, we now invoke the assumption that $|\param|(\det E)^{1/2}\lesssim1$,
so that
$|\offset^\param_E-\offset^\pzro_E|
\lesssim |\param|(\det E)^{1/2}\lesssim1$.
Then, since $g_{\tau^\param_E}(\bm{e}) \asymp |\bm{e}|^2$, there is an absolute constant $C$
such that 
the contribution in the numerator from
$|\bm{e} > C(\log(1/|\param|))^{1/2}$
will be less than $\param$.
The numerator itself is $\asymp1$,
so the additive error $|\param|$
translates into 
multiplicative error
$\exp\{O(|\param|)\}$.
Then, for $|\bm{e}|\le C(\log(1/|\param|))^{1/2}$,
we apply the estimate \eqref{e:perturb.summary.estimate}.
Finally we remove the truncation on $\bm{e}$,
giving
\begin{align*}
\scrF(\param)
&=
\f{\sum_{\bm{e}\in\Z^2}
	\exp\{-\tf{\pi}{2}g_{\tau_E^\pzro}(\bm{e}-\offset_E^\pzro) +
	 (E\param)^t (\bm{e} -\offset_E^\pzro) \}
	}
{\exp\left\{ \frac{\det E}{2\pi D}\param^t H \param \right\}
\sum_{\bm{e}\in\Z^2}
	\exp\{-\tf{\pi}{2}g_{\tau_E^\pzro}(\bm{e}-\offset_E^\pzro)\}}\\
&\qquad\qquad\times 
\exp\{ O( |\param| (\log (1/|\param|))^{1/2} ) \}.
\end{align*}
The lemma follows since  $\meanvec$ is a representative of $\offset_E^\pzro$ modulo $\Z^2$.
\epf
\elem

\bpf[Proof of Theorem~\ref{t:gaussian}]
Let $\paraminv\in\R^2$ with $|\paraminv|\lesssim1$,
and set
$\param = E^{-1}\paraminv$.
Combining \eqref{e:bip.z.perturb} with Lemmas~\ref{l:bip.free}~and~\ref{l:bip.cf.ratio} gives
\[
\frac{\zz^\param_E}{\zz^\pzro_E}
=e^{o(1)}
\f{
\exp\{ \ip{\ellhv}{\param} \det E\}\,
	\sum_{\bm{e}\in\Z^2}
	\exp\{-\tf{\pi}2g_\tau(\bm{e}-\meanvec) +
	 (E\param)^t (\bm{e}-\meanvec) \}
}
{
\exp\{ ( -\tts_0\paramx+\ttr_0\paramy )\det E \}\,
\sum_{\bm{e}\in\Z^2}
	\exp\{-\tf{\pi}{2}g_\tau(\bm{e}-\meanvec)\}
	},
\]
with
$\meanvec$
as in \eqref{e:gaussian.mean}.
Using the fact that 
\[ \langle\ellhv,\param\rangle\det E + (E\param)^t(\bm{e}-\meanvec) - ( -\tts_0\paramx+\ttr_0\paramy )\det E
 = \langle\paraminv,\bm{e}\rangle,
\]
we obtain
\begin{align*}
\E\big[ e^{ \ip{\wind \match\ominus\match_E}{\paraminv}}\big]
= \frac{\zz^\param_E}{\zz^\pzro_E}
&=e^{o(1)}
\f
{\sum_{\bm{e}\in\Z^2}
	\exp\{-\tf{\pi}{2}g_\tau(\bm{e}-\meanvec)\}
	\exp\{ \ip{\paraminv}{\bm{e}} \}
	}
{\sum_{\bm{e}\in\Z^2}
	\exp\{-\tf{\pi}{2}g_\tau(\bm{e}-\meanvec)\}}.
\end{align*}
If a sequence of probability measures have Laplace transforms that convergence pointwise
to the Laplace transform of a probability measure, then the sequence converges in distribution
to that measure \cite[Ex.~5.5]{billingsley}.
We therefore find that the winding
$\wind\match\ominus\match_E$
is asymptotically distributed as a discrete Gaussian on $\Z^2$
with parameters $\meanvec,\covmat$ as in \eqref{e:gaussian.mean} and \eqref{e:gaussian.covariance}.
\epf

\subsection{Dimers on the honeycomb graph}
\label{ss:hex}

\begin{figure}[htbp]
\begin{subfigure}[h]{0.5\textwidth}
\centering
\includegraphics[height=1.8in]{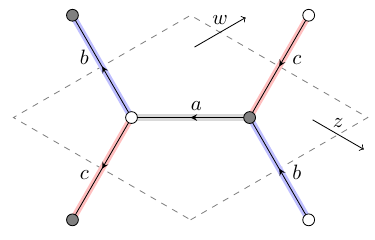}
\caption{Fundamental domain in dashed lines;
	reference matching $\match_0$ in gray;
	$\match_0$-alternating double-dimer loops
	(blue-gray horizontally,
	red-gray vertically)
	}
\end{subfigure}
\quad
\begin{subfigure}[h]{0.45\textwidth}
\centering
\includegraphics[height=1.8in]{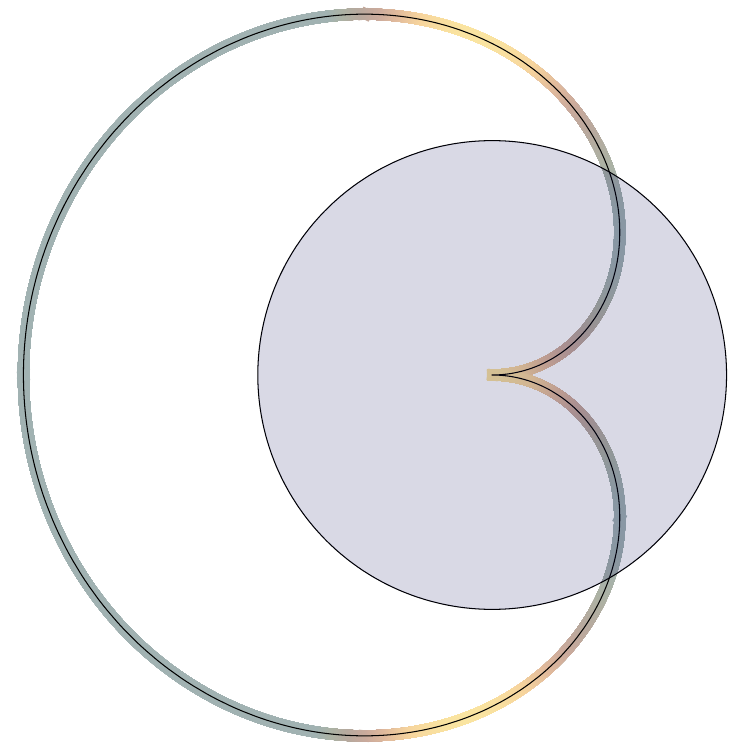}
\caption{For unweighted honeycomb graph:
radial plot $z\mapsto z|w(z)|$
	with $z\in\T$,
colored according to $\imag w(z)$
(unit disk shaded)}
\end{subfigure}
\caption{Honeycomb graph, with $Q(z,w)=a-bz-cw$}
\label{f:hex}
\end{figure}

By way of example, consider
the honeycomb graph with edge weights $a,b,c$
and bipartite fundamental domain,
$\match_0$-oriented for the reference matching $\match_0$ given by the $a$-edge (Figure~\ref{f:hex}).

The bipartite characteristic polynomial is $Q(z,w)=a-bz-cw$. If $a,b,c$ do not satisfy the triangle inequality then $Q$ is non-vanishing on $\T^2$.
If the weak triangle inequality is satisfied,
then $C\equiv\prod_{\ttr,\tts\in\set{0,1}} Q^{\ttr\tts}_+$
is non-negative (see Lemma~\ref{l:bip.arg.principle}),
and $Q$ vanishes at
$(z_0,w_0)\equiv (e^{\pi i \ttr_0},e^{\pi i \tts_0})$
and its conjugate where
\beq\label{e:hex.roots}
\ttr_0=\tf{1}{\pi}\cos^{-1}( \tf{a^2+b^2-c^2}{2ab} ) \in[0,1],\quad
\tts_0\equiv
	-\tf{1}{\pi}\cos^{-1}( \tf{a^2+c^2-b^2}{2ac} )
	\in[-1,0].
\eeq

Assume now that $a,b,c$ satisfy the strict triangle inequality,
so that $C>0$ and the conjugate zeroes of $Q$ are distinct.
These zeroes are positive nodes of $P(z,w)\equiv Q(z,w)\,Q(\slf{1}{z},\slf{1}{w})$, with Hessian
\beq\label{e:hex.hessian}
H\equiv\bpm b^2 & \tf12(a^2-b^2-c^2) \\ \tf12(a^2-b^2-c^2) & c^2
\epm,\quad
\det H\equiv \slf{C}{4}.\eeq

The following is then a direct consequence of
Theorem~\ref{t:z}\ref{t:z.b}
and Theorem~\ref{t:gaussian}
(with $\ellhv=(0,0)$):

\bcor
For the $\match_0$-oriented honeycomb graph (Figure~\ref{f:hex}),
the characteristic polynomial has zeroes at
$(z_0,w_0)=(e^{\pi i \ttr_0},e^{\pi i \tts_0})$ given by
\eqref{e:hex.roots}
and its conjugate,
with Hessian $H$ given by \eqref{e:hex.hessian}. Thus
\[\log\zz_E
=(\det E)\,\freezero
	+\logfsctwo(\phaseh_E,\phasev_E\giv\tau_E)
	+ o(1)\]
with $\tau_E,\phaseh_E,\phasev_E$ as in \eqref{e:tau.domain.phase}.
For $\match_E$ the matching of $\lat_E$ given by taking all the $a$-edges,
$\wind\match\ominus\match_0$ is asymptotically distributed as a discrete Gaussian on $\Z^2$ with parameters
\[\meanvec=
	\bpm
	x \ttr_0 + y \tts_0 \\
	-u\ttr_0 - v \tts_0
	\epm,\quad
	\covmat=
		\f{(E^t)^{-1}HE^{-1}}
		{(\det H)^{1/2}/\det E}.
	\]
\ecor

We emphasize again that while the conformal shape
$\tau_E$ depends smoothly on the entries
of the \emph{normalized\/} matrix $(\det E)^{-1/2} E$
(which has $O(1)$ entries for well-shaped tori \eqref{e:e.large}),
the domain phase $(\phaseh_E,\phasev_E)$
is highly sensitive to constant-order changes in the
\emph{non-normalized\/} entries of $E$.

\begin{figure}[h]
\includegraphics[width=.45\textwidth]{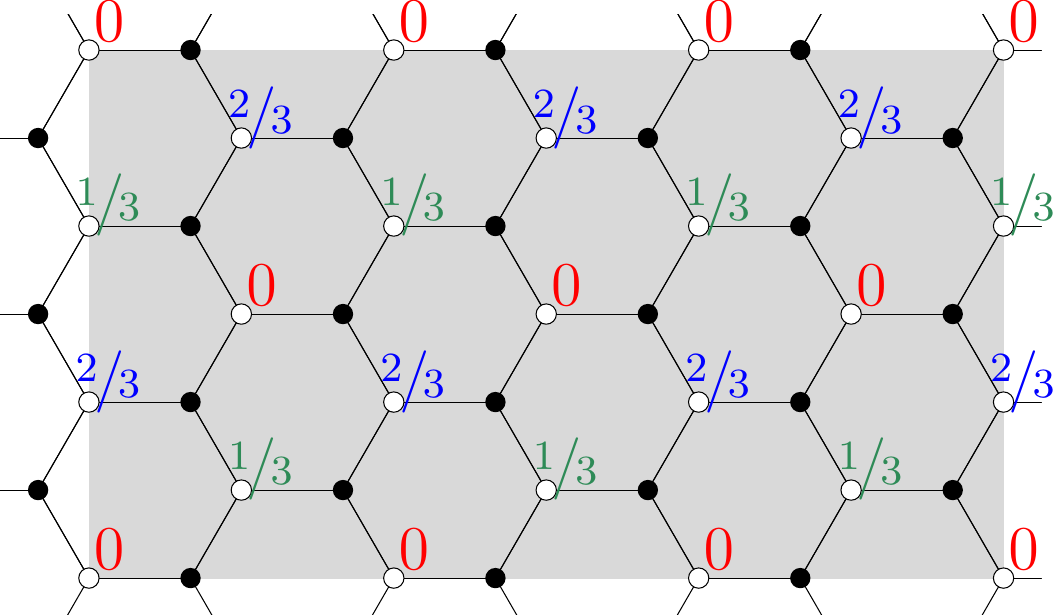}
\hfill
\includegraphics[width=.45\textwidth]{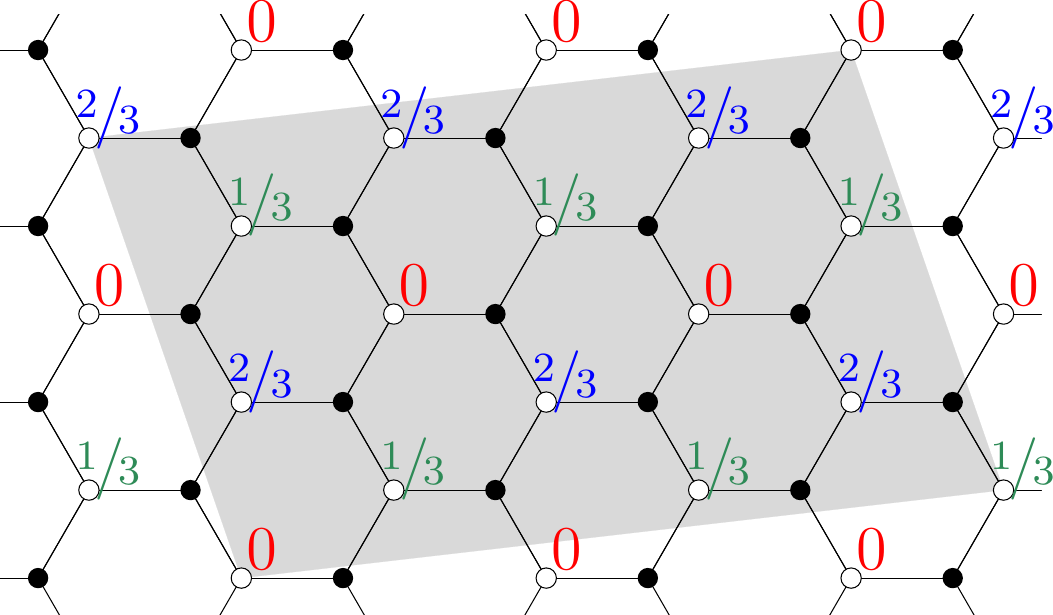}
\caption{The $4\times 3$ rectilinear torus, shown together with a skew torus}
\label{f:hex.m.by.n}
\end{figure}

This is illustrated in Figure~\ref{f:hex.logfsc} for the unweighted honeycomb graph ($a=b=c=1$).
With our choice of fundamental domain
(Figure~\ref{f:hex}),
the $m\times n$ rectilinear torus
(Figure~\ref{f:hex.m.by.n})
 studied in \cite{MR2561433}
corresponds to matrix $E$
given by $(u,v)=(m,m)$ and $(x,y)=(-n,n)$.
In the usual embedding where all hexagons are regular,
the geometric aspect ratio is $\rho\equiv n/(3^{1/2}m)$,
and it is straightforward to check that
the conformal shape $\tau$ is simply $i\rho$.
From \eqref{e:hex.roots} we have
$(\ttr_0,\tts_0) = (\slf13,-\slf13)$,
so there are multiple possibilities for the domain phase
$(\phaseh_E,\phasev_E)$.
Figure~\ref{f:hex} considers the near-rectilinear case
$\tau = i\rho + o(1)$,
and shows that
the finite-size correction lies on one of four different limiting curves
\[\rho\mapsto\logfsctwo
	( (e^{\pi i/3})^j ,(e^{\pi i/3})^k
	\giv i \rho )\]
depending on the phase $(\phaseh_E,\phasev_E)$.
Three of these
can arise from exactly rectilinear tori,
while the fourth
	(the one corresponding to both $j,k\not\equiv0\bmod3$)
arises from almost-rectilinear tori
with $\tau=i\rho+o(1)$.

\begin{figure}[h]
\begin{subfigure}[h]{0.45\textwidth}
\centering
\includegraphics[width=\textwidth]{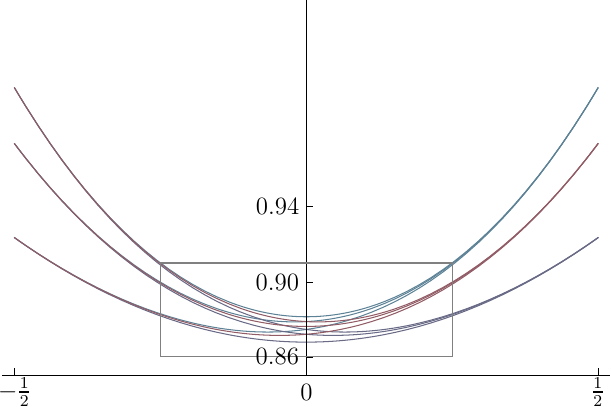}
\caption{$-\slf12\le\log\rho\le\slf12$}
\end{subfigure}
\begin{subfigure}[h]{0.45\textwidth}
\centering
\includegraphics[width=\textwidth]{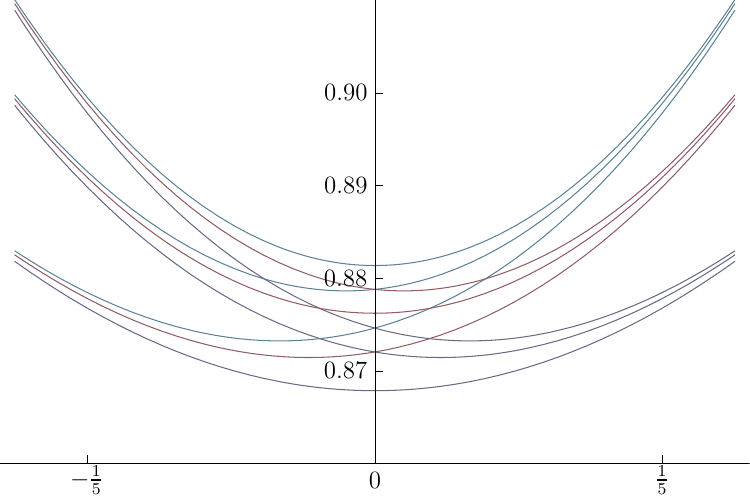}
\caption{$-\slf14\le\log\rho\le\slf14$}
\end{subfigure}
\caption{
\emph{(Honeycomb graph with weights $a=\tf12(\sqrt{5}-1)$, $b=c=1$.)}
Finite-size corrections
$\logfsctwo(\phaseh_E,\phasev_E\giv\tau)$
for near-rectilinear ($\tau=i\rho+o(1)$)
tori,
shown as a function of logarithmic
aspect ratio $\log\rho$
}
\label{f:hex.five}
\end{figure}

Arbitrarily many curves can be obtained by adjusting the weights:
for example,
if we keep $b=c=1$ but change $a$,
the conformal shape $\tau$ becomes
$\tau$ becomes $i \tf{n}{m} a(4-a^2)^{-1/2}$.
If
$a=\tf12(\sqrt{5}-1)$
then
$(\ttr_0,\tts_0)=(\slf25,-\slf25)$,
and Figure~\ref{f:hex.five}
shows the nine limiting curves
arising for near-rectilinear tori.

\section{Odd-sized fundamental domains}\label{s:odd}

In this section we briefly address the case where the fundamental domain $\lat_I$ contains an \emph{odd\/} number $k$ of vertices.
Up to now we have always assumed $k$ to be even,
which, as we review in \S\ref{ss:odd.general} below,
guarantees the existence of an $\lat_I$-periodic Kasteleyn orientation of $\lat$.
Clearly we can perform
calculations with doubled versions of the domain,
but if the ``natural'' fundamental domain of the lattice has $k$ odd
then different doublings need to be considered
to access all possible toric quotients.
For concreteness, in \S\ref{ss:square}
we illustrate with the example of the unweighted square lattice,
whose natural fundamental domain contains a single vertex.
In \S\ref{ss:odd.general} we comment on the general situation.

\subsection{Odd-sized fundamental domains in the square lattice}\label{ss:square}

\begin{figure}[hb]
\centering
\hspace*{7ex}
\begin{subfigure}[h]{.4\textwidth}
\centering
\hspace*{-7ex}\includegraphics[width=1.175\textwidth]{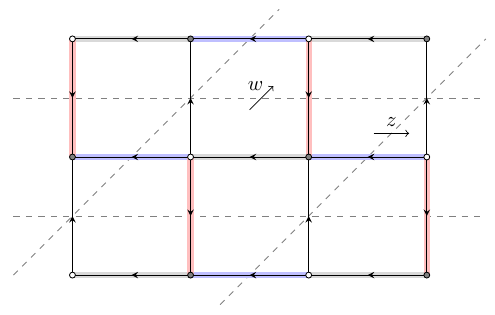}
\caption{Bipartite fundamental domain:\\
\hspace*{3ex} $Q(z,w)= 1-z-\slf{z}{w}-w$}
\label{f:square.fundamental.domains.bip}
\end{subfigure}
\hspace*{7ex}
\begin{subfigure}[h]{.4\textwidth}
\centering
\hspace*{-7ex}\includegraphics[width=1.175\textwidth]{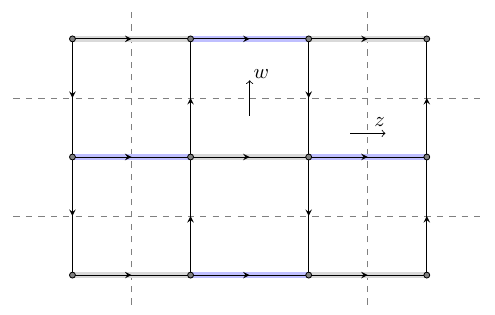}
\caption{Nonbipartite fundamental domain:\\
\hspace*{3ex}$P(z,w)= 4-z-\slf1z - w^2 - \slf{1}{w^2}$}
\label{f:square.fundamental.domains.nb}
\end{subfigure}
\caption{Two choices of fundamental domain
	for the unweighted square lattice}
\label{f:square.fundamental.domains}
\end{figure}

The natural fundamental domain $\lat_I$ of the unweighted square lattice
is the $1\times1$ torus containing a single vertex, connected to itself by one horizontal edge and one vertical edge. Clearly $\lat$ has no $\lat_I$-periodic Kasteleyn orientation with real weights; to find such an orientation we have to double the fundamental domain. Two possibilities are shown in Figure~\ref{f:square.fundamental.domains}; we note that the domain of Figure~\ref{f:square.fundamental.domains.bip} is bipartite while that of Figure~\ref{f:square.fundamental.domains.nb} is nonbipartite.

For $ad-bc$ positive, quotient by the $(a,b)$ and $(c,d)$ translations to form the $(a,b)\times(c,d)$ torus, which can have dimer covers for $ad-bc$ even. The torus can be formed from copies of the
$(2,0)\times(1,1)$
bipartite fundamental domain (Figure~\ref{f:square.fundamental.domains.bip}) if and only if
$a+b$ and $c+d$ are both even.
The bipartite characteristic polynomial has simple zeroes at $(1,\pm i)$,
so it follows from Theorem~\ref{t:z}\ref{t:z.b}
that the finite-size correction to $\zz$ is
$\fsctwo(i^\ttr,i^\tts\giv\tau)$
with $\ttr$ the value of $a$ (or $b$) modulo $2$,
and $\tts$ the value of $c$ (or $d$) modulo $2$.

The $(a,b)\times(c,d)$ torus
can be formed from copies of the
$2\times1$ nonbipartite fundamental domain
$\lat_{2,1}$ (Figure~\ref{f:square.fundamental.domains.nb})
 if and only if $a$ and $c$ are both even.
The characteristic polynomial has positive nodes at $(1,\pm1)$ with the same Hessian. It follows from Theorem~\ref{t:z}\ref{t:z.c} that
the finite-size correction to $\zz$ is
$\fscodd( (-1)^b,(-1)^d\giv\tau)$.

The case of $b,d$ both even is handled
by the $1\times2$
nonbipartite fundamental domain
$\lat_{1,2}$ ($\pi/4$-rotation of Figure~\ref{f:square.fundamental.domains.nb}).
Alternatively, by the $\pi/4$-rotational symmetry of the square lattice,
it has the same partition function as the
$(b,a)\times(-d,-c)$ torus. We therefore conclude

\bppn\label{p:square}
For the $(a,b)\times(c,d)$ torus formed from the unweighted square lattice,
the dimer partition function $\zz$ satisfies
\[
\begin{array}{l}
\log\zz - (\det E)\,\freezero  -o(1)=\\
=\begin{array}{r|llll}
		&\zro\zro&\zro\one&\one\zro&\one\one\\
\hline
\zro\zro & \logfsctwo(+1,+1\giv\tau)
		& \logfscodd(+1,-1\giv\tau)
		& \logfscodd(+1,-1\giv\tau)
		& \logfsctwo(+1,+i\giv\tau)\\
\zro\one & \logfscodd(-1,+1\giv\tau)
	& \logfscodd(-1,-1\giv\tau)
	&-\infty
	&-\infty\\
\one\zro & \logfscodd( -1,+1\giv\tau)
	&-\infty
	& \logfscodd(-1,-1\giv\tau)
	&-\infty\\
\one\one & \logfsctwo(+i,+1\giv\tau)
	&-\infty
	&-\infty
	& \logfsctwo(+i,+i\giv\tau)\\
\end{array}
\end{array}
\]
where the row index is the value of $(a,b)$ modulo $2$
while the column index is the value of $(c,d)$ modulo $2$.
\eppn

\begin{figure}[h]
\centering
\begin{subfigure}[h]{.47\textwidth}
\centering
\includegraphics[height=.6\textwidth]{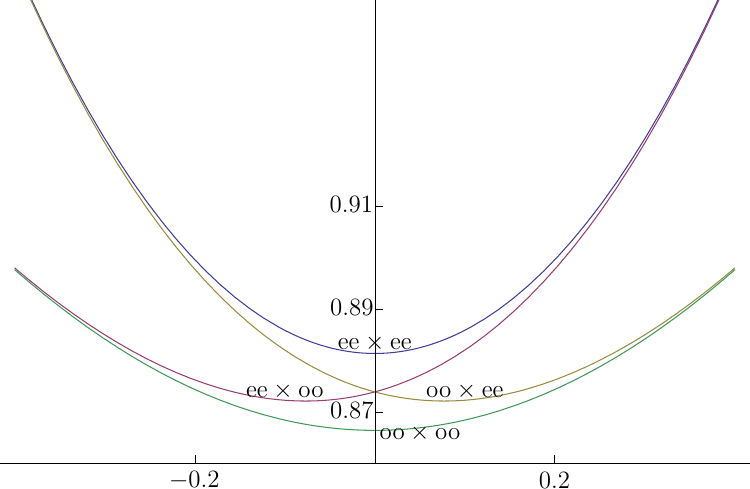}
\caption{
\parbox[t]{2.2in}{
Corrections $\logfsctwo(\ze,\xi\giv\tau)$,
obtained from bipartite domain Figure~\ref{f:square.fundamental.domains.bip}}}
\label{f:square.curves.bip}
\end{subfigure}
\begin{subfigure}[h]{.47\textwidth}
\centering
\includegraphics[height=.6\textwidth]{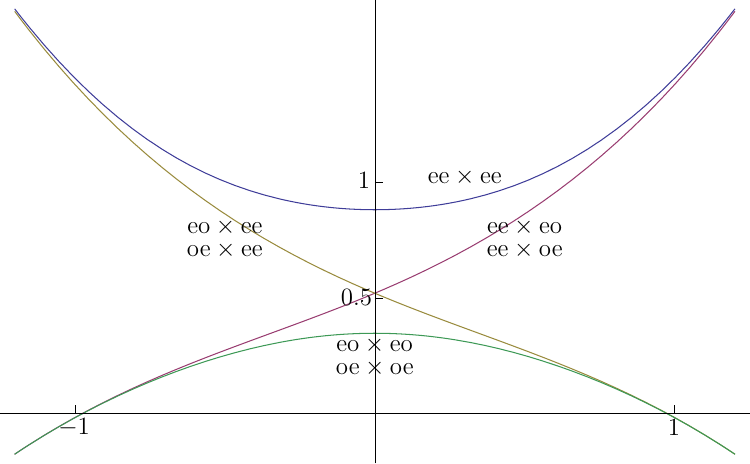}
\caption{
\parbox[t]{2.2in}{Corrections $\logfscodd(\ze,\xi\giv\tau)$,
obtained from non-bipartite domain Figure~\ref{f:square.fundamental.domains.nb}}}
\label{f:square.curves.nb}
\end{subfigure}
\caption{\emph{Unweighted square lattice dimers.}
Finite-size corrections $\logfsc$
for near-rectilinear toric quotients,
shown as a function of logarithmic aspect ratio $\log\rho$
and labelled according to parity of $(a,b)\times(c,d)$.
}
\label{f:torus.square.magnify}
\end{figure}

We see from Proposition~\ref{p:square} that for general $\tau$, the finite-size correction lies on one of seven curves.
Figure~\ref{f:square.curves.bip}
shows the four curves coming from the bipartite fundamental domain (Figure~\ref{f:square.fundamental.domains.bip}),
while Figure~\ref{f:square.curves.nb}
shows the four curves coming from the nonbipartite fundamental domain (Figure~\ref{f:square.fundamental.domains.nb}),
where the case of $a,b,c,d$ all even appears in both.

\subsection{Odd-sized fundamental domains in general graphs}\label{ss:odd.general}

The assumption of $\lat_I$ even
guarantees that $\lat$ can be given a
\emph{$\lat_I$-periodic Kasteleyn orientation}, as follows (see \cite{MR0253689}):
choose a planar spanning tree $T$ on the dual graph of $\lat_I$, and give an arbitrary orientation to any edges of $\lat_I$ not crossed by an edge of $T$. For any vertex $f\in T$ with a single neighbor $g\in T$, there is a unique way to orient the edge of $\lat_I$ crossed by the dual edge $(f,g)$ such that the face corresponding to $f$ is clockwise odd. By repeatedly pruning leaf vertices of $T$,
$\lat_I$ can be oriented such that all faces are clockwise odd except possibly the final face, corresponding to the root of $T$.
For each face $f$ let $o_f$ count the number of clockwise-oriented edges around $f$: then $\smash{\sum_f (1+o_f)}$
is even if and only if the root face is also clockwise odd.
But this sum is also simply
the number of faces and edges in the graph,
and so by Euler's formula must have the same parity
as the number $k$ of vertices.
It follows that $\lat_I$
can be oriented to give rise to a
periodic Kasteleyn orientation of $\lat$
if and only if $k$ is even.

If $k$ is odd, the above procedure produces a $\lat_I$-periodic orientation of $\lat$ which is ``almost'' Kasteleyn, in that exactly one face per fundamental domain is clockwise even. This can be resolved by doubling the fundamental domain: for example, if we put two copies of the fundamental domain side by side to make the $2\times1$ torus $\lat_{2,1}$, the orientation of the doubled graph can be corrected by choosing a dual path $\gm$ joining the two clockwise even faces, and reversing the orientation of each edge crossing the dual path $\gm$.

Recalling Definition~\ref{d:oriented},
suppose further that we are given a reference matching $\match_0$
of $\lat_{2,1}$
such that its periodic extension $\match_\infty$
does not cross any boundaries separating different copies of $\lat_{2,1}$.
We can choose the dual path $\gm$ such that its periodic extension
either does or does not cross any of these boundaries,
and we take the choice which results in all
vertical $\match_0$-alternating cycles having sign $+1$. If horizontal $\match_0$-alternating cycles also have sign $+1$ then $\lat$ is $\match_0$-oriented; otherwise reverse edges along vertical seams to complete the orientation.

\bppn
With the above orientation,
the characteristic polynomial $P(z,w)$ associated to $\lat_{2,1}$ is a polynomial in $(z,w^2)$.
Consequently, if $P(z,w)$ has two distinct real nodes then they must be of form $(z_0,\pm1)$ with the same associated Hessian.

\bpf
Express $P(z,w)=\det K(z,w)$ as a sum over permutations, so that each non-zero term corresponds to an oriented cycle configuration in $\lat_{2,1}$. Odd powers of $w$ correspond to cycle configurations winding an odd number of times in the vertical direction.

Consider the mapping on cycle configurations
induced by switching the two copies of $\lat_I$ inside $\lat_{2,1}$.
The sign of the corresponding permutation remains the same
(regardless of the parity of $k$), but the product over entries of $K(z,w)$
changes sign if and only if the configuration winds vertically an odd number of times:
if the two copies of $\lat_I$ have the same orientation except
across the dual path $\gm$ then this is immediate.
If we also reversed edges along vertical boundaries
to produce an $\match_0$-orientation of $\lat$,
the statement can be proved by considering deformations of the
edge-reversal seams.
\epf
\eppn

As demonstrated for the square lattice in \S\ref{ss:square},
if the natural fundamental domain $\lat_I$ of $\lat$
has an odd number of vertices,
any $(a,b)\times (c,d)$ torus
containing an even number of vertices
can be obtained using an
even-sized fundamental domain:
either the $2\times1$ torus $\lat_{2,1}$,
the $1\times2$ torus $\lat_{1,2}$,
or the $(2,0)\times(1,1)$ torus.

\section{Finite-size correction for characteristic polynomials}
\label{s:torus.pf}

In this section we prove Theorem~\ref{t:double.product}. For simplicity let us assume that $P(z,w)$ has a positive node located at $(1,1)$, with Hessian expansion \eqref{e:hess}. We will make no assumptions on the locations of the other (finitely many) nodes, nor will we assume the relation $P(z,w)=P(\slf{1}{z},\slf{1}{w})$ --- the theorem for a general finite collection of nodes will then follow by considering $(z,w)\mapsto P(z_0 z,w_0 w)$.

We will make use of the following standard quadrature rules (see e.g.\ \cite{\hildebrand,\miller}) to approximate sums over discrete subsets of $\T$ by contour integrals. For $f$ a smooth $\C$-valued function on $[x-h,x+h]\subset\R$, write
\[\TS\avgint^h f(x)
\equiv \tf{1}{2h} \int_{x-h}^{x+h} f(s)\,ds,\quad
\avg^h f(x)
\equiv \tf16 [f(x-h) + 4f(x) + f(x+h)].\]

\blem[quadrature rules] \label{l:quad}
For $f$ a $\C$-valued function smooth in a neighborhood of $[-h,h]$, the following hold:
\[\begin{array}{rlll}
\TS\tf1{2h}\int_{-h}^h f(s)\,ds \hspace{-6pt}
&= \tf12[f(-h)+f(h)]
	-\tf{4 h^2}{12} f''(\xi),
&\text{some }\xi\in(-h,h)
& \text{(trapezoid rule);}\\
\avgint^h f(0) \hspace{-6pt}
&= \avg^h f(0)
	- \f{h^4}{180} f^{(4)}(\xi),
&\text{some } \xi\in(-h,h)
& \text{(Simpson's rule).}
\end{array}\]
\elem

\noindent
If $g$ is a $\C$-valued function
smooth in a neighborhood of
$[x-h,x+h]\times[y-h,y+h]$,
we write $\smash{\avg^h_j g(x,y)}$
to indicate $\avg^h$ applied to $g$ in the $j$-th coordinate,
and $\avg^h g(x,y)$ to indicate the composition of $\avg^h_1,\avg^h_2$ applied to $g$. Similarly we write $\avgint^h g(x,y)\equiv \avgint^h_1\avgint^h_2 g(x,y)$; Lemma~\ref{l:quad} then implies
\[
\abs{\avg^h g(x,y)-\avgint^h g(x,y)}
\le h^4 \sup\set{
	\abs{(\pd_x^4 + \pd_y^4 ) g(x',y')}:
	\abs{x'-x}\vee\abs{y'-y}\le h
	}.
\]
We turn now to the computation of $P_E(\ze,\xi)$
with $(\ze,\xi)\equiv
(e^{2\pi i\phi},e^{2\pi i\psi})$.
From the double product formula \eqref{e:productformula},
this is the product of $P(e^{2\pi ir},e^{2\pi is})$
over $(r,s)$ in the intersection of the unit square
$\bm{H}\equiv \smash{[-\tf12,\tf12)^2}$ with the grid of points $E^{-1}[(\phi+\Z)\times(\psi+\Z)]$.
Equivalently,
define $F\equiv n E^{-1}$
for a positive scaling parameter $n$ 
within a constant factor of
$(\det E)^{1/2}$,
and let
$\pskew(r,s)\equiv \log P(e^{2\pi ir'},e^{2\pi is'})$
with
$(r',s')$ the vector obtained by applying the matrix $F$ to the vector $(r,s)$: then
\beq\label{e:double.product.sum.log}
\begin{array}{l}
\TS \log P_E(\ze,\xi)
= \sum_{(r,s) \in \gridlat} \pskew(r,s)
= \sum_{(r,s) \in \gridlat} \avg^h \pskew(r,s),\\
\text{where }
h\equiv n^{-1}\text{ and }
\bm{L}\equiv L^{\phi\psi}_E
\equiv (F^{-1}\bm{H})
	\cap h[(\phi+\Z)\times(\psi+\Z)]
\end{array}
\eeq
(so $\gridlat$ is a subset of a rectilinear lattice with spacings $h$).

\brmk
For practical purposes the double product \eqref{e:productformula}
is easily computed as follows: for $\ze=\xi=1$, take the product of $P(e^{2\pi ir},e^{2\pi is})$ over
\[\mx{r\\s}
= (\det E)^{-1}
\mx{ yj-vk\\-xj+uk }
\quad\text{ over }
0\le j < \f{\det E}{\mathrm{gcd}(x,y)}, \
0\le k < \mathrm{gcd}(x,y).\]
(To see that this is valid, note that for each fixed $k\in\Z$,
all $j$'s map to $(r,s)$ pairs
which are distinct modulo $\Z^2$,
since clearly $(\det E)^{-1}(yj,-xj)\in\Z^2$
if and only if $(\det E)^{-1}\mathrm{gcd}(x,y)j\in\Z^2$.
As $(j,k)$ ranges over all $\Z^2$,
$(r,s)$ must take exactly $\det E$ distinct values,
and this would fail if any repeats occur
in the stated range of $k$.)
\ermk

For convenience we hereafter choose $n$ to be an integer within a constant factor of $(\det E)^{1/2}$, sufficiently small such that $\bm{H}$ is contained in $F^{-1}\bm{H}$, and such that the restriction of $\pskew$ to $\bm{H}$ is non-vanishing except at the origin --- for example this can be accomplished with
\beq\label{e:n.h.Fmat}
n\equiv h^{-1}\equiv
	\max\set{ j\in\Z_{\ge1}:
	\bm{H}\subseteq j^{-1}E\bm{H}
	\text{ and } \pskew|_{\bm{H}\setminus(0,0)} > -\infty },\quad
	F\equiv n E^{-1}.
\eeq
If $P$ has expansion \eqref{e:hess} at the origin with Hessian matrix $H$,
then $\Pskew\equiv e^{\pskew}$ at the origin
has the transformed Hessian matrix
$\hess_\square  \equiv F^t H F$,
whose entries we shall denote
$\As,\Bst,\At$,
with $\Disc\equiv\sqrt{\det\hess_\square}=\sqrt{\As\At-\Bst^2}>0$.
A second-order approximation to $\Pskew$ near the origin is then given by
\beq\label{e:Prect}
\begin{array}{l}
\Prect(z,w)
\equiv \be(z)w+\be(\slf{1}{z})/w-2\gm(z),\quad\text{where}\\
\be(z) \equiv -\At+\Bst(1-z)\quad
	\text{and}\quad
\gm(z) \equiv (\As-\Bst) \tf12(z+\slf{1}{z}) - (\As+\At-\Bst).
\end{array}
\eeq
Let $\prect\equiv\log\Prect$:
we consider both $\pskew,\prect$ as functions of $(r,s)\in\R^2$;
note however that
$\pskew$ has period $F^{-1}\bm{H}$
while $\prect$ has period $\bm{H}$.

The polynomial
$\tf12 w \Prect(z,w)$ is quadratic in $w$, with
discriminant
$\gm(z)^2-\be(z)\be(\slf{1}{z})$
which is quadratic in $\smash{x\equiv\tf12(z+z^{-1})}$
with real coefficients. In particular,
for $z\in\T$, $x=\real z$ and the discriminant is real-valued.
The discriminant is minimized over all $x\in\R$
at $x_\star= 1 + [\Disc/(\As-\Bst)]^2>1$,
hence minimized over $z\in\T$ at $z=1$
where it takes value zero.
The function $\gm$ is linear in $x$, and is easily checked
(using $\As\At>\Bst$) to be negative at $x\in\set{\pm1}$, therefore $\gm(z)<0$ for all $z\in\T$.
If $z\in\T$ with $\be(z)=0$, then clearly $P(z,w)=-2\gm(z)>0$ for all $w$.
Fixing $z$ with $\be(z)\ne0$,
$\Prect(z,w)$ has roots (in $w$) given by
\beq\label{e:torus.ptilde.roots}
w_\pm(z)
\equiv \be(z)^{-1}[ \gm(z)\mp\de(z) ],\quad
\de(z)\equiv\sqrt{\gm(z)^2-\be(z)\be(\slf{1}{z})}\ge0.\eeq
For $z\in\T$,
$w_+(z)\overline{w_-(z)} = w_+(z)w_-(\slf{1}{z})=1$,
and $|w_-(z)|\le1\le|w_+(z)|$
with
strict inequality except at $z=1$ where $w_\pm$ both evaluate to $1$.
In particular, together with \eqref{e:n.h.Fmat} this shows that
$\Pskew/\Prect$
is bounded and non-vanishing on $\bm{H}$.
For $z=e^{2\pi ir}$ with $|r|$ small,
Taylor expanding \eqref{e:torus.ptilde.roots} gives
\beq\label{eq:torus.ptilde.roots.exp}
w_\pm(z)
=1 - \tf{\Bst}{\At}(2\pi ir)
	\pm \tf{\Disc}{\At} (2\pi \abs{r})
		+ O(r^2)
=1+\taux(2\pi ir)
	\pm\tauy(2\pi \abs{r}) + O(r^2).
\eeq

The following lemma shows that
in large toric graphs $\lat_E$,
any finite-size corrections in the
asymptotic expansion of $P_E(\ze,\xi)$
(with $\ze,\xi\in\T$)
depend \emph{only\/} on the second-order behavior of
the fundamental domain polynomial
$P(z,w)$ around its nodes
---
thus, for the purposes of calculating this correction,
we may replace $\pskew$ near the node with its
approximation $\prect$. The precise statement is as follows:

\blem\label{l:torus.localize}
In the above setting,
let $\mu\equiv\mu_n\equiv n^{-4/5}$
and $\gridlat_\mu \equiv \gridlat \cap [-\mu,\mu]^2$.
It holds for any
$\gridlat_\mu\subseteq
	\bm{V}\subseteq \gridlat\cap\bm{H}$ that
\[\TS
\log P_E(\ze,\xi) - (\det E)\,2\freezero +O(n^{-2/5})
= \log\Pi_{\bm{V}} \equiv
	\TS\sum_{(r,s)\in \bm{V}}
	(\avg^h-\avgint^h) \prect(r,s)
	\]

\bpf
The proof is an application of Simpson's rule
(Lemma~\ref{l:quad}).

\smallskip\noindent\emph{Estimates on derivatives.}
For any smooth function $a$ defined on an interval of $\R$,
\beq\label{e:log.four.deriv}
\TS
(\log a)^{(4)}
=-6 (L_1 a)^4
	+12 (L_1 a)^2 (L_2 a)
	-3 (L_2 a)^2
	-4 (L_1 a)(L_3 a)
	+ L_4 a
\eeq
where $L_ja\equiv a^{(j)}/a$.
If we take $a_\square,a_\circ$
to be $\pskew, \prect$
regarded as a function of $r$ only or $s$ only,
we find $|L_j a_\square|+|L_j a_\circ|\lesssim t^{(2-j)\vee0}/t^2$
where $t\equiv\|(r,s)\|$.
Substituting into \eqref{e:log.four.deriv} gives
\beq\label{e:fourth.deriv}
|\pd_s^4 \pskew |+|\pd_t^4 \pskew |
	+|\pd_s^4 \prect |+|\pd_t^4 \prect |
	\lesssim r^{-4}\quad\text{with }
	t\equiv \|(r,s)\|.
\eeq

\smallskip\noindent
\emph{Application of Simpson's rule.} From \eqref{e:double.product.sum.log},
\[\TS\log P_E(\ze,\xi)-(\det E)\,2\freezero
= \sum_{(r,s)\in\gridlat} (\avg^h-\avgint^h) \pskew(r,s).\]
If $P$ were non-vanishing on $\T^2$, Simpson's rule would give that the right-hand side is $O(n^{-2})$. In the current setting this false,
however, combining Simpson's rule and \eqref{e:fourth.deriv}
shows that the total contribution
 from points outside $\gridlat_\mu$ is small:
\[\TS
\sum_{(r,s)\in \gridlat\setminus\gridlat_\mu}
	|(\avg^h-\avgint^h) \pskew(r,s)|
\lesssim n^{-4} \sum_{(r,s)\in\gridlat\setminus\gridlat_\mu} t^{-4}
\lesssim (n\mu)^{-2} = n^{-2/5}.\]
We also claim that
\[\TS
\sum_{(r,s)\in\gridlat \cap \bm{H}}
	|(\avg^h-\avgint^h) (\pskew-\prect)(r,s)|\lesssim n^{-2/5}.\]
Indeed the contribution from points outside $\gridlat_\mu$ is
is (crudely) $\lesssim n^{-2/5}$
	by the same reasoning as above,
recalling \eqref{e:n.h.Fmat}
	that $\bm{H}\subseteq F^{-1}\bm{H}$.
Near the origin
we do not apply Simpson's rule, and instead note that
 $|(\pskew-\prect)(r,s)|\lesssim t$,
so that the contribution from $\gridlat_\mu$
is $\lesssim n^2 \mu^3 = n^{-2/5}$.
Combining these estimates gives the result.
\epf
\elem

Write $\gridlat\equiv\gridlat^1\times\gridlat^2$ with $\gridlat^j$ the projection of $\gridlat$ onto the $j$-th coordinate, and similarly $\gridlat_\mu\equiv \gridlat^1_\mu\times\gridlat^2_\mu$.
Hereafter we take
$\bm{V}\equiv \bm{V}_\mu \equiv \gridlat^1_\mu \times \gridlat^2\subset\gridlat\cap\bm{H}$,
and compute the quantity
$\Pi_{\bm{V}}$
defined in the statement of Lemma~\ref{l:torus.localize}.
First note that since $\prect$ has period $\bm{H}$, $\Pi_{\bm{V}}$ may be simplified by summing over the $s$-coordinate: define
\beq\label{e:torus.avg.int.prect}
\begin{array}{rl}
\avgprect(r)
	\hspace{-6pt}&
	\equiv n^{-1}\sum_{s\in\bm{L}^2}\prect(r,s)\\
\intprect(r)
	\hspace{-6pt}&
	\equiv \int_\T \prect(r,s)\,ds
	=\log|\be(e^{2\pi ir})| + \log|w_+(e^{2\pi ir})|,
\end{array}
\eeq
where $\intprect$ was evaluated by
a standard contour deformation argument,
recalling \eqref{e:torus.ptilde.roots}
and our choice of branch cut for the logarithm.
Then
\beq\label{e:torus.V}
\begin{array}{rl}
\log\Pi_{\bm{V}}
\hspace{-6pt}
&= n\sum_{ r \in \bm{L}^1_\mu}
	[ \avg^h\avgprect(r)
		- \avgint^h\intprect(r) ]\\
&=
\underbrace{\TS n\sum_{ r \in \bm{L}^1_\mu}
	 \avg^h
	 (\avgprect-\intprect)(r)}_{\equiv \log\Pi_\vartheta}
+ \underbrace{\TS n\sum_{ r \in \bm{L}^1_\mu}
	(\avg^h-\avgint^h)\intprect(r)}_{\equiv \log\Pi_\mathrm{cts}}.
\end{array}
\eeq

\blem\label{l:torus.q}
Let $-\ze = \smash{e^{2\pi i\phimin}}$
with $2\pi\phimin\in(-\pi,\pi]$.
Then
\[\log \Pi_\cts = (-\tf16 + 2 \phimin^2)\log q + O(n^{-1})
\quad\text{with }
\q\equiv\abs{\qtau} = e^{-\pi\tauy}.\]

\bpf
Abbreviate $f\equiv\intprect$, regarded as a $\Z$-periodic function of $s\in\R$. It follows from \eqref{e:torus.avg.int.prect} that $f$ is analytic except at $r\in\Z$ where the roots $w_+$ and $w_-$ cross one another.
Recalling \eqref{e:n.h.Fmat}, let
$r_\star\equiv h\phimin$,
so that $r_-\equiv r_\star-\tf{h}{2}$
and $r_+\equiv r_\star + \tf{h}{2}$,
are the unique pair of adjacent points in $\gridlat_\mu$
with $r_-\le 0<r_+$.
Let
$t_-\equiv r_--h$, $t_+\equiv r_++h$:
\[\begin{array}{rl}
n^{-1} \log\Pi_\mathrm{cts}
\hspace{-6pt}
&= O(n^{-3})
	+ (\avg^h-\avgint^h) f(r_-)
	+ (\avg^h-\avgint^h) f(r_+)
	\quad\text{(Simpson's rule)}\\
&= O(n^{-3})
	+ \tf16 f(t_-)
	+ \tf56 f(r_-)
	+ \tf56 f(r_+)
	+ \tf16 f(t_+)\\
&\qquad- \tf{1}{2h}\int_{t_-}^{r_-} f\,ds
	- \tf{1}{2h}\int_{r_+}^{t_+} f\,ds
	- \tf1h\int_{r_-}^{r_+} f\,ds\\
&= O(n^{-2})-\tf1{12} [f(t_-)+f(t_+)]
	+ \tf7{12} [f(r_-)+f(r_+)]\\
&\qquad - \tf{|r_-|}{2h} [ f(r_-)+f(0) ]
	- \tf{|r_-|}{2h} [ f(r_-)+f(0) ]
	\quad\text{(trapezoid rule)}.
\end{array}\]
From \eqref{e:torus.avg.int.prect} and
 \eqref{eq:torus.ptilde.roots.exp}, near $s=0$ we have
$f(s)
= \log \AAw + 2\pi \tauy |s| + O(s^2)$. Substituting into the above
and simplifying gives
\[\log\Pi_\mathrm{cts}+O(n^{-1})
= 2\pi\tauy [ \tf13- \tf1{2h^2}[r_-^2+r_+^2] ]
= -\pi\tauy [-\tf16 +2 (r_\star/h)^2],\]
concluding the proof.
\epf
\elem

Recall
$\Prect(z,w) = \be(z) w^{-1} (w-w_+(z))(w-w_-(z))$,
with
$w_+(z)\overline{w_-(z)}=1$ for $z\in\T$
(cf.~\eqref{e:torus.ptilde.roots}). Combining with \eqref{e:torus.avg.int.prect} gives
	(with $z\equiv e^{2\pi ir}$)
\[\begin{array}{rl}
\exp\{n\avgprect(r)\}\hspace{-6pt}
&= \prod_{w^n=\xi}\Prect(z,w)
= [-\be(z) w_+(z)]^n
	(1 - \xi w_+(z)^{-n}) (1 - \xi^{-1} w_-(z)^n)\\
&= \exp\{ n\intprect(r) \}\,
	(1 - \xi w_+(z)^{-n}) (1 - \xi^{-1} w_-(z)^n),
\end{array}\]
where we used that
$-\be(z)w_+(z) = |\be(z)w_+(z)|$ for $z\in\T$.
It is clear from \eqref{eq:torus.ptilde.roots.exp}
that
$|w_+(z)|^{-n}=|w_-(z)|^{-n} \le
	\exp\{ -\Om[(\log n)^2] \}$
for $z\in\T$ with
$|z-1| \gtrsim n^{-1}(\log n)^2$, so
we can ignore the effect of $\avg^h$
in the definition \eqref{e:torus.V} of $\Pi_\vartheta$, giving
\[\TS
\Pi_\vartheta\exp\{ o(n^{-2}) \}
=
\exp\{
	n \sum_{r\in\gridlat^1_\mu}
	( \avgprect-\intprect )(r)
	\}
=
\prod_{r\in\gridlat^1_\mu}
	|1 - \xi^{-1} w_-(e^{2\pi ir})^n|^2.\]
The following lemma computes $\Pi_\vartheta$.
Up to now the error estimates hold \emph{uniformly\/} over $\ze,\xi\in\T$, even allowing for dependence on $n$.
In the following, the error blows up if $(\ze,\xi)$ approaches too closely to a singularity of $P_E$.

\blem\label{l:torus.theta}
Let $(e^{2\pi i\phi},e^{2\pi i\psi})\equiv(\ze,\xi)$, $(e^{2\pi i\phimin},e^{2\pi i\psimin})\equiv(-\ze,-\xi)$, and write $\bm{r}$ for the Euclidean distance between $(\ze,\xi)$ and $(1,1)$. Then
\[\Pi_\vartheta
=\exp\{ O( n^{-2/5} \bm{r}^{-1} )\}
\Big| \f{\vth_{\zro\zro}(
	\phimin \tau-\psimin \giv\tau) }{G(\qtau)} \Big|^2\]

\bpf
For $r\equiv j/n \in\gridlat_\mu$ with $\mu\equiv n^{-4/5}$,
it follows from \eqref{eq:torus.ptilde.roots.exp} that
\[w_-( e^{2\pi i r} )^n
= \exp\{
	2\pi i\taux j
	- 2\pi\tauy|j|
	+ O(n^{-3/5}) \}.\]
Thus the closest approach of the points
$\xi^{-1}w_-(e^{2\pi ir})^n$
to $1$
as $r$ varies over $\gridlat_\mu$ is asymptotically lower bounded by
\[\liminf_{n\to\infty}
	\inf_{r\in\gridlat_\mu}
	|1-\xi^{-1}w_-(e^{2\pi ir})^n|
\ge | 1-\exp\{ 2\pi i
	(|\psi|-|\taux\phi|)
	-2\pi\tauy|\phi| \} |
\gtrsim \bm{r}.\]
Combining with the preceding estimate gives
\[
\TS \Pi_\vartheta \exp\{ O( n^{-2/5} \bm{r}^{-1} )\}
=
	\prod_{r\equiv j/n\in\gridlat_\mu}
	|1-\xi^{-1}
		\exp\{ 2\pi i \taux j-2\pi\tauy|j| \}
		|^2.
\]
Clearly we can replace the product over $\gridlat_\mu$
by the product over $\gridlat_\infty\equiv h[\phi+\Z]$
with no effect on the overall $O( n^{-2/5} \bm{r}^{-1} )$ error bound.
From straightforward computation, the product over $\gridlat_\infty$ is exactly the right-hand side: that is
\[\Pi_\vartheta\,\exp\{ O( n^{-2/5} \bm{r}^{-1} )\}
	= |\bm{\pi}_\vartheta|^2\]
where,  writing
$\pi(j)\equiv
1-\xi^{-1}\exp\{ 2\pi i \taux j-2\pi\tauy|j| \}$,
\[
\begin{array}{rl}
\bm{\pi}_\vartheta
	\hspace{-6pt}
	&\equiv
	\prod_{j\in\N-1/2}
	\pi(\phimin+j) \overline{\pi(\phimin-j)} \\
&=\prod_{j\in\N-1/2}
	( 1 + \qtau^{2j}\,\exp\{2\pi i ( \tau\phimin-\psimin )\})
	( 1 + \qtau^{2j}\,\exp\{-2\pi i ( \tau\phimin-\psimin )\}),
\end{array}
\]
which equals
$G(\qtau)^{-1} \vth_{\zro\zro}(\tau\phimin-\psimin\giv\tau)$
by \eqref{e:jacobi.product}.
\epf
\elem

\bpf[Proof of Theorem~\ref{t:double.product}]
Recall that we assumed throughout this section that $P$ has a node at $(1,1)$. By the argument of Lemma~\ref{l:torus.localize} applied to $(z,w)\mapsto P(z_j z,w_j w)$,
$P_E(\ze,\xi)$ is (up to $\exp\{O(n^{-2/5})\}$ multiplicative error)
$e^{(\det E)\,2\freezero}$ times a product of factors $\Pi_{\bm{V}}^j$,
one for each node $(z_j,w_j)$ ($1\le j\le\ell$)
 of $P$ on $\T^2$.
We decomposed $\Pi_{\bm{V}}=\Pi_\cts\Pi_{\vth}$
	\eqref{e:torus.V};
combining Lemmas~\ref{l:torus.q}~and~\ref{l:torus.theta}
then gives
\[\TS
P_E(\ze,\xi)
=\exp\{ O(n^{-2/5}\bm{r}^{-1}) \}\,
	\exp\{(\det E)\,2\freezero \}\,
	\prod_{j=1}^\ell  \cf(\ze,\xi\giv\tau_j)^2,
	\]
with $\bm{r}\equiv\min_{1\le j\le \ell}
	\|(\ze,\xi)-(z_j^u w_j^v,z_j^x w_j^y)\|$,
	concluding the proof.
\epf

\appendix

\newcommand{\discriminant}{\bm{\de}}

\section{Dimers on the Fisher and 3.4.6.4 graph}
\label{s:fisher.hst}

Recall the quantities $\ka_a,\ka_b,\ka_c,\ka_\circ$ defined in \eqref{e:kappa}. Since the weights $a,b,c$ are assumed to be strictly positive, clearly $-\ka_\circ<\ka_a,\ka_b,\ka_c$;
also, any two elements of $\set{\ka_a,\ka_b,\ka_c}$ have positive sum, proving that no two of the $\ka's$ can vanish simultaneously. The vanishing of any $\ka$ imposes some further constraints:
\beq\label{e:ka.constraints}
\begin{array}{l}
\ka_\circ=0 \text{ implies }
	a=\tf{b+c}{bc-1} \text{ so }
	bc>1, \text{ and (by symmetry) }
	ac>1,ab>1;\\
\ka_c=0 \text{ implies }
	c=a+b+abc>a+b, \text{ and }
	c=\tf{a+b}{1-ab}
	\text{ so } ab<1,
\end{array}\eeq
and the constraints arising from $\ka_a=0$ or $\ka_b=0$ are symmetric to that of $\ka_c=0$.

\subsection{Fisher graph}

\begin{figure}[h]
\begin{subfigure}[h]{0.55\textwidth}
\centering
\includegraphics[width=\textwidth]{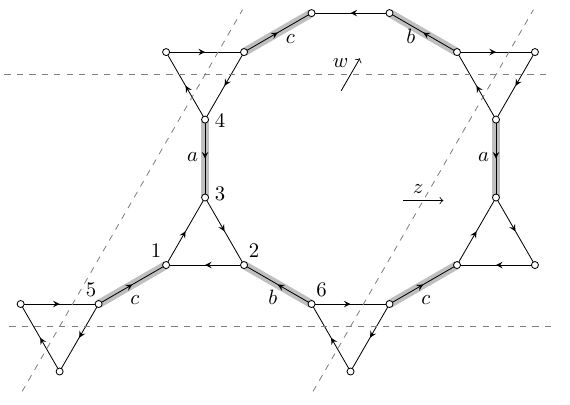}
\end{subfigure}
\begin{subfigure}[h]{0.4\textwidth}
\centering
\[
\text{\footnotesize $\begin{pmatrix}
0 & -1 & 1 & 0 & -c & 0 \\
 1 & 0 & -1 & 0 & 0 & -b \\
 -1 & 1 & 0 & -a & 0 & 0 \\
 0 & 0 & a & 0 & -w & \slf{w}{z} \\
 c & 0 & 0 & \slf{1}{w} & 0 & \slf{-1}{z} \\
 0 & b & 0 & \slf{-z}{w} & z & 0
\end{pmatrix}$}
\]
\end{subfigure}
\caption{Fisher or 3.12.12 graph
with fundamental domain inside the dashed lines,
shown with corresponding matrix
$K(z,w)$.}
\label{f:fisher}
\end{figure}

The fundamental domain together with the matrix $K(z,w)$ is shown in Figure~\ref{f:fisher}. Applying \eqref{e:torus.table} to the fundamental domain gives
\[
\bpm
-\pf K(+1,+1)\\
+\pf K(+1,-1)\\
+\pf K(-1,+1)\\
+\pf K(-1,-1)\\
\epm
=\Signs
\left(
\begin{array}{ll}
\zz^{\zro\zro} \hspace{-6pt}&=abc\\
\zz^{\one\zro} \hspace{-6pt}&=a\\
\zz^{\zro\one} \hspace{-6pt}&=b\\
\zz^{\one\one} \hspace{-6pt}&=c
\end{array}
\right)
=
\bpm \ka_\circ \\ \ka_a \\ \ka_b \\ \ka_c \epm
\]
(which can also be verified by direct calculation).
The characteristic polynomial is
\[\begin{array}{rl}
P(z,w) \hspace{-6pt}&\equiv \det K(z,w)
	= \be(z) w + \be(1/z) 1/w -2\gm(z),\quad\text{where}\\
\be(z) \hspace{-6pt}&=
	a(c-b)(1+bc)+ab(1-c^2)(1+\slf1z),\\
\gm(z) \hspace{-6pt}&=
	-(1-a^2)bc x_1
	-\tf12(a^2+b^2+c^2+(abc)^2)
	\quad \text{with }x_1\equiv \tf12(z+\slf1z).
\end{array}\]
Note that it is clear from the lattice symmetry that the polynomial must transform in a simple manner under permutations of the weights $a,b,c$; indeed we can also write
\[\begin{array}{l}
P(z,w)
=a^2+b^2+c^2+(abc)^2
	+2bc(1-a^2) x_1
	+2ac(1-b^2) x_2
	+2ab(1-c^2) x_3 \\
\quad\text{with }
x_1\equiv\tf12(z+\slf{1}{z}), \
x_2\equiv\tf12(w+\slf{1}{w}), \text{ and }
x_3\equiv\tf12(\slf{z}{w}+\slf{w}{z}).
\end{array}
\]
As $(z,w)$ varies over the unit torus,
$(x_1,x_2,x_3)$ traces out a Cayley surface in $\R^3$.

\bpf[Proof of Proposition~\ref{p:fisher.crit} (Fisher graph)]
We now show that the spectral curve can only intersect the unit torus at a single real node. The argument is similar to the one for the polynomial $\Prect(z,w)$ defined in \eqref{e:Prect}.
Throughout the proof we write $x$ for $x_1\equiv \tf12(z+\slf1z)$;
there should be no confusion with the entry $x$ appearing in \eqref{e:E}.
Since $\gm$ is linear in
	$x$, evaluating at the extremes
\[\begin{array}{l}
-2\gm(+1) = a^2(1-bc)^2+(b+c)^2>0,\\
-2\gm(-1) = a^2(1+bc)^2+(b-c)^2>0
\end{array}
\]
proves that $\gm$ is negative on all of $\T$.
If at any $z\in\T$ we have $\be(z)=0$ then $P(z,w)=-2\gm(z)>0$, so clearly the spectral curve cannot intersect the unit torus at this value of $z$.

We claim that the discriminant $\discriminant(z)\equiv\gm(z)^2-\be(z)\be(\slf{1}{z})$ is non-negative for all $z\in\T$, and can vanish only at $z=\pm1$, corresponding to the vanishing of one of the $\ka$'s.
If $a=1$ then $\discriminant(z)$ is linear in $x\equiv\tf12(z+\slf1z)$, and evaluating at the extremes $x=\pm1$ shows clearly that $\discriminant$ is strictly positive on $\T$.
If $a\ne1$ then $\discriminant(z)$ is
convex quadratic in~$x$;
from the symmetry $(ab)^{-2} P(z,w)|_{(a,b,c)}=P(-z,-w)|_{1/a,1/b,c}$
we hereafter assume $a<1$.
To prove the claim it suffices to show either
that the global minimum $\discriminant^\star$
	of $\discriminant(z)$ over $x\in\R$ is non-negative,
or that the global minimizer $x^\star$ has absolute value $\ge1$.
Indeed, assume further that $bc\ne1$: then
$\discriminant$ is minimized over all $x\in\R$ at
\[\begin{array}{rl}
x^\star
\hspace{-6pt}
&=\DS-\f{1+a^2}{1-a^2}
\f{b^2+c^2}{2bc}
\f{1-\f{1+(bc)^2}{b^2+c^2} a^2}{1-a^2}
\equiv
- \f{1+a^2}{1-a^2} R(a),\quad\text{with value}
	\vspace{6pt}\\
\discriminant^\star
\hspace{-6pt}&=\DS C (\bm{U}-a)(a^2-\bm{u}^2)
	\quad\text{where (assuming $a\ne1$, $bc\ne1$)}\\
&\quad
C\equiv a^2(a+b+c+abc)(1+bc)\tf{(1-bc)^2}{(1-a^2)^2}>0, \
\bm{U}\equiv  \tf{b+c}{1+bc}, \text{ and }
	\bm{u}\equiv |\tf{b-c}{1-bc}|.
\end{array}\]
One of the following occurs (recalling $a<1$):
\bnm[1.]
\item If $\bm{U}\ge1$
	(equivalently $b^2+c^2 \ge 1+(bc)^2$)
	then $x^\star<-1$.
\item If $\bm{U}<1$, then
\[
\bm{u}^2
=\f{b^2+c^2-2bc}{1+(bc)^2-2bc}
<\f{b^2+c^2}{1+(bc)^2}
<\f{b^2+c^2+2bc}{1+(bc)^2+2bc}
=\bm{U}^2<1.\]
In the range $\bm{U}\le a<1$, $\abs{R(a)}=-R(a)$ is increasing in $a$, hence bounded below by $-R(\bm{U})$ which can be calculated to simply equal $1$, thereby implying $x^\star<-1$. In the range $a\le\bm{u}$, $\abs{R(a)}=R(a)$ is increasing in $a$, hence bounded below by $R(\bm{u})$ which again equals $1$, implying $x^\star>1$. Finally in the intermediate range
$\bm{u}<a<\bm{U}$ it is clear that $\discriminant^\star>0$.
\enm
Lastly, in the case $bc=1$,
$\discriminant^\star<0$,
but
$|R(a)|=R(a)$ is increasing in the regime $a<1$,
so it is
bounded below by $R(0)>1$
which implies $x^\star<-1$.
Combining these cases concludes the characterization of the cases where
the spectral curve may intersect the unit torus.
It is straightforward to check
(using \eqref{e:ka.constraints})
that if $P$ has a real node it must be a positive node,
concluding the proof of the proposition.
\epf

\subsection{3.4.6.4 graph}

\begin{figure}[h]
\begin{subfigure}[h]{0.55\textwidth}
\centering
\includegraphics[width=\textwidth]{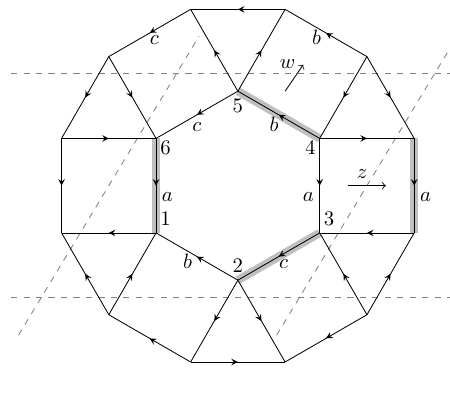}
\end{subfigure}
\begin{subfigure}[h]{0.4\textwidth}
\centering
\[
\text{\footnotesize$\bpm
0 & -b & \slf{1}{z} & 0 & \slf{-1}{w} & -a \\
 b & 0 & -c & \slf{1}{w} & 0 & \slf{z}{w} \\
 -z & c & 0 & -a & \slf{-z}{w} & 0 \\
 0 & -w & a & 0 & b & z \\
 w & 0 & \slf{w}{z} & -b & 0 & c \\
 a & \slf{-w}{z} & 0 & \slf{-1}{z} & -c & 0
\epm$.}
\]
\end{subfigure}
\caption{The 3.4.6.4 or rhombitrihexagonal tiling graph
with fundamental domain inside the dashed lines,
shown with
corresponding matrix $K(z,w)$}
\label{f:hst}
\end{figure}

The fundamental domain for the 3.4.6.4 graph together with the matrix $K(z,w)$ is shown in Figure~\ref{f:hst}.
Applying \eqref{e:torus.table} to the fundamental domain gives
\[
\bpm
-\pf K(+1,+1)\\
+\pf K(+1,-1)\\
+\pf K(-1,+1)\\
+\pf K(-1,-1)\\
\epm
=\Signs
\left(
\begin{array}{ll}
\zz^{\zro\zro} \hspace{-6pt}&=2c\\
\zz^{\zro\one} \hspace{-6pt}&=2b\\
\zz^{\one\zro} \hspace{-6pt}&=2a\\
\zz^{\one\one} \hspace{-6pt}&=2abc
\end{array}
\right)
=
\bpm 2\ka_c \\ 2\ka_b \\ 2\ka_a \\ 2\ka_\circ \epm\]
(which can also be verified by direct calculation).
The characteristic polynomial $P(z,w)=\det K(z,w)$ has the factorization
\[\begin{array}{rl}
P(z,w)
	\hspace{-6pt}&= -P_\zro(z,w) \, P_\one(z,w)
	= -P_\zro(z,w)\,[P_\zro(z,w)-4\rho], \\
P_\tts(z,w)
	\hspace{-6pt}&\equiv
	\be(z) w + \be(\slf1z)\slf1w - 2\gm_\tts(z)
	\quad\text{where}\\
\be(z)
	\hspace{-6pt}&\equiv a(b/z+c),
\quad \rho\equiv [(1+a^2)(1+b^2)(1+c^2)]^{1/2},\\
\gm_\tts(z)
	\hspace{-6pt}&\equiv
		- bcx_1 + 1 - (-1)^\tts \rho
	\quad\text{for }\tts\in\set{0,1},
	\text{ with } x_1\equiv \tf12(z+\slf1z).
\end{array}\]
The polynomial transforms simply under permutations of $a,b,c$:
\[
\begin{array}{l}
-\tf12 P_\zro(z,w)+\rho
=-\tf12 P_\one(z,w)-\rho
=1-bc x_1 -ac x_2 - ab x_3\\
\quad\text{with }
x_1\equiv\tf12(z+\slf{1}{z}), \
x_2\equiv\tf12(w+\slf{1}{w}), \text{ and }
x_3\equiv\tf12(\slf{z}{w}+\slf{w}{z}).
\end{array}
\]

\bpf[Proof of Proposition~\ref{p:fisher.crit} (3.4.6.4 graph)]
Since $\gm_\zro,\gm_\one$ are linear in $x\equiv \tf12(z+\slf1z)$, evaluating at the extremes $x=\pm1$ shows that $\gm_\zro$ is negative on all of $\T$ while $\gm_\one$ is positive (where we have used the easy bound $\rho>1+bc$). If at any $z\in\T$ we have $\be(z)=0$ then clearly the spectral curve cannot intersect the unit torus at this value of $z$.

Now consider the discriminants
$\discriminant_\tts(z) \equiv \gm_\tts(z)^2-\be(z)\be(\slf1z)$
($\tts=\zro,\one$)
which are convex quadratic in $x$:
$\discriminant_\tts$ is minimized over all $x\in\R$
at
\[\begin{array}{rl}
x_\tts^\star
	\hspace{-6pt}&=
	(bc)^{-1} (1+a^2 - (-1)^\tts \rho),
\quad\text{with value}\\
\discriminant_\tts^\star
	\hspace{-6pt}&=
	-a^2 (2+a^2+b^2+c^2 - (-1)^\tts 2 \rho).
\end{array}
\]
Again the result will follow by showing that
for both $\tts=\zro,\one$,
either $\discriminant_\tts^\star\ge0$
or $|x_\tts^\star|\ge1$.
Clearly $x_\one^\star>1$ so it remains to consider $\tts=\zro$.
Suppose
\beq\label{e:hst.xmin}
-1<x^\star_\zro<1,
\quad\text{so that}\quad
1+a^2-bc<\rho<1+a^2+bc.
\eeq
The global minimum $\discriminant_\zro^\star$ has the same sign as
\[(2\rho)^2-(2+a^2+b^2+c^2)^2
= (2abc)^2
	+[ (b+c)^2-a^2 ] [a^2-(b-c)^2],\]
which is clearly positive for $|b-c|\le a\le b+c$.
Also, the lower bound \eqref{e:hst.xmin} on $\rho$ implies
$\discriminant_\zro^\star	> a^2 [ a^2-(b+c)^2]$,
so we also have $\discriminant_\zro^\star>0$ for $a\ge b+c$.
Lastly, we observe that $a<|b-c|$ contradicts the upper bound
\eqref{e:hst.xmin} on $\rho$:
the function
$g(a)\equiv (1+a^2+bc)^2-\rho^2$
is convex quadratic in $A\equiv a^2$,
and evaluating at the extremes $A=(b-c)^2$ and $A=0$ shows that
$g(a)\le0$ for all $a<|b-c|$, giving the contradiction.
\epf

We remark that after submitting our article, David Cimasoni found an interesting
alternate proof that the spectral curve intersects the unit torus in the same
way for the Fisher graph and the 3.4.6.4 graph.

\newcommand{\arXiv}[1]{Preprint, \href{http://arxiv.org/abs/#1}{\texttt{arXiv:#1}}}

\bibliography{refs}
\bibliographystyle{alphaabbr}

\end{document}